\documentclass[useAMS,usenatbib] {mn2e}

\usepackage{times}
\usepackage{graphics,epsfig}
\usepackage{graphicx}
\usepackage{amsmath}
\usepackage{amssymb}
\usepackage{comment}
\usepackage{supertabular}
\usepackage[flushleft]{threeparttable}
\usepackage[titletoc]{appendix}

\newcommand{\mb}  {\ensuremath{\rm M_{B}}}

\newcommand{\nh}  {\ensuremath{\rm N_{HI}}}

\newcommand{\kms}{km~s$^{-1}$}

\newcommand{\ssfr}{\ensuremath{\Sigma_{\rm SFR}}}
\newcommand{\ha}{\ensuremath{{\rm H}\alpha}\ }

\newcommand{\ms}{\ensuremath{\rm M_{\odot}}}

\newcommand{\acc}{\ensuremath{\rm atoms~cm^{-2}}}
\newcommand{\ac}{\ensuremath{\rm cm^{-2}}}

\newcommand{\msyrkpc}{\ensuremath{\rm M_{\odot} yr^{-1} kpc^{-2} }}
\newcommand{\HI}{H{\sc i }}
\newcommand{\tb}{\ensuremath{\rm T_{B}}\ }

\newcommand{\mnras}{MNRAS}
\newcommand{\apj}{ApJ}
\newcommand{\apjs}{ApJS}
\newcommand{\apjl}{ApJL}
\newcommand{\aap}{A\&A}

\newcommand{\aj}{AJ}

\newcommand{\araa}{ARA\&A}

\newcommand{\bain}{BAN}

\title {Cold \HI in faint dwarf galaxies}
\author [Patra et al.]{
Narendra Nath Patra$^{1}$ \thanks {E-mail:narendra@ncra.tifr.res.in},
	Jayaram N. Chengalur$^{1}$, 
	Igor D. Karachentsev$^{2}$,
	Serafim S. Kaisin$^2$, \and Ayesha Begum$^{3}$ \\ 
	$^{1}$ NCRA-TIFR, Post Bag 3, Ganeshkhind, Pune 411 007, India \\
	$^{2}$ Special Astrophysical Observatory, Nizhniy Arkhyz, Karachai-Cherkessia 369167, Russia \\
	$^{3}$ IISER-Bhopal,ITI Campus (Gas Rahat)Building,Govindpura, Bhopal - 23, India.
}
\date {}
\begin {document}
\maketitle
\label{firstpage}

\begin{abstract}
We present the results of a study of the amount and distribution of cold atomic gas, as well its correlation with recent star formation in a sample of extremely faint dwarf irregular galaxies. Our sample is drawn from the Faint Irregular Galaxy GMRT Survey (FIGGS) and its extension, FIGGS2. We use two different methods to identify cold atomic gas. In the first method, line-of-sight \HI spectra were decomposed into multiple Gaussian components and narrow Gaussian components were identified as cold \HI. In the second method, the brightness temperature (\tb) is used as a tracer of cold \HI. We find that the amount of cold gas identified using the \tb method is significantly larger than the amount of gas identified using Gaussian decomposition. We also find that a large fraction of the cold gas identified using the \tb method is spatially coincident with regions of recent star formation, although the converse is not true. That is  only a small fraction of the regions with recent star formation are also covered by cold gas. For regions where the star formation and the cold gas overlap, we study the relationship between the star formation rate density and the cold \HI column density. We find that the star formation rate density has a power law dependence on the \HI column density, but that the slope of this power law is significantly flatter than that of the canonical Kennicutt-Schmidt relation. 
\end{abstract}


\section{Introduction}

The ISM in galaxies is multi-phase, consisting of  molecular, atomic and ionised phases. Early observations of the HI 21cm emission/absorption spectra in our galaxy showed that the atomic gas itself has multiple phases \citep{clark65,radhakrishnan72}, a clumpy dense phase with low spin temperature and a more diffuse widely distributed phase with a high spin temperature.
Theoretically it has been established that in steady state (i.e. when the heating rate is balanced by the cooling rate) the gas settles into one of two stable phases, a cold dense phase (the ``Cold Neutral Medium'' or CNM) and a warm diffuse phase (the ``Warm Neutral Medium'' or WNM) \citep{field65,field69b,wolfire95a,wolfire03}. In these models, gas at temperatures intermediate between that of the stable CNM and WNM is subject to thermal instability and is expected to settle into one of the two stable phases.

Gas in the CNM has small velocity dispersion and large 21cm optical depths, and is easily detected in absorption. The properties of the CNM in our galaxy are hence  quite well established by emission/absorption studies. On the other hand the WNM has a very low optical depth, large velocity dispersion, is extremely challenging to detect in absorption, and its properties hence remain observationally not very established. Theoretical studies however indicate that the  timescales for gas that has been disturbed away from the stable WNM conditions to settle into the WNM can be quite large. Sensitive emission absorption studies done over the last decade, aimed partly at detecting the WNM in absorption \citep[e.g.][]{heiles03a,heiles03b,kanekar03b,roy13a,roy13b} have resulted in the somewhat surprising conclusion  that a reasonable fraction of the atomic ISM in our galaxy could be out of thermal equilibrium, i.e. have temperatures that are intermediate between that of the CNM and WNM. Similarly numerical simulations of gas in a turbulent medium \citep[e.g.][]{audit05} have also shown that in steady state a reasonable fraction of the gas could have temperatures in the unstable range.

In the case of nearby dwarf galaxies, most earlier studies of the phase structure of the gas have been done in the context of the classical two phase models and have focused on using the velocity dispersion to distinguish between the CNM and WNM. \citet{younglo97,young03,begum06b,warren12} all analysed the spatially resolved  line-of-site \HI emission spectra of a number of nearby dwarf galaxies to look for narrow velocity dispersion gas. The narrow velocity dispersion gas was identified via multi-Gaussian fitting to the observed spectra. Components with small velocity dispersion were identified as arising from the CNM, whereas components with large velocity dispersion were identified with the WNM. 

In our own galaxy, it has observationally been established that large optical depths (implying presence of the CNM) can be found along all lines of sight where the HI 21cm brightness temperature exceeds some threshold value \citep{braun92,roy13a}. This can be understood in terms of a maximum brightness temperature that the WNM can produce.
As discussed in more detail below, because of its low optical depth, there is a limit on how large a brightness temperature one can get from physically reasonably sized path lengths through the WNM.
The presence of high brightness temperature gas can hence be used to locate dense cold gas in external galaxies \citep[e.g.][]{braun97}. Since the CNM is clumpy, this requires sensitive high spectral and velocity resolution observations. In this paper we use these two different methods to detect cold \HI in our sample of galaxies and try to see its correlation with star formation. 

\section{Sample \& Data reduction}

The sample of galaxies we use is drawn from the Faint Irregular
Galaxy GMRT Survey (FIGGS) \citep{begum08c}, and FIGGS2 survey (Patra et al. 2015 (submitted)). These two samples contain a total of 73  galaxies with HI interferometric images.  As described in more detail below, for detecting CNM the galaxies needed to be imaged at a specific linear resolution.  Out of the original sample of 73 galaxies, 11 galaxies could not be imaged at the required resolution, mainly because of signal to noise ratio related issues.  The final sample hence  consists of 62 galaxies.  The median HI mass and median blue magnitude (M$_{\rm B}$) of the sample are $10^{7.3}$\ms~and -12.7 respectively. Some general properties of the sample galaxies  are listed in Tab.~\ref{tab:sample}. The columns of the table are:
col(1): Name of the galaxy,
col(2): Other major surveys as part of which the galaxy was observed in \HI,
col(3): Right Ascention  (J2000),
col(4): Declination (J2000),
col(5): the distance to the galaxy in Mpc (taken from \citep{karachentsev13})
col(6): the method by which the distance is estimated,
col(7): The absolute blue magnitude (\mb) (again taken from \citep{karachentsev13})
col(8): The log of \HI mass of the galaxy,
col(9): Optical inclination of the galaxy,
column(10) indicates if the galaxy is included in the sample used for detecting the CNM via the Gaussian Decomposition method (see~Sec.\ref{sec:GDmethod}), 
column(11) indicates if the galaxy is included in the sample for detecting the CNM via the  Brightness Temperature (\tb) method (see~Sec.\ref{sec:TBmethod}), 
col(12) the velocity resolution used to image the galaxy at $\sim$ 100 pc (as required for the \tb method).

As discussed above we use two different methods to try and identify gas in the CNM phase, viz. the Gaussian Decomposition method and Brightness Temperature (\tb) method. As shown in Sec.~\ref{sec:GDmethod}, the Guassian  Decomposition method requires spectra with high signal to noise ratio. For  this we use data cubes at a uniform linear resolution of $\sim 400$pc. Data cubes at higher linear resolutions have a signal to noise ratio that is too low for the Gaussian decomposition method to work reliably. On the other hand, as discussed in  Sec.~\ref{sec:TBmethod}, the brightness temperature method is well suited to  high resolution data cubes, and does not require as high signal to noise  ratio spectra. We hence use data cubes made at $\sim \ 100$pc linear resolution for finding cold gas using the brightness temperature method. Some galaxies from the parent sample did not have sufficient signal to noise ratio (SNR) in the low resolution $\sim 400$pc data cubes, while some galaxies did not have 
sufficient snr in the $\sim 100$pc data cubes. These galaxies were hence dropped from the sub-sample used for that particular method. Details of  the actual galaxies that were used for each method are also given in  Tab.~\ref{tab:sample}.

\begin{table*}
\caption[Sample Galaxies]{General properties of our sample galaxies.}
\begin{threeparttable}
\begin{tabular}{lccccccccccc}
\hline
 Name & Other surveys$^*$ & RA & DEC & Dist & Method & \mb & $\log M_{HI}$ & $\rm i_{opt}$ & Gaussian & \tb & $\Delta$v (\tb) \\
  & & (J2000) & (J2000) & (Mpc) & & (mag) & (\ms) & ($^o$) &  &  & (\kms) \\ 
\hline
DDO226 & $\rm -$ & $004303.8$ & $-221501$ & $4.92$ & $\rm TRGB$ & $-13.6$ & $7.53$ & $\rm 90$ & $\rm Y$ & $\rm N$ & $-$ \\ 

DDO006 & $\rm 2$ & $004949.3$ & $-210058$ & $3.34$ & $\rm TRGB$ & $-12.4$ & $7.04$ & $\rm 90$ & $\rm N$ & $\rm N$ & $3.3$ \\ 

UGC00685 & $\rm -$ & $010722.3$ & $+164102$ & $4.51$ & $\rm TRGB$ & $-14.3$ & $7.74$ & $\rm 51$ & $\rm Y$ & $\rm Y$ & $6.6$ \\ 

KKH6 & $\rm -$ & $013451.6$ & $+520530$ & $3.73$ & $\rm TRGB$ & $-12.4$ & $6.63$ & $\rm 62$ & $\rm Y$ & $\rm Y$ & $3.3$ \\ 

AGC112521 & $\rm 3$ & $014107.9$ & $+271926$ & $6.58$ & $\rm TRGB$ & $-11.6$ & $6.82$ & $\rm 67$ & $\rm N$ & $\rm N$ & $-$ \\ 

KK14 & $\rm 3$ & $014442.7$ & $+271716$ & $7.20$ & $\rm mem$ & $-12.1$ & $7.56$ & $\rm 90$ & $\rm N$ & $\rm N$ & $-$ \\ 

KK15 & $\rm 3$ & $014641.6$ & $+264805$ & $9.04$ & $\rm TRGB$ & $-11.9$ & $7.24$ & $\rm 90$ & $\rm N$ & $\rm N$ & $-$ \\ 

KKH11 & $\rm -$ & $022435.0$ & $+560042$ & $3.00$ & $\rm mem$ & $-13.3$ & $7.66$ & $\rm 68$ & $\rm Y$ & $\rm Y$ & $1.7$ \\ 

KKH12 & $\rm -$ & $022727.0$ & $+572916$ & $3.00$ & $\rm mem$ & $-13.0$ & $7.23$ & $\rm 90$ & $\rm N$ & $\rm N$ & $1.6$ \\ 

KKH34 & $\rm -$ & $055941.2$ & $+732539$ & $4.61$ & $\rm TRGB$ & $-12.3$ & $6.76$ & $\rm 66$ & $\rm N$ & $\rm Y$ & $3.3$ \\ 

ESO490-017 & $\rm -$ & $063756.6$ & $-255959$ & $4.23$ & $\rm TRGB$ & $-14.5$ & $7.55$ & $\rm 46$ & $\rm N$ & $\rm N$ & $-$ \\ 

KKH37 & $\rm 2$ & $064745.8$ & $+800726$ & $3.39$ & $\rm TRGB$ & $-11.6$ & $6.70$ & $\rm 55$ & $\rm N$ & $\rm N$ & $-$ \\ 

UGC03755 & $\rm -$ & $071351.8$ & $+103119$ & $7.41$ & $\rm TRGB$ & $-15.7$ & $8.25$ & $\rm 63$ & $\rm N$ & $\rm N$ & $-$ \\ 

DDO043 & $\rm 1$ & $072817.2$ & $+404613$ & $5.73$ & $\rm BS$ & $-13.9$ & $7.85$ & $\rm 53$ & $\rm N$ & $\rm N$ & $-$ \\ 

KK65 & $\rm -$ & $074231.2$ & $+163340$ & $8.02$ & $\rm TRGB$ & $-14.3$ & $7.70$ & $\rm 66$ & $\rm N$ & $\rm N$ & $-$ \\ 

UGC04115 & $\rm -$ & $075701.8$ & $+142327$ & $7.73$ & $\rm TRGB$ & $-14.3$ & $8.34$ & $\rm 66$ & $\rm Y$ & $\rm N$ & $-$ \\ 

KK69 & $\rm -$ & $085250.7$ & $+334752$ & $7.70$ & $\rm mem$ & $-12.2$ & $7.51$ & $\rm 46$ & $\rm N$ & $\rm N$ & $-$ \\ 

KKH46 & $\rm -$ & $090836.6$ & $+051732$ & $5.60$ & $\rm h$ & $-11.9$ & $7.28$ & $\rm 34$ & $\rm N$ & $\rm N$ & $-$ \\ 

UGC04879 & $\rm -$ & $091602.2$ & $+525024$ & $1.36$ & $\rm TRGB$ & $-11.9$ & $5.98$ & $\rm 66$ & $\rm N$ & $\rm N$ & $-$ \\ 

UGC05186 & $\rm -$ & $094259.8$ & $+331552$ & $8.30$ & $\rm TF$ & $-13.4$ & $7.38$ & $\rm 90$ & $\rm N$ & $\rm N$ & $-$ \\ 

UGC05209 & $\rm -$ & $094504.2$ & $+321418$ & $6.56$ & $\rm h$ & $-13.1$ & $7.15$ & $\rm 18$ & $\rm N$ & $\rm Y$ & $3.3$ \\ 

UGC05456 & $\rm -$ & $100719.7$ & $+102144$ & $5.60$ & $\rm TRGB$ & $-15.1$ & $7.71$ & $\rm 72$ & $\rm Y$ & $\rm N$ & $-$ \\ 

LeG06 & $\rm -$ & $103955.7$ & $+135428$ & $10.40$ & $\rm mem$ & $-11.9$ & $6.85$ & $\rm 57$ & $\rm N$ & $\rm N$ & $-$ \\ 

KDG073 & $\rm 2$ & $105257.1$ & $+693245$ & $3.70$ & $\rm TRGB$ & $-10.8$ & $6.51$ & $\rm 38$ & $\rm N$ & $\rm N$ & $-$ \\ 

UGC06145 & $\rm -$ & $110535.0$ & $-015149$ & $10.70$ & $\rm mem$ & $-13.9$ & $7.97$ & $\rm 66$ & $\rm N$ & $\rm N$ & $-$ \\ 

UGC06456 & $\rm 1$ & $112800.6$ & $+785929$ & $4.35$ & $\rm TRGB$ & $-14.1$ & $7.77$ & $\rm 69$ & $\rm N$ & $\rm Y$ & $1.6$ \\ 

UGC06541 & $\rm 1$ & $113329.1$ & $+491417$ & $3.89$ & $\rm TRGB$ & $-13.6$ & $6.97$ & $\rm 65$ & $\rm N$ & $\rm Y$ & $3.3$ \\ 

KK109 & $\rm -$ & $114711.2$ & $+434019$ & $4.51$ & $\rm TRGB$ & $-10.3$ & $6.54$ & $\rm 55$ & $\rm N$ & $\rm N$ & $-$ \\ 

DDO099 & $\rm 2$ & $115053.0$ & $+385250$ & $2.64$ & $\rm TRGB$ & $-13.5$ & $7.74$ & $\rm 90$ & $\rm N$ & $\rm N$ & $3.3$ \\ 

ESO379-007 & $\rm -$ & $115443.0$ & $-333329$ & $5.22$ & $\rm TRGB$ & $-12.3$ & $7.51$ & $\rm 49$ & $\rm Y$ & $\rm Y$ & $1.7$ \\ 

ESO321-014 & $\rm -$ & $121349.6$ & $-381353$ & $3.18$ & $\rm TRGB$ & $-12.7$ & $7.21$ & $\rm 83$ & $\rm N$ & $\rm N$ & $-$ \\ 

UGC07298 & $\rm -$ & $121628.6$ & $+521338$ & $4.21$ & $\rm TRGB$ & $-12.3$ & $7.28$ & $\rm 67$ & $\rm Y$ & $\rm Y$ & $1.7$ \\ 

VCC0381 & $\rm -$ & $121954.1$ & $+063957$ & $4.71$ & $\rm h$ & $-11.7$ & $7.14$ & $\rm 26$ & $\rm N$ & $\rm N$ & $-$ \\ 

KK141 & $\rm -$ & $122252.7$ & $+334943$ & $7.78$ & $\rm h$ & $-12.9$ & $7.20$ & $\rm 45$ & $\rm N$ & $\rm N$ & $-$ \\ 

IC3308 & $\rm -$ & $122517.9$ & $+264253$ & $12.80$ & $\rm TF$ & $-15.5$ & $8.57$ & $\rm 77$ & $\rm N$ & $\rm N$ & $-$ \\ 

KK144 & $\rm -$ & $122527.9$ & $+282857$ & $6.15$ & $\rm h$ & $-12.5$ & $7.86$ & $\rm 81$ & $\rm N$ & $\rm N$ & $-$ \\ 

DDO125 & $\rm 2$ & $122741.8$ & $+432938$ & $2.74$ & $\rm TRGB$ & $-14.3$ & $7.48$ & $\rm 66$ & $\rm N$ & $\rm Y$ & $1.7$ \\ 

UGC07605 & $\rm -$ & $122839.0$ & $+354305$ & $4.43$ & $\rm TRGB$ & $-13.5$ & $7.32$ & $\rm 49$ & $\rm Y$ & $\rm Y$ & $3.3$ \\ 

KK152 & $\rm -$ & $123324.9$ & $+332105$ & $6.90$ & $\rm TF$ & $-13.0$ & $7.54$ & $\rm 83$ & $\rm N$ & $\rm N$ & $-$ \\ 

UGCA292 & $\rm 1,2$ & $123840.0$ & $+324600$ & $3.61$ & $\rm TRGB$ & $-11.8$ & $7.44$ & $\rm 52$ & $\rm N$ & $\rm Y$ & $1.7$ \\

BTS146 & $\rm -$ & $124002.1$ & $+380002$ & $8.50$ & $\rm TF$ & $-12.2$ & $6.97$ & $\rm 67$ & $\rm N$ & $\rm N$ & $-$ \\ 

LVJ1243+4127 & $\rm -$ & $124355.7$ & $+412725$ & $6.09$ & $\rm h$ & $-11.8$ & $7.02$ & $\rm 83$ & $\rm N$ & $\rm N$ & $-$ \\ 

KK160 & $\rm -$ & $124357.4$ & $+433941$ & $4.31$ & $\rm TRGB$ & $-10.9$ & $6.59$ & $\rm 47$ & $\rm N$ & $\rm N$ & $-$ \\ 

UGC08055 & $\rm -$ & $125604.0$ & $+034841$ & $13.00$ & $\rm TRGB$ & $-13.7$ & $8.89$ & $\rm 36$ & $\rm N$ & $\rm N$ & $-$ \\

GR8 & $\rm 1,2$ & $125840.4$ & $+141303$ & $2.13$ & $\rm TRGB$ & $-12.0$ & $6.89$ & $\rm 27$ & $\rm N$ & $\rm Y$ & $1.7$ \\ 

UGC08215 & $\rm -$ & $130803.6$ & $+464941$ & $4.55$ & $\rm TRGB$ & $-12.3$ & $7.27$ & $\rm 52$ & $\rm Y$ & $\rm Y$ & $3.3$ \\ 

DDO167 & $\rm 1$ & $131322.8$ & $+461911$ & $4.19$ & $\rm TRGB$ & $-12.7$ & $7.19$ & $\rm 67$ & $\rm N$ & $\rm N$ & $-$ \\ 

KK195 & $\rm -$ & $132108.2$ & $-313147$ & $5.22$ & $\rm TRGB$ & $-11.8$ & $7.56$ & $\rm 78$ & $\rm N$ & $\rm N$ & $-$ \\ 

KK200 & $\rm -$ & $132436.0$ & $-305820$ & $4.63$ & $\rm TRGB$ & $-12.0$ & $6.84$ & $\rm 60$ & $\rm Y$ & $\rm Y$ & $1.7$ \\ 

UGC08508 & $\rm 1,2$ & $133044.4$ & $+545436$ & $2.69$ & $\rm TRGB$ & $-13.1$ & $7.30$ & $\rm 63$ & $\rm N$ & $\rm Y$ & $1.7$ \\ 

UGCA365 & $\rm -$ & $133630.8$ & $-291411$ & $5.25$ & $\rm TRGB$ & $-13.3$ & $7.26$ & $\rm 87$ & $\rm N$ & $\rm N$ & $-$ \\ 

UGC08638 & $\rm -$ & $133919.4$ & $+244633$ & $4.27$ & $\rm TRGB$ & $-13.7$ & $7.08$ & $\rm 55$ & $\rm Y$ & $\rm N$ & $-$ \\ 

DDO181 & $\rm 2$ & $133953.8$ & $+404421$ & $3.01$ & $\rm TRGB$ & $-13.2$ & $7.33$ & $\rm 65$ & $\rm N$ & $\rm N$ & $-$ \\ 

IC4316 & $\rm -$ & $134018.1$ & $-285340$ & $4.41$ & $\rm TRGB$ & $-13.9$ & $7.05$ & $\rm 59$ & $\rm N$ & $\rm N$ & $-$ \\ 

DDO183 & $\rm 2$ & $135051.1$ & $+380116$ & $3.22$ & $\rm TRGB$ & $-13.2$ & $7.31$ & $\rm 90$ & $\rm N$ & $\rm N$ & $1.7$ \\ 

KKH86 & $\rm 2$ & $135433.6$ & $+041435$ & $2.59$ & $\rm TRGB$ & $-10.3$ & $5.91$ & $\rm 51$ & $\rm N$ & $\rm N$ & $-$ \\ 

UGC08833 & $\rm 2$ & $135448.7$ & $+355015$ & $3.08$ & $\rm TRGB$ & $-12.2$ & $7.00$ & $\rm 30$ & $\rm N$ & $\rm Y$ & $1.7$ \\ 

DDO187 & $\rm 1,2$ & $141556.5$ & $+230319$ & $2.20$ & $\rm TRGB$ & $-12.4$ & $7.06$ & $\rm 46$ & $\rm N$ & $\rm Y$ & $3.3$ \\ 

PGC051659 & $\rm -$ & $142803.7$ & $-461806$ & $3.58$ & $\rm TRGB$ & $-13.1$ & $7.78$ & $\rm 90$ & $\rm N$ & $\rm N$ & $1.7$ \\ 

KK246 & $\rm -$ & $200357.4$ & $-314054$ & $7.83$ & $\rm TRGB$ & $-13.7$ & $8.07$ & $\rm 87$ & $\rm N$ & $\rm N$ & $-$ \\ 

DDO210 & $\rm 1$ & $204651.8$ & $-125053$ & $0.94$ & $\rm TRGB$ & $-11.1$ & $6.42$ & $\rm 72$ & $\rm N$ & $\rm N$ & $1.6$ \\ 

KKH98 & $\rm 2$ & $234534.0$ & $+384304$ & $2.45$ & $\rm TRGB$ & $-10.8$ & $6.45$ & $\rm 67$ & $\rm N$ & $\rm Y$ & $3.3$ \\ 

\hline
\end{tabular}
\label{tab:sample}
\begin{tablenotes}
\item[*] Survey notations , 1 = LITTLE-THINGS \citep{hunter12} , 2 = VLA-ANGST \citep{ott12} , 3 = SHIELD \citep{cannon11}
\end{tablenotes}
\end{threeparttable}

\end{table*}

\section{Searching for the CNM using Gaussian decomposition}
\label{sec:GDmethod}

Classical models for the phase structure of the atomic ISM in the Milkyway and nearby galaxies indicate the existence of two stable phases with kinetic temperatures $\sim 5000-8000$~K, i.e. the ``Warm Neutral Medium'' (WNM) and $\sim 50-300$~K, i.e. the 
``Cold Neutral Medium'' (CNM) respectively \citep{wolfire95b}. Recent observations and models suggest that in addition, there could be a significant amount of gas with kinetic temperatures that lie between $500-5000$~K, i.e. in the range that is thermally unstable \citep[see e.g.][]{heiles03,kanekar03a,roy13,kim14}. We follow \citet{chengalur13} in calling this as the ``Unstable Neutral 
Medium'' (UNM).

 In the Gassian decomposition method, one decomposes the line of sight velocity profile into multiple Gaussians, and identifies components with velocity widths corrresponding to kinetic temperatures in the CNM range as the cold neutral medium. Gas at a temperature of 200~K would have a one dimensional thermal velocity dispersion of $\sim 1.28$~\kms~or a FWHM of  $\sim 3.1$~\kms. However, most previous studies of the CNM in dwarf galaxies \citep[see e.g.][]{younglo97,blok06,warren12} have used significantly larger 
velocity dispersions  (up to $\sim$ 6 \kms) to identify CNM gas. We note that a velocity dispersion of $\sim 6$~\kms~corresponds to kinetic temperatures of $\sim$ 4400~K. While this temperature is significantly higher than the maximum expected for the CNM, it is nonetheless lower than the expected kinetic temperature for the WNM. In the absense of non thermal contributions to the velocity width, gas with velocity dispersion $\lesssim 6$~\kms~would correspond to gas in phases other than the WNM, i.e. the CNM or UNM.

In the presence of non-thermal motions however, the situation is more complex (see \citet{roy13a} for a detailed discussion). The 
simplest case is if  one assumes that the gas is iso-thermal and that the only non-thermal motions are micro-turbulence and that the net effect of these motions is to keep the profile shape as a Gaussian, albiet with a larger velocity width than that corresponding to the kinetic temperature.  In this case, the velocity width of the profile provides an upper limit to the kinetic temperature, and gas with velocity dispersion of $\sim 6$~\kms~ must be at a kinetic temperature lower than that corresponding to the WNM. However if the gas has a range of temperatures, or if because of bulk motions the line profile is intrinsically non-Gaussian, then it is not clear what the components derived from a Gaussian decomposition physcially correspond to. In this section, we follow earlier studies (i.e. those cited above, as well as galactic studies such as \citet[][]{shuter64,heiles03}) in assuming that the velocity width 
obtained from Gaussian decomposition can be used as a proxy for the kinetic temperature, but we return to the issues raised above in 
Sec.~\ref{sec:TBmethod}.

\subsection{Implementation }

\begin{figure}
\begin{center}
\resizebox{90mm}{!}{\includegraphics{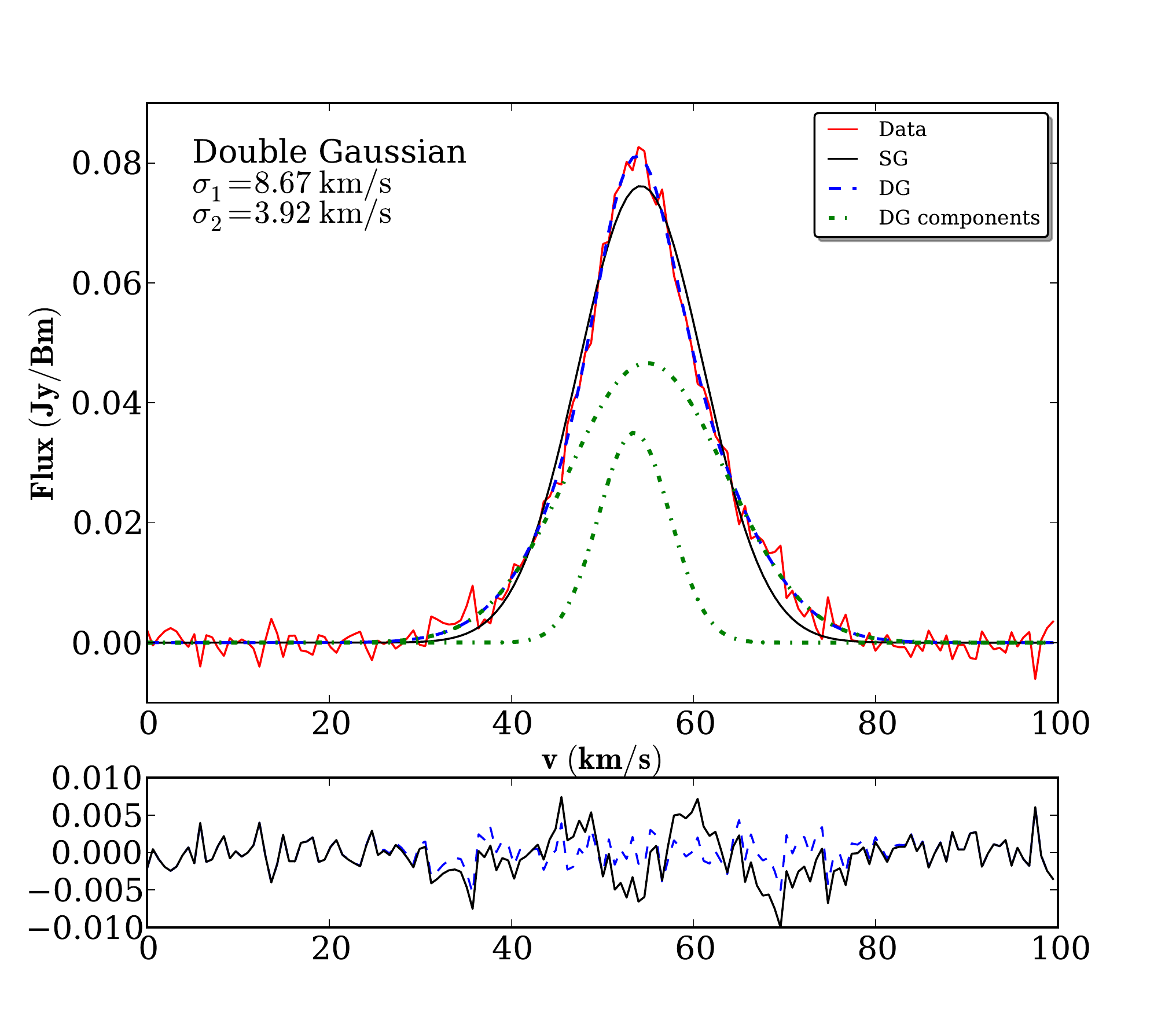}}
\end{center}
\caption{Example fits at SNR=20. The (simulated) input data (red solid noisy line),
a single Gausian fit (black solid smooth line), double Gausian fit (blue dashed line) as
well as the inidividual components of the double Gaussian fit (dotted lines)
are shown. The residuals after subtracting the fits from the data are also 
shown (dashed line residual from the double Gaussian fit, solid line 
residual from the single Gausian fit). The routine detects a double 
gaussian fit as the best fit. The velocity dispersions for the double 
Gaussian fit are $\sigma_1 = 8.67$ \kms and  $\sigma_2 = 3.92$ \kms, 
which match the true input values (8 \kms and 4 \kms) very well.}
\label{fig:gaussexample}
\end{figure}

Line-of-sight \HI spectra were extracted from the 400~pc image cubes and decomposed into multiple Gaussian components using a
python routine {`\tt multigauss'} that was developed by us for this purpose. {`\tt multigauss'} uses a non-linear optimization routine as implemented using the  {`\tt lmfit'} python package.

Starting with a single Gaussian component, Gaussian components were 
successively added to the model till there is no improvement in the fit  residuals as interpreted by a single-tailed F-test.  This is similar to the procedure followed earlier by \cite{warren12}. A single-tailed F-test estimates the probability (or confidence level) that two fit residuals are different just by random chance. Two residuals, one with N components and another with N+1 components are compared by single-tailed F test and the question asked is: what is the confidence level with which one can reject the hypothesis that 
 the residual of fit with N+1 components is better than the residual of fit with N components purely by chance. This confidence level is referred as F-test confidence level of rejecting the null hypothesis.

 Narrow components  returned by the routine were accepted as representing cold gas if the following critea are met. (1)~The component must have a velocity dispersion 
$\mathrm{\sigma \ \lesssim \ 6}$ \kms, (2)~the amplitude of the narrow 
component must be at least 3 times the per channel rms noise, and (3)~the velocity dispersion of the component must be greater than the 
velocity resolution of the data by at least the standard error returned by the fitting routine. Conditions (2) and (3) are meant to reduce the possibility of falsely identifying a narrow noise spike as a genuine narrow component. There was no restriction placed on the central velocity of the components.  These criteria are very similar to those used in past studies, e.g. \citep{younglo97,blok06,warren12}. One important difference between the current routine and those used in previous studies is that previous studies limited the maximum number of Gaussian components per spectrum to 2, where as in the current decomposition the total number of components to be fit
has been left as a free parameter. All of our spectra are however well 
described by single or double Gaussian, with no spectra having more than two components in the best fit model.

In Fig.~\ref{fig:gaussexample} we show an example of the decomposition done by the {\tt 'multigauss'} routine. The input simulated spectrum has an  SNR = 20 with two Gaussian components of width 4 \kms and 8 \kms. From Fig.~\ref{fig:gaussexample} one can see that the routine correctly finds a double Gaussian fit better describes the spectra than a single Gaussian fit. Our routine also correctly recovers the velocity widths of two different input components. See Figure caption for more details.

\subsubsection{Characterization of the routine}

\begin{figure*}
\begin{center}
\begin{tabular}{cc}
\resizebox{9cm}{!}{\includegraphics{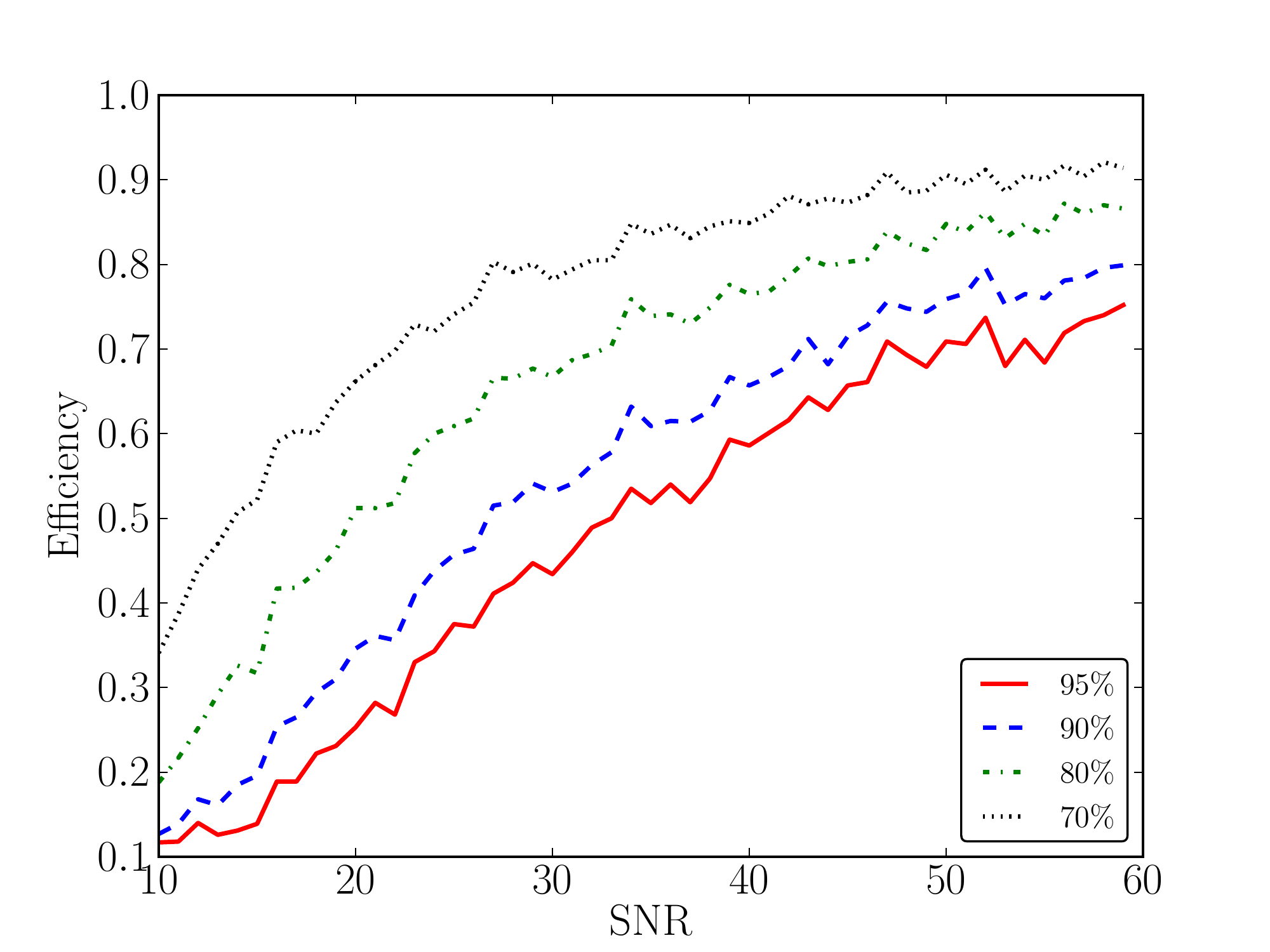}} 
\resizebox{9cm}{!}{\includegraphics{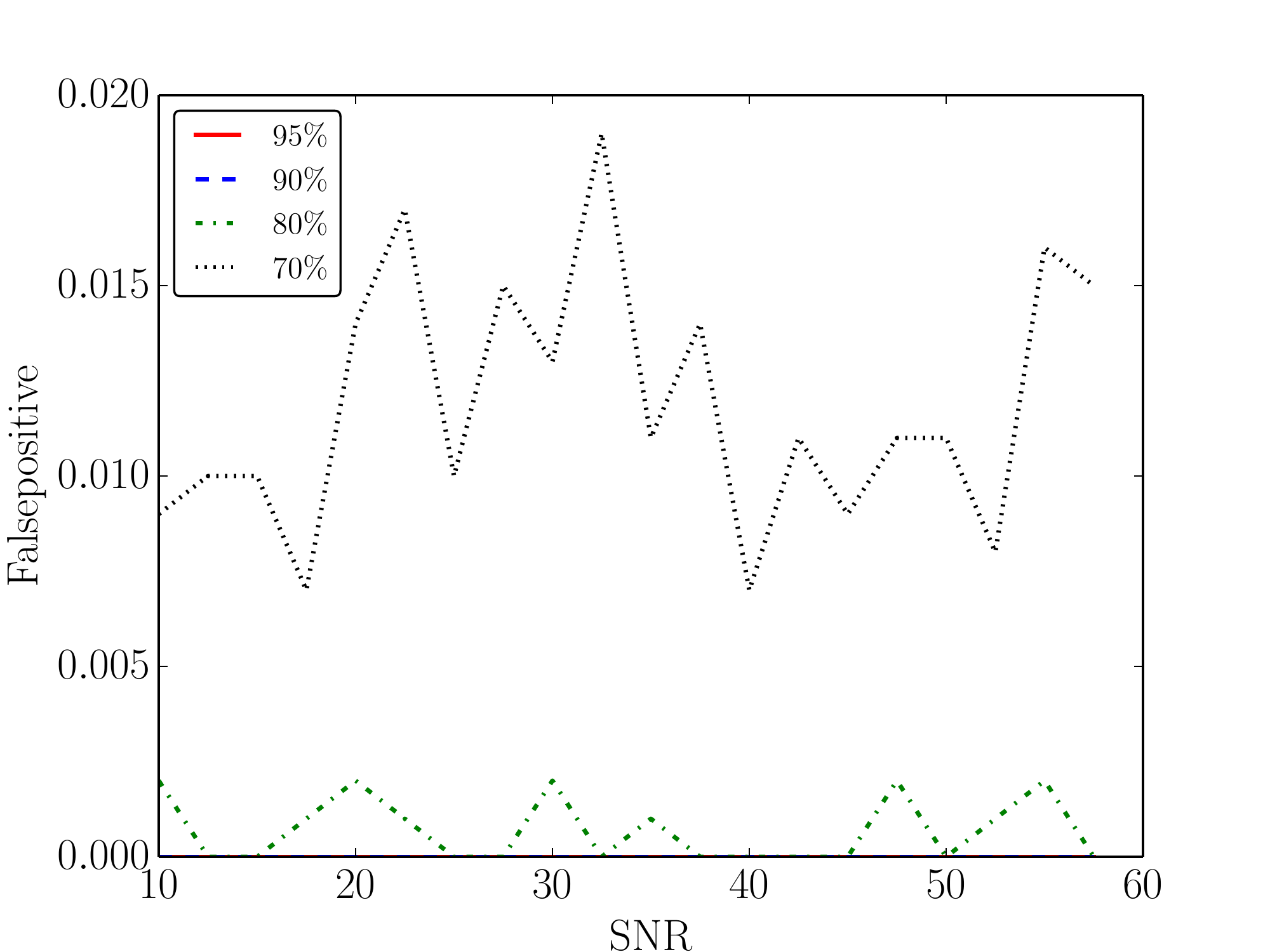}} \\
\end{tabular}
\end{center}
\caption{The recovery efficiency of our routine as a function of 
            the signal to noise ratio (SNR) at different F-test confidence 
            levels.
         Right panel: False detection rate of our routine as a function of 
             the SNR at different F-test confidence limits.}
\label{fig:efficiency}
\end{figure*}

To characterize the detection efficiency of our routine {\tt multigauss}, 50,000 model spectra were simulated having global SNR distributed uniformly between 10 $\textless$ SNR $\textless$ 60. Each spectrum consisted of a narrow component with $\Delta v ~\textless ~\sigma ~\textless ~6.0 $ \kms, (where $\Delta v$ is set 1.65 \kms, i.e. corresponding to the velocity resolution of the observed data) and a broad component with $8.0 ~\textless ~\sigma ~\textless ~12.0 $ \kms. Each of the components was required to have a SNR (i.e. peak amplitude) more than 3. Adopting a conservative approach, the 
centroids of two components were set to be the same, as any offset between these two would lead to a higher detection efficiency. As mentioned above, no spectra required more than two components for the 
best fit; the simulated spectra were hence restricted not to have more than two components. These synthetic spectra then passed through the {\tt multigauss} routine. In Figure~\ref{fig:efficiency} (left panel), we plot the recovery efficiency of our routine as a function of SNR. Different curves represent different confidence levels of F-test. 

 To characterize the false detection rate of our routine, we simulated 1000 spectra with only one component with velocity dispersion uniformly distributed between $8.0 \ \textless ~\ \sigma ~\ \textless \ 12.0 $ \kms, added gaussian random noise, and then passed them through {\tt multigauss}. The false positive detection rate measured in this way is shown in Fig.\ref{fig:efficiency} (right panel).  We note that the false positive rate is always less than a few percent for our working range of SNR (i.e. $> 10$).  This is similar to the results found in previous studies \citep[e.g.][]{warren12}. The detection efficiency and detection accuracy  are summarized in Table~\ref{tab:characterization}. The columns in the table are as follows:
Col. (1) the confidence level of the F-test. 
Col. (2) the amplitude ratio of the input and recovered narrow component. 
Col. (3) the mean offset between the centroids of the input and recovered  narrow component. 
Col. (4) The average of the absolute value of the difference between input and the recovered velocity dispersion of the narrow component.
We note that the offset in the recovered quantities are in all cases small compared to the scatter between runs. As such there are no significant trends in the recovered quantities.
\begin{table}
\caption{Characterization of the efficiency and the accuracy of the decomposition routine}
\begin{center}
\begin{tabular}{|l|c|c|c|c|}
\hline
Confidence level & $\mathrm{A_{sim}/A_{extr}}$ & $\mathrm{\Delta b \ (km/s)}$ & $\mathrm{\Delta \sigma \ (km/s)}$ \\
\hline
95\% & 0.97 $\pm$ 0.17 & 0.00 $\pm$ 0.26 & -0.12 $\pm$ 0.57 \\
90\% & 0.98 $\pm$ 0.21 & 0.00 $\pm$ 0.35 & -0.07 $\pm$ 0.57 \\
80\% & 0.99 $\pm$ 0.29 & -0.01 $\pm$ 0.58 & 0.00 $\pm$ 0.60 \\
70\% & 1.02 $\pm$ 0.36 & 0.00 $\pm$ 0.74 & -0.06 $\pm$ 0.69 \\
\hline
\end{tabular}
\end{center}
\label{tab:characterization}
\end{table}


For the rest of the analysis below we adopt a conservative F-test confidence 
level, which we set to 95\%. At this confidence level, the detection 
rate is comparatively small, but the false positive rate is negligible. 
Further, in the cases where a narrow component is recovered, the parameters 
of this Gaussian are also fairly accurately recovered. Using this 
conservative approach implies that the number of identified narrow components
is a lower limit to the true number of such components. Once again the 
approach adopted here is similar to that taken in earlier studies  
\citep[e.g.][]{warren12}, to allow an easy comparison of our results with 
those from earlier work.

\begin{figure*}
\begin{center}
\begin{tabular}{cc}
\resizebox{7.5cm}{!}{\includegraphics{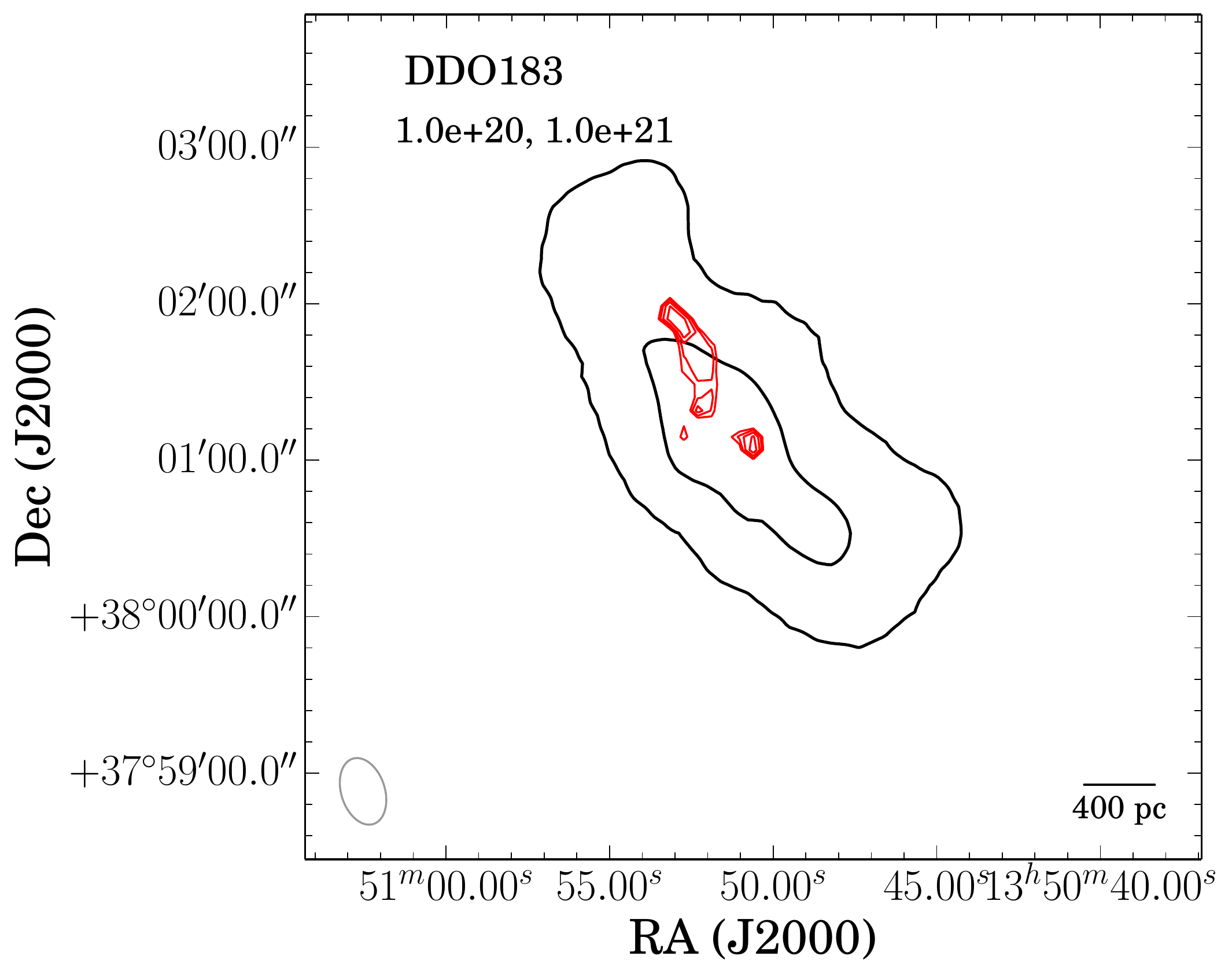}} &
\resizebox{7.5cm}{!}{\includegraphics{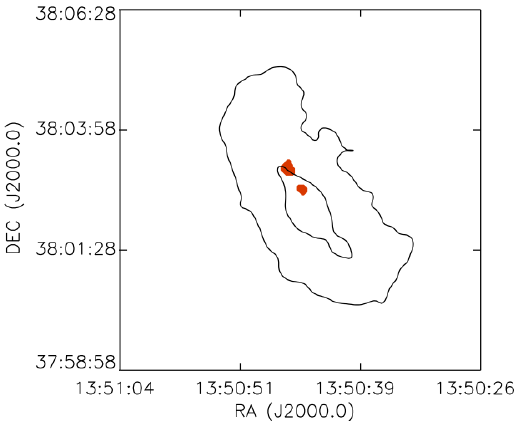}} \\
\end{tabular}
\end{center}
\caption{Left panel: Recovered cold \HI in DDO183 using {\tt 'multigauss'}
                     routine.
         Right panel: Recovered cold \HI using Gaussian decomposition method
                      by \citet{warren12}}
\label{fig:comp_warren}
\end{figure*}

To compare our routine with previous work we show in Fig.~\ref{fig:comp_warren} the regions with narrow velocity components in the galaxy DDO183 as identified using the {\tt 'multigauss'} routine (left panel) as well as the regions identified  by \citet{warren12} (right panel). The black contours in both the panels show the \HI column density at $10^{20}$ and $10^{21}$ \acc. The red contours (left panel) and red region (right panel) marks the locations of detected narrow velocity dispersion \HI. Fig.~\ref{fig:comp_warren} shows that our routine compares reasonably well with previous work. Narrow velocity dispersion \HI is identified at similar locations in spite of the data being taken with two different telescopes (GMRT and JVLA respectively) at two different spatial resolutions (400~kpc and 200~kpc respectively).

We also passed some of the publicly available data from LITTLE-THINGS \citep{hunter12} survey through the {\tt 'multigauss'} routine. The narrow velocity dispersion  detected by {\tt 'multigauss'} matches well with the published results by \citep{warren12}.

\subsection{Results}

 Our parent sample consist of 62 galaxies.
 For galaxies with high inclinations, the observed spectrum is a 
blend of the emission from gas distributed along a relatively long line 
of sight. We hence exclude galaxies with inclination more than $70^o$ 
from the sample. As discussed above we also require a 
peak SNR of $> 10$, and galaxies for which the data cube contained no spectra
with SNR greater than this threshold were also dropped from the sample. After 
these cuts, we are left with a total of 13 galaxies. 

The data for these 13 galaxies were passed through the {\tt multigauss} routine,
which detected narrow components in 5 of them. In Table~\ref{tab:cldhi} we 
summarise the inputs and the results of Gaussian decomposition method. The
columns are as follows:
Col. (1) Name of the galaxy, 
Col. (2) The log of \HI mass of the galaxy, 
Col. (3) The absolute blue magnitude of the galaxy,
Col. (4) The number of lines-of-sight which met the SNR criteria (i.e. SNR $>$10),
Col. (5) Column density at SNR $=$ 10,
Col. (6) the average SNR of the spectra with SNR greater than the threshold,
Col. (7) shows the average column density of the lines-of-sight having SNR 
         greater than 10. 
Col. (8) The peak column density in the galaxy, and 
Col. (9) The amount of narrow velocity dispersion \HI detected.

\begin{table*}
\begin{center}
\caption{Summary of cold \HI search using Gaussian decomposition method}\label{tab:cldhi}
\begin{tabular}{|l|c|c|c|c|c|c|c|c|c|}
\hline
Name & $\rm \log M_{HI}$ & $\rm M_B$ & N & $\rm CD_{10}$ & $\rm \textless SNR \textgreater $ & $\rm \textless CD \textgreater $ & $\rm CD_{peak}$ & $\rm Cold \ M_{HI}$ & $\rm \Delta \ Cold \ M_{HI}$  \\
 & $\rm (M_{\odot})$ & & & ($\rm \times 10^{19}$ \ac) & & ($\rm \times 10^{19}$ \ac) & ($\rm \times 10^{19}$ \ac)  & $\rm (\times 10^{5} \ M_{\odot})$ & $\rm (\times 10^{5} \ M_{\odot})$  \\

\hline
DDO226 & 7.53 & -13.6 & 2 & 10.70 & 10.07 & 10.80 & 10.80 & $\rm -$ & $\rm -$ \\
UGC00685 & 7.74 & -14.3 & 220 & 15.30 & 11.81 & 17.20 & 22.30 & $\rm 8.72$ & $\rm 2.46$ \\
KKH6 & 6.63 & -12.4 & 19 & 9.63 & 10.81 & 10.40 & 11.10 & $\rm -$ & $\rm -$ \\
KKH11 & 7.66 & -13.3 & 716 & 6.01 & 12.65 & 7.22 & 11.50 & $\rm 6.11$ & $\rm 3.72$ \\
UGC04115 & 8.34 & -14.3 & 8 & 18.70 & 10.18 & 19.00 & 19.50 & $\rm -$ & $\rm -$ \\
UGC05456 & 7.71 & -15.1 & 27 & 12.90 & 10.58 & 13.70 & 14.50 & $\rm -$ & $\rm -$ \\
ESO379-007 & 7.51 & -12.3 & 11 & 10.90 & 10.74 & 11.60 & 12.30 & $\rm -$ & $\rm -$ \\
UGC07298 & 7.28 & -12.3 & 1 & 10.40 & 10.01 & 10.40 & 10.40 & $\rm -$ & $\rm -$ \\
UGC07605 & 7.32 & -13.5 & 105 & 11.80 & 12.03 & 13.60 & 18.10 & $\rm 0.17$ & $\rm 0.04$ \\
UGC08215 & 7.27 & -12.3 & 13 & 14.00 & 10.56 & 14.70 & 15.60 & $\rm 1.91$ & $\rm 0.98$ \\
KK200 & 6.84 & -12.0 & 3 & 14.20 & 10.37 & 14.70 & 15.00 & $\rm 0.11$ & $\rm 0.06$ \\
UGC08638 & 7.08 & -13.7 & 16 & 13.40 & 10.30 & 13.80 & 14.20 & $\rm -$ & $\rm -$ \\
\hline
\end{tabular}
\end{center}
\end{table*}

\begin{figure}
\begin{center}
\resizebox{85mm}{!}{\includegraphics{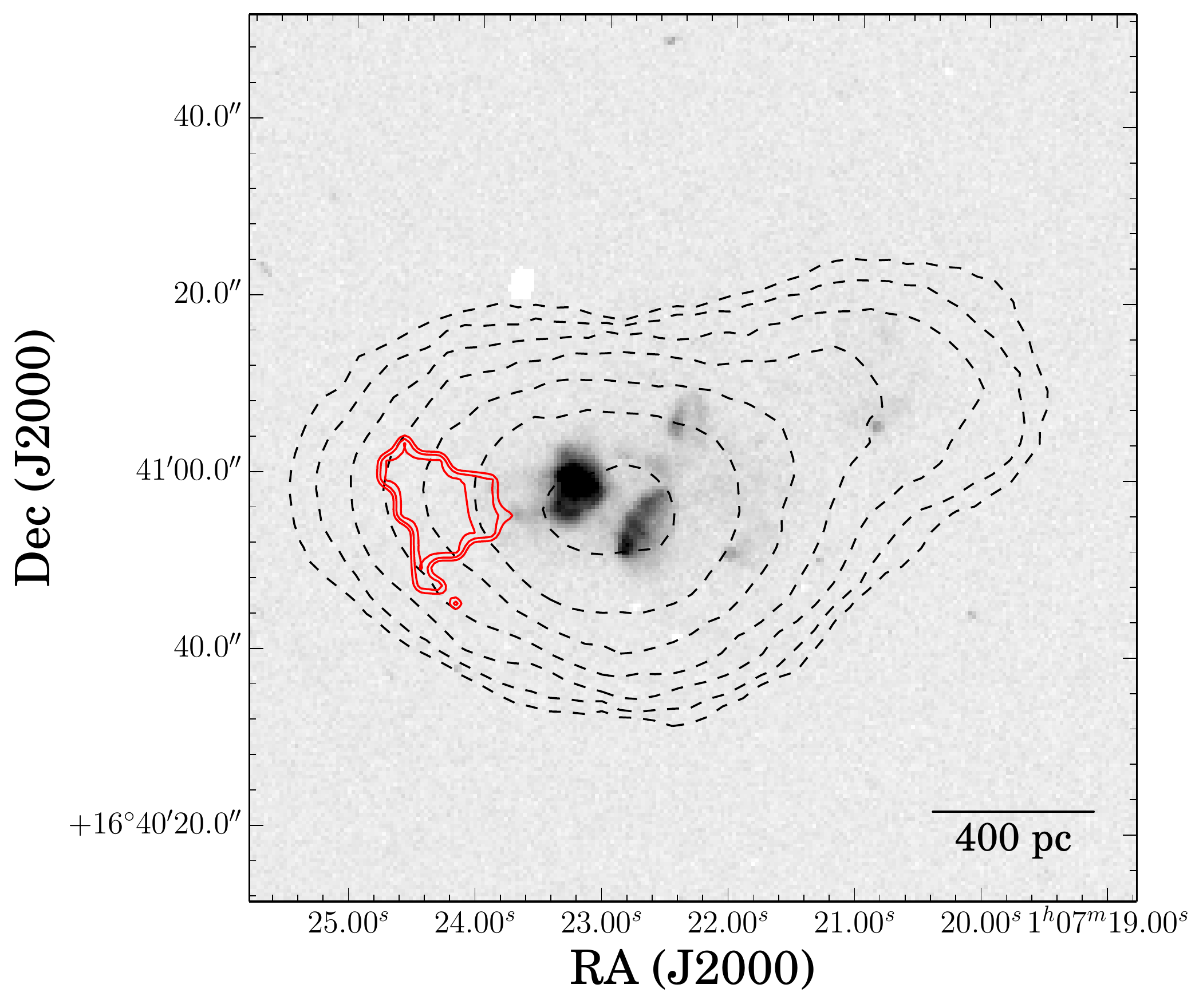}} \\
\end{center}
\caption{Gray scale represents the H$\alpha$ map of U685. Black broken contours represents total \HI column density and the levels are $(1,1.4,2,2.8,4,5.6,8,11.2,16,22.4,...) \ 5 \times 10^{20}$ \acc. Red solid contours show the cold \HI as recovered by Gaussian method. The levels are $(1,1.4,2,2.8,4,5.6,8,11.2,16,22.4,...) \ 6.2 \times 10^{20}$ \acc. The spatial resolution of the grey scale and the overlays are $\sim$ 400 pc.}

\label{fig:u685_gauss}
\end{figure}

Figure~\ref{fig:u685_gauss} shows the narrow velocity dispersion \HI detected 
in the galaxy UGC~685. The dashed line contours show the integrated \HI 
emission, while the $\rm H\alpha$ emission is shown in grey-scales. The 
regions where narrow velocity dispersion \HI were detected is delineated by 
solid lines. 

We note that the narrow velocity dispersion  \HI as detected by the Gaussian decomposition method doesn't coincide with the highest column density regions or with the $\rm H\alpha$ emission. The $\rm H\alpha$ emission originates mainly from regions with recent (i.e. $3 -10$~Myr) star-formation. For all of our sample galaxies, the narrow velocity dispersion \HI is offset from the regions of current star formation. If the CNM is associated with the gas in which the star formation occurs, one would expect that there would be a good correlation between the detected CNM and the H$\alpha$ emission. We note that previous studies \citep{young96,warren12} also observed similar offset between the gas with small velocity dispersion and the regions of current star formation. One possible explanation is that in this context, the Gaussian Decomposition method is not best suited for identifying the CNM in dwarf galaxies. We explore this possibility, as well as other techniques to identify the CNM in the rest of this paper.

\subsection{Drawbacks of Gaussian decomposition method}

As described above we used (following earlier studies of the CNM in nearby dwarf galaxies) Gaussian decomposition to identify narrow velocity dispersion gas. This method has several drawbacks. The first is that, as is well known, Gaussians do not form a orthogonal basis set. Thus any Gaussian decomposition suffers from the problem of not being unique. Further as mentioned above, it is not clear that the components recovered from a given decomposition correspond to separate physical entities. Finally the shapes of the line profiles obtained using observations with finite spatial resolution (as well as integration along the line of sight) are determined by a number of physical processes, including turbulent motions and bulk velocities in addition to the thermal velocity dispersion. In view of these drawbacks, we examine in the next section, an alternate method 
for identifying cold atomic gas in dwarf galaxies.

\section{The Brightness temperature method}
\label{sec:TBmethod}

The observed brightness temperature of  \HI 21cm emission (defined as 
$\rm {T_B = \frac{I_{\nu}c^2}{2k \nu ^2}}$ where \tb is the brightness 
temperature in Kelvin, `$\rm I_{\nu}$' is the specific intensity, `k' is Plank's constant and `c' is the speed of light), has long been used to constrain the physical conditions in the atomic ISM 
\citep[e.g.][]{schmidt57,rohlfs71,baker75,braun92}. In detail, the observed brightness temperature depends on the distribution of the spin temperature,  optical depth as well as spatial ordering of all the gas along the line of sight. In the context of multi-phase models however, the presense of high brightness temperature gas is generally indicative of the presense of a cold dense phase along the line of sight. This is because in these models the warm gas generally has too low an optical depth to contribute more than a few degrees Kelvin to the brightness temperature. Consistent with this, Galactic 
studies show that, there is no detectable \HI absorption along lines of sight where the emission brightness temperature, \tb $\textless$ $3-10$K and almost  all line-of-sights with \tb $\textgreater$ $3-10$K have detectable absorption \citep{braun92,roy13}. \citet{braun92} and \citet{braun97} also show that the \HI emission observations of gas in the Milkyway and M31 can be well fit with models in which the brightness temperature is correlated to the opacity. An alternative method for searching for the CNM would then be to look for gas with brightness temperature much larger than say $\sim 10$K. Indeed this method has previously been used to identify cold, optically thick gas in external galaxies by \citet{braun92,braun97,braun12}. We refer to  this the brigthness
temperature method (or \tb method) for detecting the CNM.

Since gas in the CNM phase is expected to be clumpy, a high resolution 
\HI image is a primary requirement to identify the high density cold \HI in the brightness temperature method. In low resolution images the high \tb regions will be smoothed over, reducing ones ability to detect the compact peaks. The first step in trying this method on our sample galaxies was hence to re-image all of the galaxies at a spatial resolution of $\sim$ 100~pc. Since for many galaxies the SNR is modest at this high resolution, the data were smoothed up to a velocity resolution of $\sim$ 5~\kms~wherever it was necessary (see Tab.~\ref{tab:sample}). We note that the parameters that we use here (i.e. spatial resolution of $\sim 100$~pc and a velocity resolution of $\sim 5$~\kms) are similar to those used by \citep{braun97} to identify high brightness temperature gas.

\begin{figure}
\begin{center}
\resizebox{85mm}{!}{\includegraphics{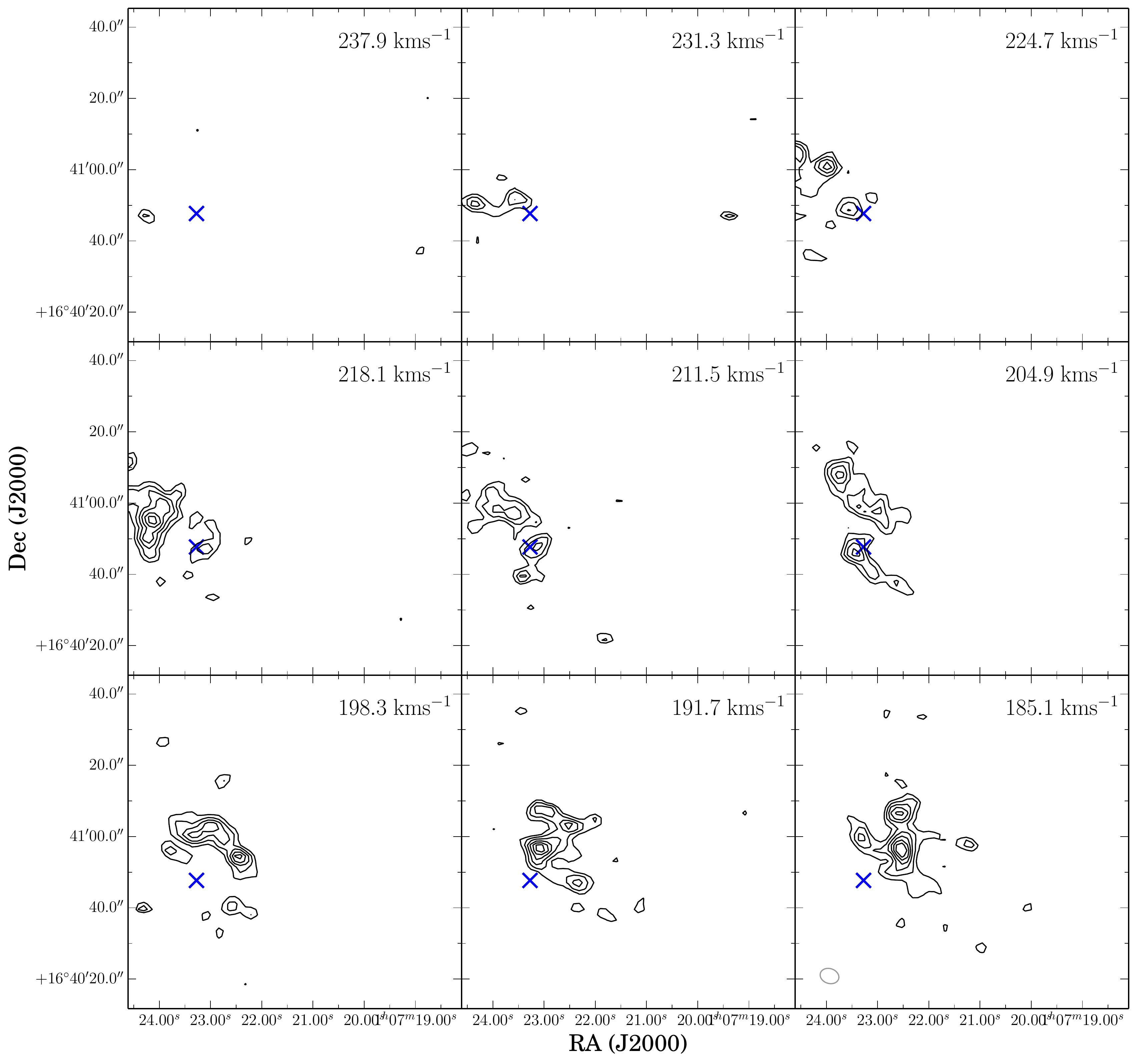}} \\
\end{center}
\caption{The \HI channel maps of the galaxy UGC~685 at a spatial resolution 
of $\sim$ 100 pc. The cross mark spots an emission region, which is seen
over several contigous channels (viz. a velocity range from 231 \kms to 204 \kms). Our moment map parameters were tuned to pick up only
such clearly coherent structures.}
\label{fig:chmap}
\end{figure}

We then inspected all of the data cubes to make sure that any emission
that we detect is present at more than a 2$\sigma$ level in
several contiguous channels (of typically 1.6\kms velocity width).
We show for example in Fig.~\ref{fig:chmap} the channel maps for UGC~685 at a spatial resolution of $\sim$ 100 pc. The cross traces the location of one of the detected emission regions. As can be seen there is emission in velocities from 231 to 204 \kms.  

In the two-phase models of the ISM \citep{wolfire95b} at low metalicity and low dust content (as would be appropriate for our dwarf galaxies), the density of the WNM is expected to be in the range of $0.19-1.9$ atoms~cm$^{-3}$ and the spin temperature in the range of $5900-8400$ K \citep{wolfire95a}. For a region of size
$\sim 100$~pc, the expected optical depth would then be $\lesssim 0.0031$, and the brightness temperature $\lesssim 18$~K.  We hence conservatively set a threshold brightness temperature cutoff of $\tb > 50$~K to identify gas that arises from phases other than the classical WNM. For some galaxies, the per channel $2\sigma$ rms corresponded to \tb larger than 50~K, for these galaxies the $2\sigma$ level was instead used as a cut-off. The actual cut-off used per galaxy is listed in Tab.~\ref{tab:tb_cldhi}. Once we have identified 
gas with high brightness temperature (which we henceforth refer to as ``cold'' gas), the next step is to determine the total \HI mass producing this emission. Two different approaches were adopted to calculate the cold \HI mass from the emission brightness. In first case we assume that the emission is optically thin, in which case the total column density can be trivially computed from the moment map. This provides a lower limit to the cold \HI mass. Our second approach was to assume that the emitting gas has  a spin-temperature of 500~K, which is a robust upper limit in the two phase models \citep{wolfire95b} for low
metalicity gas. We calculated both estimates of the \HI mass for all of galaxies in our sample.

\subsection{Results}

The GMRT has a hybrid configuration \citep{swarup91}, and for most of our observations the UV coverage allows one to make images at spatial resolutions as high as $\sim  4^{\prime \prime}$. For the analysis below we wish to use a uniform linear resolution, which we require to lie within the range $100-150$pc. For 24 of the 62 galaxies in our sample  either there was insufficient signal to noise ratio, or the galaxies were too distant for us to make images with $100-150$pc resolution. These  galaxies were dropped from the sample. A highly inclined galaxy would provide a much longer line-of-sight, leaving a possibility of producing high brightness temperature out of low density gas. To avoid these circumstances we further exclude another 10 galaxies from the sample which have inclination more than 70$^o$. High resolution images were made for the remaining 28 galaxies, and high brightness temperature emission (i.e. ``cold \HI'')  was detected in 19 of them, corresponding to a detection efficiency of $\sim 67\%$. 

\begin{table*}
\begin{center}
\caption{Summary of cold \HI search using \tb method.}\label{tab:tb_cldhi}
\begin{tabular}{|l|c|c|c|c|c|c|c|}
\hline
Name & $\rm \log M_{HI}$ & $\rm M_B$ & $\rm (T_B)_{cut}$ & $\rm (M_{HI})_{thin}$ & $\rm \Delta (M_{HI})_{thin}$  & $\rm (M_{HI})_{500K} $ & $\rm \Delta (M_{HI})_{500K} $ \\
 & $\rm (M_{\odot})$ & & (K) & $\rm (\times 10^{5} M_{\odot})$ & $\rm (\times 10^{5} M_{\odot})$ & $\rm (\times 10^{5} M_{\odot})$ & $\rm (\times 10^{5} M_{\odot})$ \\
\hline
UGC00685 & 7.74 & $-14.3$ & 64 & 315.00 & 102.00 & 354.00 & 104.00 \\
KKH6 & 6.63 & $-12.4$ & 67 & 22.90 & 7.55 & 25.70 & 7.76 \\
KKH11 & 7.66 & $-13.3$ & 50 & 125.00 & 34.20 & 133.00 & 34.50 \\
KKH34 & 6.76 & $-12.3$ & 69 & 5.41 & 1.75 & 6.12 & 1.80 \\
UGC05209 & 7.15 & $-13.1$ & 55 & 14.80 & 5.01 & 16.20 & 5.11 \\
UGC06456 & 7.77 & $-14.1$ & 104 & 19.10 & 6.93 & 22.60 & 7.26 \\
UGC06541 & 6.97 & $-13.6$ & 58 & 31.40 & 10.30 & 34.70 & 10.60 \\
ESO379-007 & 7.51 & $-12.3$ & 51 & 78.70 & 26.20 & 85.60 & 26.60 \\
UGC07298 & 7.28 & $-12.3$ & 64 & 14.60 & 4.80 & 16.30 & 4.92 \\
DDO125 & 7.48 & $-14.3$ & 60 & 3.79 & 1.31 & 4.17 & 1.34 \\
UGC07605 & 7.32 & $-13.5$ & 53 & 63.60 & 19.40 & 70.30 & 19.80 \\
UGCA292 & 7.44 & $-11.8$ & 102 & 173.00 & 58.30 & 209.00 & 61.00 \\
GR8 & 6.89 & $-12.0$ & 50 & 4.64 & 0.77 & 4.90 & 0.77 \\
UGC08215 & 7.27 & $-12.3$ & 57 & 34.30 & 10.70 & 38.20 & 10.90 \\
KK200 & 6.84 & $-12.0$ & 53 & 23.10 & 7.45 & 25.30 & 7.59 \\
UGC08508 & 7.30 & $-13.1$ & 50 & 113.00 & 32.30 & 123.00 & 32.80 \\
UGC08833 & 7.00 & $-12.2$ & 50 & 40.40 & 10.70 & 43.50 & 10.90 \\
DDO187 & 7.06 & $-12.4$ & 139 & 24.80 & 8.34 & 33.40 & 8.91 \\
KKH98 & 6.45 & $-10.8$ & 50 & 4.76 & 1.25 & 5.07 & 1.27 \\
\hline
\end{tabular}
\end{center}
\end{table*}

In Table~\ref{tab:tb_cldhi} we summarise the results of the cold \HI detected using the \tb method. The columns of the table are:
Col.~(1)~the galaxy name. 
Col.~(2)~log of the \HI mass (in solar units)
Col.~(3)~the absolute blue magnitude 
Col.~(4)~the \tb cutoff used to identify cold \HI. We set the cut off to 50K 
or 2$\sigma$, which ever is larger.
Col.~(5)~the mass of  cold \HI estimated assuming the emission to be
optically thin.
Col.~(6)~error in the cold \HI mass (optically thin)
Col.~(7)~the mass of cold \HI estimated assuming the emission to be originated from a gas of spin temperature 500 K.
Col.~(8)~error in the cold \HI mass (assuming a 500 K spin temperature gas) 

\begin{figure}
\begin{center}
\resizebox{85mm}{!}{\includegraphics{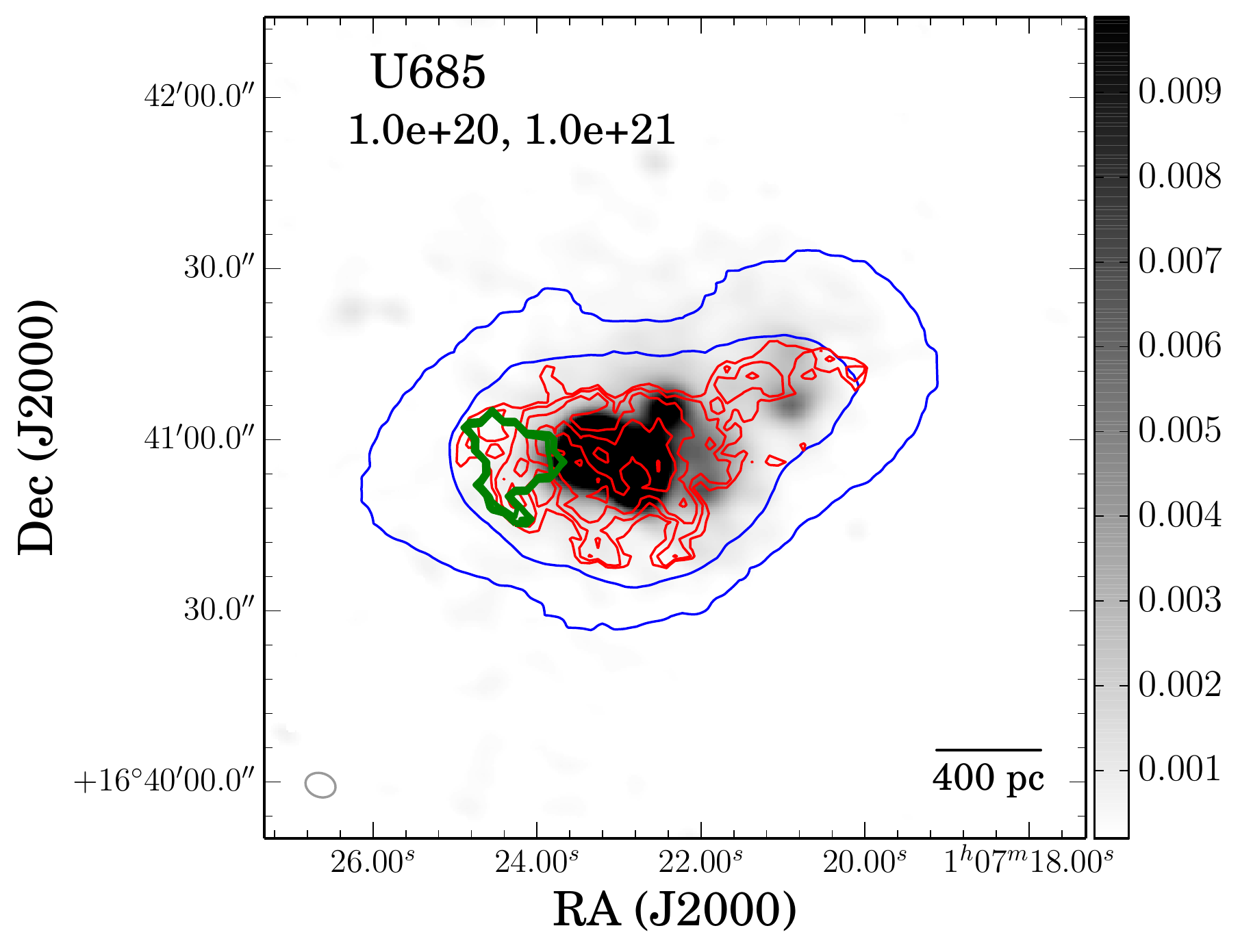}}
\end{center}
\caption{Cold \HI as recovered by two different methods shown in contours overlaid on $\rm H\alpha$ map in gray scale. The red solid contours represent cold \HI recovered by \tb method and the green thick contours are the same for Gaussian decomposition method. The contours levels are $(1,1.4,2,2.8,4,5.6,8,11.2,16,22.4,...) 
\ 1.5 \times 10^{21}$ \acc for cold \HI detected by \tb method (red solid contours) and $(1,1.4,2,2.8,4,5.6,8,11.2,16,22.4,...) \ 6.2 \times 10^{20}$ \acc for cold \HI recovered by Gaussian decomposition method (green thick contours). The spatial resolution of $\rm H\alpha$ data and the \tb data is $\sim$ 100 pc.}
\label{fig:tb_gauss}
\end{figure}

\begin{figure}
\begin{center}
\resizebox{85mm}{!}{\includegraphics{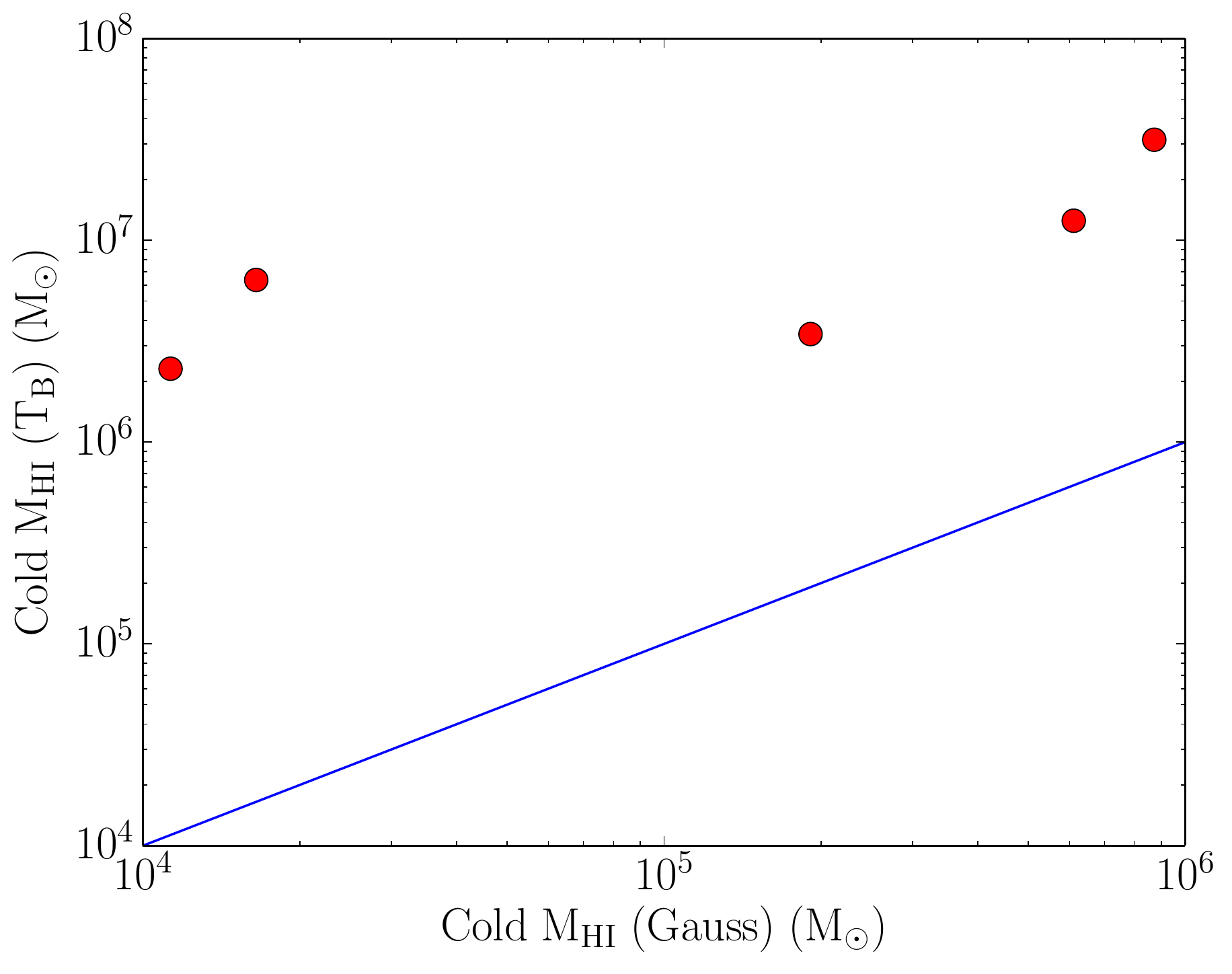}}
\end{center}
\caption{Comparison of cold \HI mass estimated in the two different methods as discussed in the text. The blue solid line represent the 1:1 limit. From the figure it can be seen clearly that \tb method recovers more cold \HI as compared to Gaussian decomposition method, however no obvious trend between the detected cold \HI in \tb method and in Gaussian decomposition method can be found.}
\label{fig:comp_tb_gauss}
\end{figure}

\begin{figure}
\begin{center}
\resizebox{85mm}{!}{\includegraphics{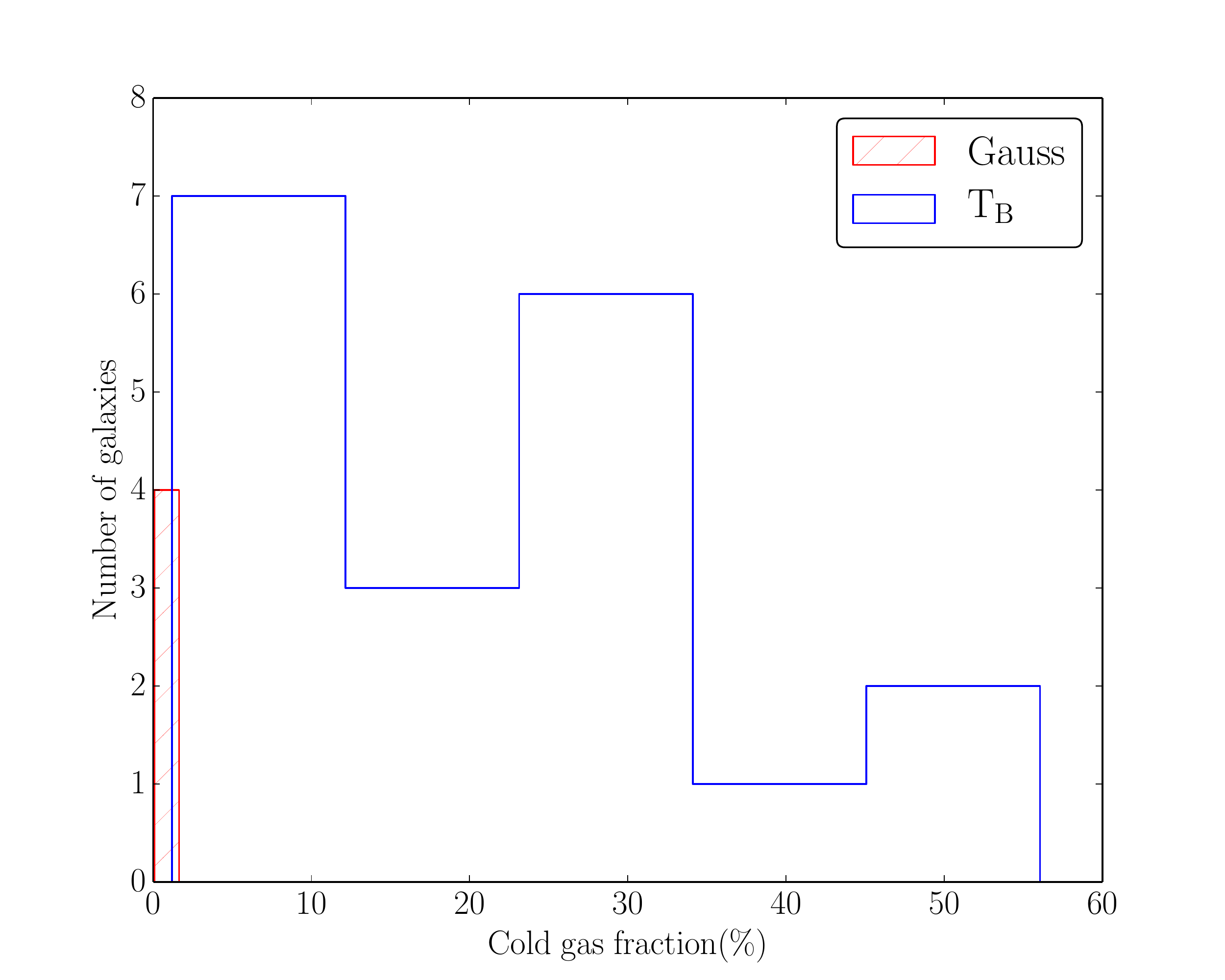}} \\
\end{center}
\caption{ Histograms of cold gas fraction as detected by two different methods, i.e. Gaussian decomposition method and \tb method. As can be seen the \tb method detects much more cold gas as compared to Gaussian decomposition method.
}
\label{fig:cld_frac}
\end{figure}

In Fig.~\ref{fig:tb_gauss} we show the cold \HI as detected by the \tb method as well as the Gaussian decomposition method overlayed  on the \ha\  emission for UGC~685. The red solid contour represents the cold \HI identified using the \tb method, whereas the green thick contours show the cold \HI identified by the Gaussian decomposition method. From the plot one can see that the \tb method identifies gas that correlates with the \ha\ emission better than the Gaussian decomposition method. For an easy comparison of our data and imaging quality as well as the locations of detected cold \HI using different methods, we overplot \HI, cold \HI (Gaussian and \tb method) with different star formation tracers (\ha and FUV) in Appendix (Fig. \ref{ovr_h1_g} to Fig. \ref{ovr_fuv_tb}). In Fig.~\ref{fig:comp_tb_gauss} we compare the cold \HI recovered by both the methods. As can be seen \tb method recovers significantly more `cold' gas than the Gaussian decomposition method.  In Fig.~\ref{fig:cld_frac} we plot the histograms of detected cold gas fraction in both the methods. Once again it is clear that the \tb method identifies much more cold gas than the Gaussian Decomposition method. 

\begin{figure*}
\begin{center}
\resizebox{170mm}{!}{\includegraphics{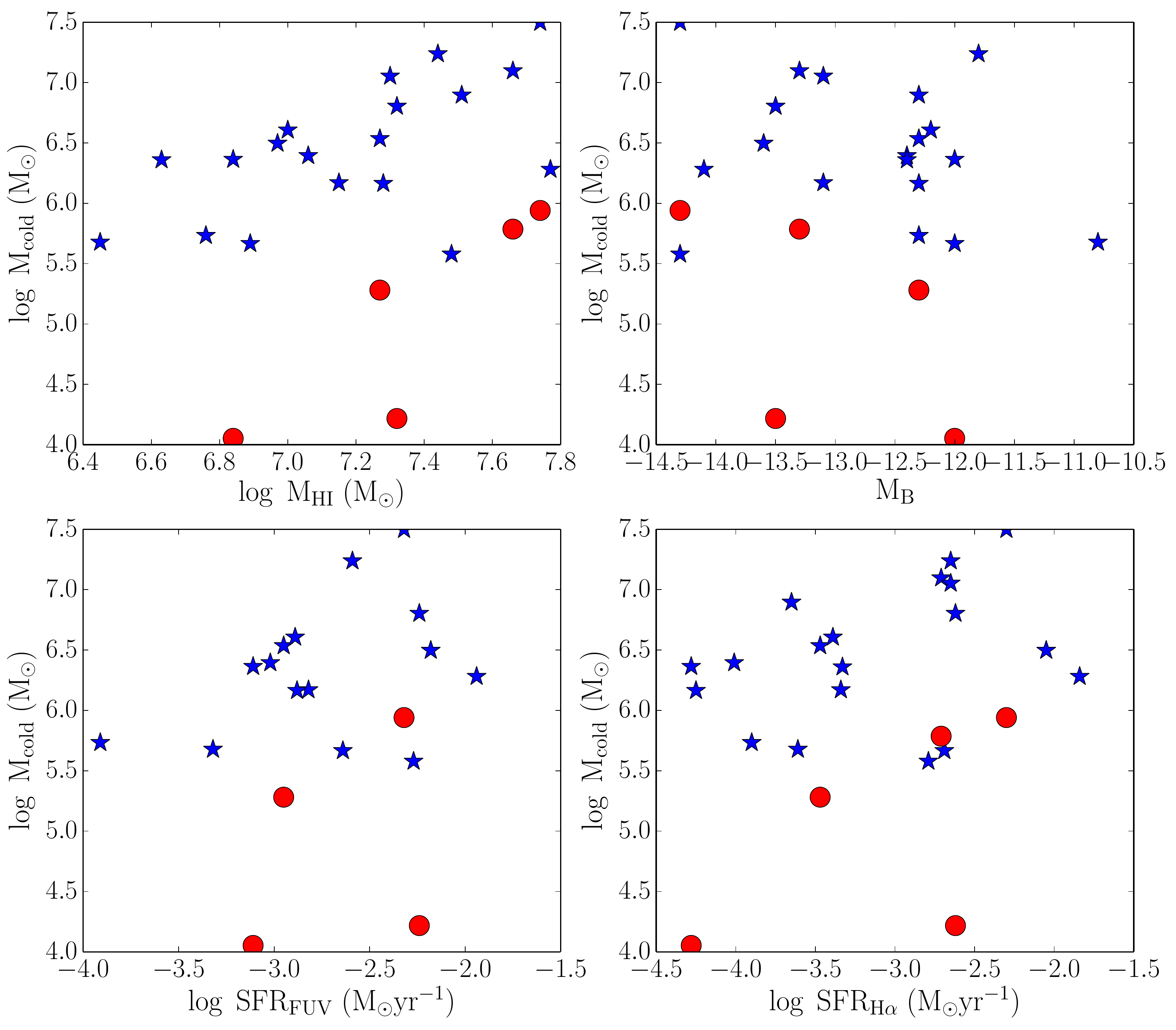}} \\
\end{center}
\caption{ In all the panels, filled circles represents cold \HI recovered by the Gaussian decomposition method while filled stars represents cold \HI recovered by \tb method (optically thin approximation). 
Top left    : The scatter plot of cold \HI mass with total \HI mass ($\rm M_{HI}$) 
Top right   : The scatter plot of cold \HI mass with the absolute blue magnitude 
              ($\rm M_B$).
Bottom left : The scatter plot of cold \HI mass with star formation rate, as
              deduced from the FUV emission ($\rm SFR_{FUV}$).
Bottom right: The scatter plot of cold \HI mass with star formation rate, as
              deduced from the \ha star-formation rates ($\rm SFR_{H \alpha}$).            See
              the text for more details on the derivation of the star
              formation rates.
}
\label{fig:global}
\end{figure*}

To investigate the connection between different global properties of galaxies and the detected cold \HI, we plot the amount of cold \HI as a function of different global parameters in Fig.~\ref{fig:global}. 
The global \ha and FUV SFR used in the plots were taken from the Updated Nearby Galaxy Catalogue (UNGC) \citep{karachentsev13}. From the  plots it is clear that brighter galaxies with more \HI mass and higher star formation rates contain more cold \HI. The correlation coefficients of the various prameters are listed in of Tab.~\ref{tab:global}. The error bars have been estimated by bootstrap resampling. As can be seen from the table, the cold \HI mass (as measured using the T$_{\rm B}$ method does not appear to correlate with any of the other global parameters except the total \HI mass and perhaps the FUV star formation rate. For cold gas measured using the Gaussian method, the number of galaxies in the sample are in general too small for any trends to be visible.

\begin{table*}
\begin{tabular}{lcccc}
\hline
Method & $\rm M_{HI} \ vs \ M_{cold}$ & $\rm M_B \ vs \ M_{cold}$ & $\rm SFR_{FUV} \ vs \ M_{cold}$ & $\rm SFR_{H \alpha} \ vs \ M_{cold}$ \\
\hline
$\rm Gauss $ & 0.87 $\pm$ 0.22 & -0.55 $\pm$ 0.44 & 0.29 $\pm$ 0.66 & 0.60 $\pm$ 0.43 \\
$\rm T_B $ & 0.57 $\pm$ 0.19 & -0.26 $\pm$ 0.29 & 0.37 $\pm$ 0.22 & 0.32 $\pm$ 0.18 \\
\hline
\end{tabular}
\caption{Correlation between the cold \HI mass and other global properties}
\label{tab:global}
\end{table*}

\begin{figure}
\begin{center}
\resizebox{85mm}{!}{\includegraphics{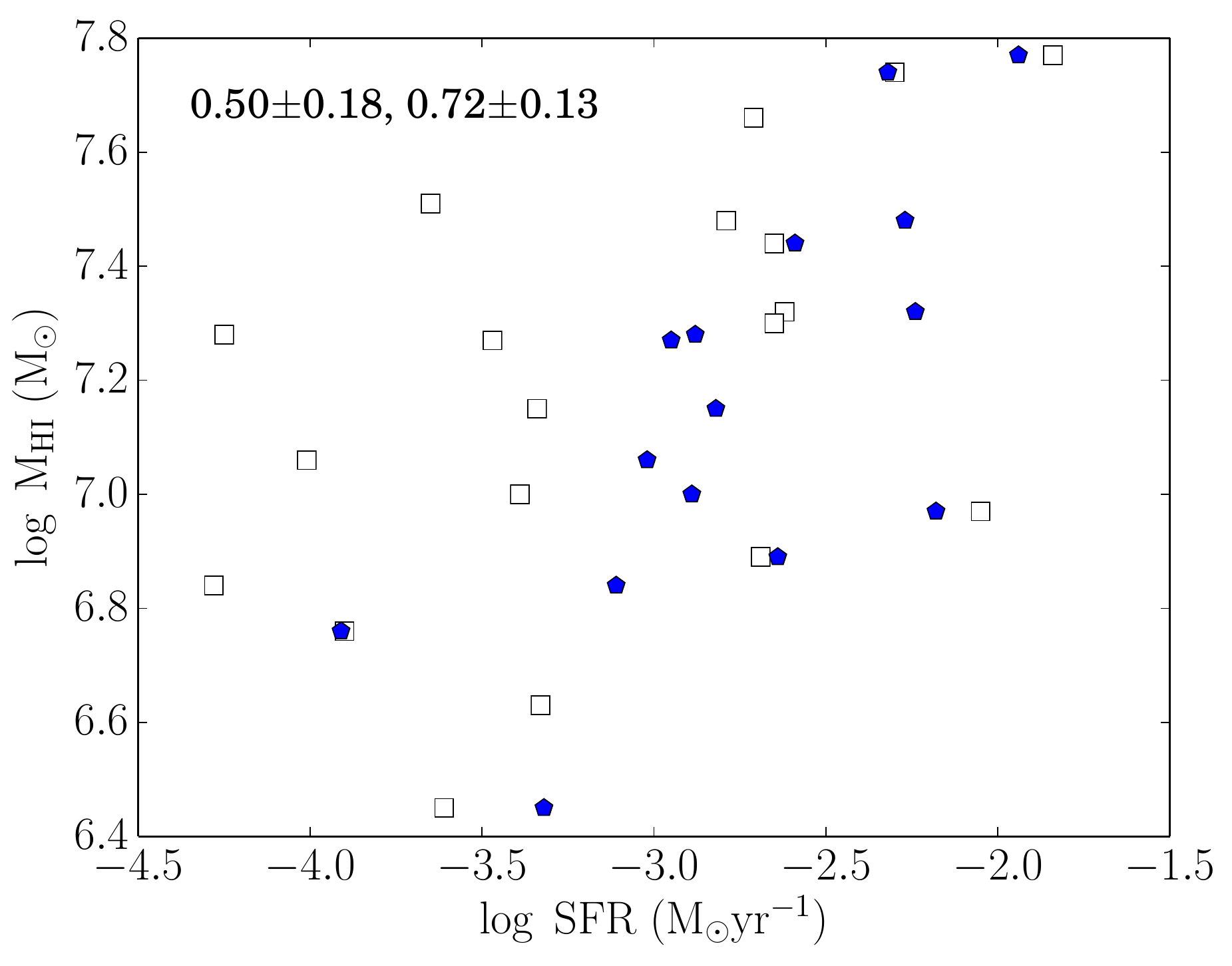}} \\
\end{center}
\caption{ Total \HI mass is plotted against global star-formation rates.
The empty squares represents \ha data and blue pentagons represents FUV 
data. The numbers quoted on the top left shows the correlation coefficients 
between total \HI and global star-formation rate for \ha and FUV respectively.
}
\label{fig:corrtot}
\end{figure}

In Fig.~\ref{fig:corrtot}, the total \HI is plotted against total global star formation rates. The numbers quoted on the top left of the figure represents the correlation coefficients for \ha, and FUV data respectively. Comparing with Fig.~\ref{fig:global} one can see that the total \HI correlates better with global star-formation  than cold \HI. 

%

\begin{figure}
\begin{center}
\resizebox{85mm}{!}{\includegraphics{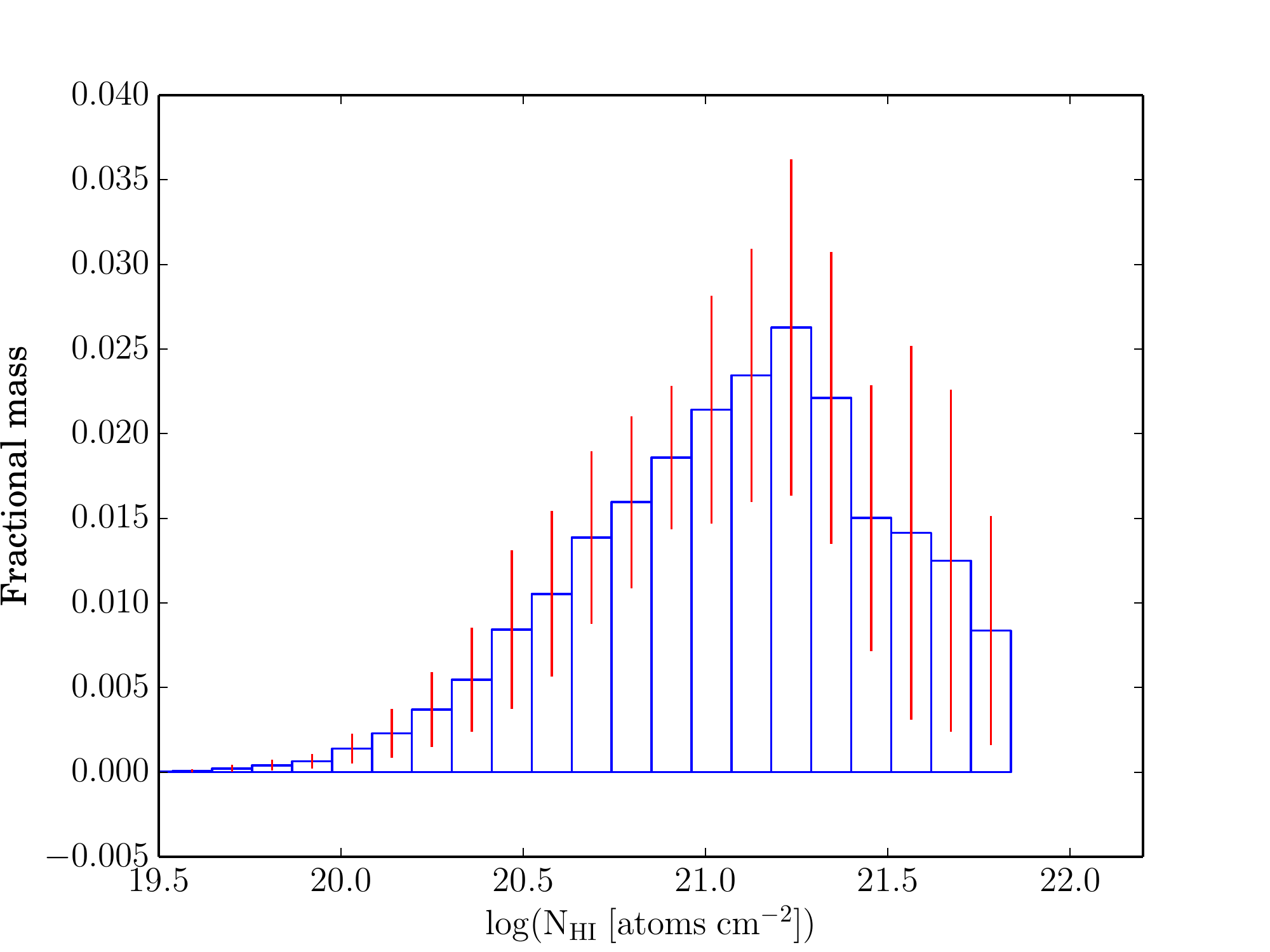}} \\
\end{center}
\caption{The fractional \HI mass in the cold phase (as determined by the \tb method) as a function of \nh. At $\log$ \nh $\sim 21.8$ cold \HI fraction drops to zero, consistent with the expected threshold column density for conversion of low metallicity (Z$\rm \sim 0.1 ~Z_{\odot}$) gas into the molecular phase. \citep{krumholz09b}.
}
\label{fig:mnhfrac}
\end{figure}

Conversion of atomic gas to the molecular phase is expected to lead to a maximum column density of the atomic phase. At column densities larger than this value the shielding against Lyman-Werner band photons which could dissociate the H$_2$ molecule (both self-shielding, as well as shielding due to dust) is expected to be insufficient so that the gas transitions from the atomic to the molecular phase. In our own galaxy for example, the molecular fraction becomes significant above column densities of $\log(\nh) \sim 20.7$ \citep{savage77,gillmon06a,gillmon06b}. If the gas to dust ratio scales with the metalicity one expects that the threshold column density would increase with decreasing metalicity (since lower metalicity would correspond to lower dust content and hence lower shielding).  In Fig.~\ref{fig:mnhfrac} the fractional cold \HI mass (i.e. the ratio of the cold \HI mass detected over the entire sample to the total \HI mass of the galaxies in the sample) is plotted as a function the column density of the cold \HI. As can be seen the fraction gradually increases with increasing \nh\  but there is an abrupt cut off near \nh $\sim \ 10^{21.8}$ \acc. At low column densities, one would expect that the gas would predominantly be in the WNM phase, and indeed in our own galaxy  \cite{kanekar11} show that there appears to be a threshold density of $\sim 10^{20.3}$ \acc\ for the formation of the CNM. \cite{kanekar11} suggest that this threshold arises because one needs a critical amount of shielding before the gas can cool down to the CNM phase, while the numerical models of \cite{kim14} indicates that this threshold can be understood in terms of vertical equilibrium in the local Milkyway disk. In our observations, since low \nh\ would correspond to lower brightness temperature and hence a lower signal to noise ratio, the fall off that we see at low column densities could partly be a sensitivity related issue. What is more interesting is the peak \HI column density of $\sim 10^{21.8}$ that is seen in our data. \citet{krumholz09b} present detailed calculations of the atomic to molecular transition in gas of different metalicities. At the metalicity of our sample galaxies (i.e. $\sim$ 0.1 $\rm Z_{\odot}$),  the saturation column density that is predicted by their model (Fig 1. in \cite{krumholz09b}) is in excellent agreement with the maximum column density seen in Fig.~\ref{fig:mnhfrac}. 

\section{CNM and star formation}
\subsection{The subsample and data preparation}
In this section we investigate the correlation between CNM and star formation. We used {\tt GALEX} FUV data and \ha data mostly drawn from the Russian BTA telescope and the LVL survey \citep{dale09} for estimating star formation rates. CNM was detected via the Gaussian decomposition method only in five galaxies and the recovery efficiency is also very low. The cold \HI detected using the Gaussian decomposition method is also spread over only a small number of independent beams. In contrast, from the \tb method, we detected cold \HI in nineteen galaxies with significant number of independent beams per galaxy.

We used both \ha and/or FUV data for estimating recent star formation for our sample galaxies. We detected cold \HI using Gaussian decomposition method in 5 galaxies. Out of which, good quality \ha data were available for two galaxies (UGC7605 and UGC0685). On the other hand good quality FUV data were available for all the galaxies except KKH11. For \tb method, our primary subsample consist of 19 galaxies for which cold \HI was detected. Out of these 19 galaxies, good quality \ha data were available for 9 galaxies. Three galaxies (ESO379-007, UGC8508 and KKH11) did not have {\tt GALEX} observations and another four galaxies (KK200, KKH34, KKH6 and UGC5209) were observed as part of the all sky survey, and hence their FUV images have poor SNR. We exclude these seven galaxies from our sample which leaves us to a subsample of total 12 galaxies with good quality FUV {\tt GALEX} data, as well as detected cold \HI in \tb method.

\begin{table*}
\caption{Properties of the sub-sample galaxies selected for \tb method.}
\label{table1_tblocal}
\begin{threeparttable}
\begin{tabular}{|l|c|c|c|c|c|c|c|c|}
\hline
Name & RA & DEC & $\rm \log M_{HI}$ & $\rm M_B$ & incl  & $\rm Metallicity$ & $\rm \log SFR_{H \alpha}$ & $\rm \log SFR_{FUV}$ \\
 & (J2000) & (J2000) & $\rm (M_{\odot})$ &  & $\rm (^o)$ & $\rm (Z_{\odot})$ & $\rm (M_{\odot}yr^{-1})$ & $\rm (M_{\odot}yr^{-1})$ \\

\hline

UGC00685$^{1,2,3,4}$ & $010722.3$ & $+164102$ & $7.74$ & $-14.3$ & $51$ & $0.20^\dagger$ & $-2.30$ & $-2.32$ \\ 

UGC06456$^{2,3,4}$ & $112800.6$ & $+785929$ & $7.77$ & $-14.1$ & $69$ & $0.10^\ddagger$ & $-1.84$ & $-1.94$ \\ 

UGC06541$^{2,4}$ & $113329.1$ & $+491417$ & $6.97$ & $-13.6$ & $65$ & $0.13$ & $-2.05$ & $-2.18$ \\ 

UGC07298$^{2,3,4}$ & $121628.6$ & $+521338$ & $7.28$ & $-12.3$ & $67$ & $0.06$ & $-4.25$ & $-2.88$ \\ 

DDO125$^{2,3,4}$ & $122741.8$ & $+432938$ & $7.48$ & $-14.3$ & $66$ & $0.12$ & $-2.79$ & $-2.27$ \\ 

UGC07605$^{1,2,3,4}$ & $122839.0$ & $+354305$ & $7.32$ & $-13.5$ & $49$ & $0.09$ & $-2.62$ & $-2.24$ \\ 

UGCA292$^{2,3,4}$ & $123840.0$ & $+324600$ & $7.44$ & $-11.8$ & $52$ & $0.05$ & $-2.54$ & $-2.59$ \\ 

GR8$^{2,4}$ & $125840.4$ & $+141303$ & $6.89$ & $-12.0$ & $27$ & $0.09$ & $-2.69$ & $-2.64$ \\ 

UGC08215$^{1,2,4}$ & $130803.6$ & $+464941$ & $7.27$ & $-12.3$ & $52$ & $0.06$ & $-3.47$ & $-2.95$ \\ 

KK200$^{1,4}$ & $132436.0$ & $-305820$ & $6.84$ & $-12.0$ & $60$ & $0.06$ & $-4.28$ & $-3.11$ \\ 

UGC08508$^{2,3}$ & $133044.4$ & $+545436$ & $7.30$ & $-13.1$ & $63$ & $0.08$ & $-2.65$ & $-$ \\ 

UGC08833$^{2,3,4}$ & $135448.7$ & $+355015$ & $7.00$ & $-12.2$ & $30$ & $0.06$ & $-3.39$ & $-2.89$ \\ 

DDO187$^{2,4}$ & $141556.5$ & $+230319$ & $7.06$ & $-12.4$ & $46$ & $0.11^\dagger$ & $-4.01$ & $-3.02$ \\ 

KKH98$^{2,3,4}$ & $234534.0$ & $+384304$ & $6.45$ & $-10.8$ & $67$ & $0.04$ & $-3.61$ & $-3.32$ \\ 

\hline
\end{tabular}
\begin{tablenotes}
\item[1,2] Galaxies with detected cold \HI in Gaussian decomposition and \tb method respectively
\item [3,4] Galaxies with available \ha and FUV map respectively
\item [] References - $\dagger$ : \citep{marble10}; $\ddagger$ : \citep{moustakas06}
\end{tablenotes}
\end{threeparttable}
\end{table*}

In Table~\ref{table1_tblocal} we describe the general properties of this sample. The columns are as follows: 
Col.~(1)~Galaxy name, 
Col.~(2)~and ~3) the equatorial coordinates (J2000), 
Col.~(4)~and~(5) log of \HI mass and absolute blue magnitude, 
Col.~(6)~inclination taken from \citep{karachentsev13}. They have been computed from the axial ratio of the optical disc, and the typical error is $\sim 10\%$.
Col.~(7)~metallicity in solar unit,
Col.~(8)~and~(9) represents the global \ha and FUV star-formation rates. 
Galaxies with detected cold \HI in Gaussian decomposition and brightness temperature methods are marked with superscript $^1$ and $^2$ respectively. Galaxies with available \ha data are marked with superscript $^3$ whereas $^4$ represents galaxies with available FUV star formation maps. The data in table~\ref{table1_tblocal} were taken from \citep{karachentsev13}.

\subsubsection{Estimation of the star formation rate}
\label{sec:ssfrerr}
FUV data were downloaded from the {\it GALEX} site and were used to prepare $\rm SFR_{FUV}$ maps for our sample galaxies. Only FUV $\rm (1350-1750 \AA)$ data were used to derive the star formation as a part of the NUV band $\rm (1750-2800 \AA)$ lies outside the range for which calibrations for conversion to the star formation rate are available. Emission from Galactic foreground stars was identified by manual inspection of the images, and removed from the FUV map using the task \scalebox{.7}{BLANK} in \scalebox{.7}{AIPS}. The foreground stars were identified and cross matched using the SDSS object identifier tool. The FUV maps were then smoothed to the resolution of the \HI maps using the task \scalebox{.7}{SMOTH} in \scalebox{.7}{AIPS}. The geometries of the smoothed FUV maps were then aligned with that of the \HI maps using the task \scalebox{.7}{OHGEO}. The FUV maps are in units of counts per second which can be converted into FUV flux using the calibration provided in {\it GALEX} site

\begin{align}
&m_{GALEX} = -2.5 \log (counts \ s^{-1}) \\
&m_{AB} = m_{GALEX} + 18.82
\end{align}

Using the above equations FUV flux maps were generated and corrected for the galactic dust extinction using the dust map of \citep{schlafly11}. The formulae given by \citep{cardelli89} were used to extrapolate the extinction to the FUV band.  For the internal dust correction, we adopt the approach described in \citet{roychowdhury14}. Briefly, 25 $\mu$m {\it Spitzer} observations of FIGGS galaxies were used to derive a numerical relation between the FUV star formation rate and the 24 $\mu$m flux. The 25 $\mu$m flux can be given as:

\begin{align}
\log F_{25\mu m} (Jy) = 1.78~ \log SFR_{FUV}(M_{\odot}yr^{-1}) + 2.62
\end{align}

This corrected emergent FUV luminosity is then corrected for internal dust extinction using relation given by \citep{hao11}

\begin{align}
L_{FUV,corr} = L_{FUV, obs} + 3.89~ L_{25 \mu m}
\label{eq:fuv_corr}
\end{align}

The FUV luminosity of equation~\ref{eq:fuv_corr} can then be converted into the star formation rate using the calibration given by \citep{kennicutt12,hao11}

\begin{align}
\log \dot{M}_* (M_{\odot}yr^{-1}) = \log L_x - \log C_x
\end{align}

where $L_x = \nu L_{\nu}$ is in the units of $ergs~s^{-1}$ and $\log C_x = 43.35$. This calibration assumes an ongoing star formation for $\sim~10^8$ yr and a Salpeter IMF with solar metallicity. Studies of the star formation rate of nearby dIrr galaxies \citep{weisz12} indicate that the assumption that the star formation rate has been approximately constant for the last $\sim 10^8$~yr is a reasonable one; however we do need to correct the star formation rate estimated above for the metallicity.

The typical metallicity of the dwarf galaxies in our sample is Z$\sim 0.1$ \citep{roychowdhury14}. Metallicity measurements are available only for a few of our sample galaxies \citep[e.g.][]{marble10,moustakas06}. For the rest of our galaxies, we use the \mb-Z metallicity relation for dIrs from \citep{ekta10} to estimate their metallicity. In Table~\ref{table1_tblocal} we list the metallicity estimated for our sample galaxies. The star formation rate estimated above  hence needs to be corrected to account for this.  \citet{raiter10} calculate the emergent FUV fluxes in sub-solar metallicity environment using an evolutionary synthesis model assuming a continuous star formation for $\rm 10^8$ yrs and a Salpeter IMF. They show that the emergent FUV flux increases by $\sim$ 11\%, 19\%, 27\% \& 32\% for 0.4, 0.2, 0.05 and 0.02 times solar metallicity respectively. For each sample galaxy we do a linear interpolation between these tabulated values to get the metallicity correction to the FUV flux.

The star formation rate density was also computed using \ha data largely taken from the 6m BTA telescope in Russia. For a detailed analysis of \ha data see \citep{karachentsev07,kaisin08}. The \ha data were processed exactly the same way (foreground subtraction, smoothing to \HI map resolution, geometrical alignment) as the FUV data. To derive the star-formation rate from \ha map we used Kennicutt's calibration, equation (4) with $L_x$ in the units of $ergs~s^{-1}$ and $\log C_x = 41.27$. This calibration also assumes a Salpeter IMF with a continuous star-formation for $\rm 3-10~ Myr$. We note that at low star formation rates, the \ha emission becomes unreliable as a tracer of the star formation rate largely because of stochasticity in the number of high mass stars observed at any instant \citep{dasilva14}. The actual star formation rate as deduced from the \ha emission for our sample should hence be treated with caution.

A detailed discussion of the error estimate for the star formation rates can be found in \citet{roychowdhury14}, but for completeness we give a brief outline here.  The total error in \ssfr~is calculated as the quadrature sum of the individual errors. Firstly, we adopt a 10\% calibration error in FUV flux and a 15\% calibration error in \ha flux measurement. An additional error of $\sim$ 50\% in star formation rate is expected due the variation in IMF and star formation history \citep{leroy12,leroy13}. Any other systematic errors are expected be significantly smaller than the above errors. The total estimated error in the \ssfr~for our sample galaxies is $\sim 50-55\%$.

\subsubsection{Measurement of the cold \HI column density}

To compute the face on column density of cold gas and surface density of star formation rate at the same spatial location we deproject the observed column densities using the inclination angles given in Table \ref{table1_tblocal}. We also multiply the \HI surface density by a factor 1.34 to account for the presence of helium. In several plots we average over several regions lying within a given $\rm \Sigma_{HI}$, or $\rm \Sigma_{SFR}$ bin in order to improve the signal to noise ratio and reduce the effects of stochasticity in the star formation rate.

\subsection{Results}
\subsubsection{Cold gas with associated star formation}

\begin{figure}
\begin{center}
\resizebox{90mm}{!}{\includegraphics{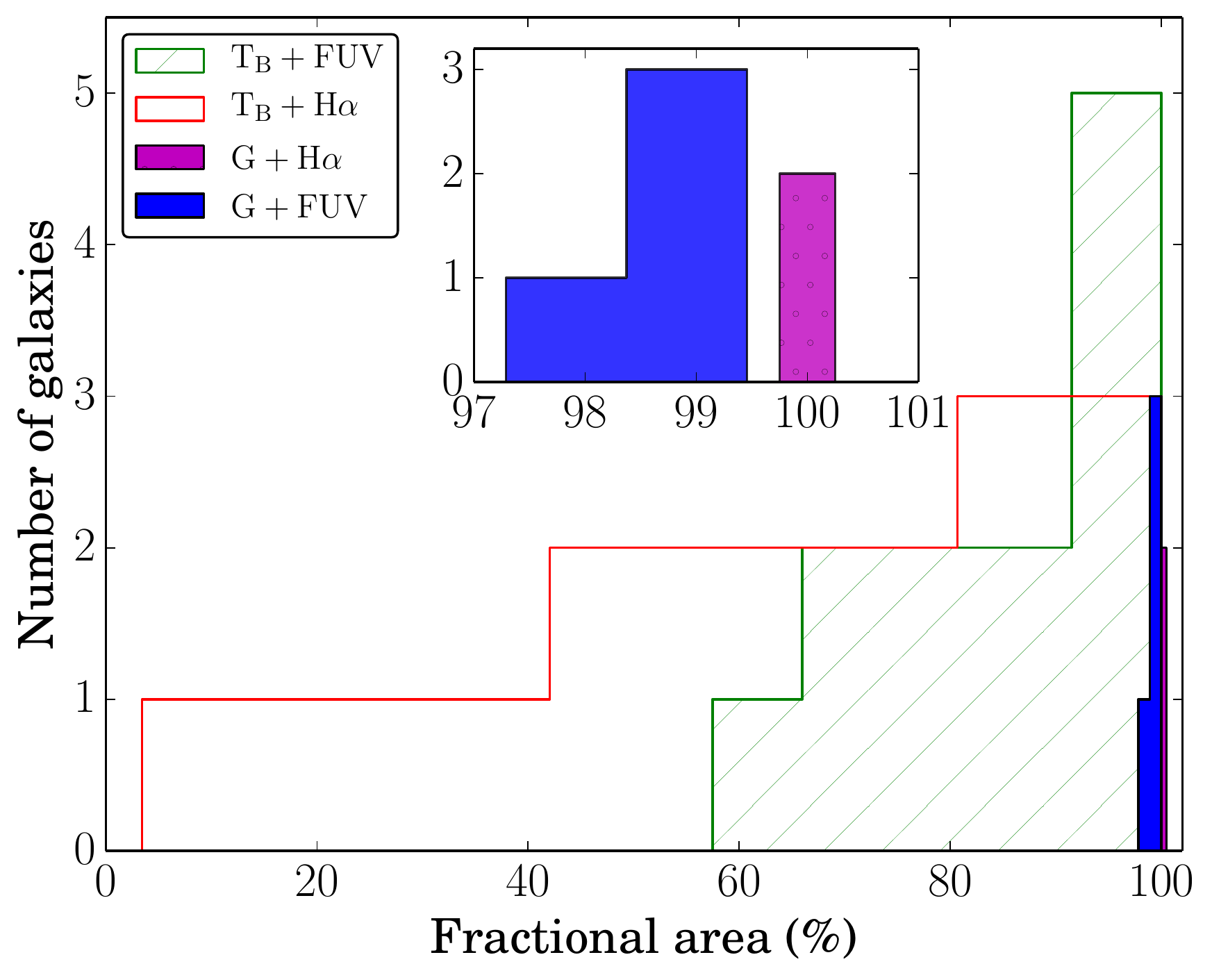}}
\end{center}
\caption{Histogram of areal fraction of cold \HI with associated recent star formation. The filled magenta and blue histograms represents fractional cold \HI detected in Gaussian decomposition method with associated star formation as traced by \ha and FUV respectively. These histograms are zoomed in at the top middle portion of the figure for clarity. The red stepped and green hatched histograms represents fractional cold \HI as detected in \tb method with associated recent star formation as traced by \ha and FUV respectively. See text for discussion.}
\label{gasfrac}
\end{figure}

\begin{figure}
\begin{center}
\resizebox{90mm}{!}{\includegraphics{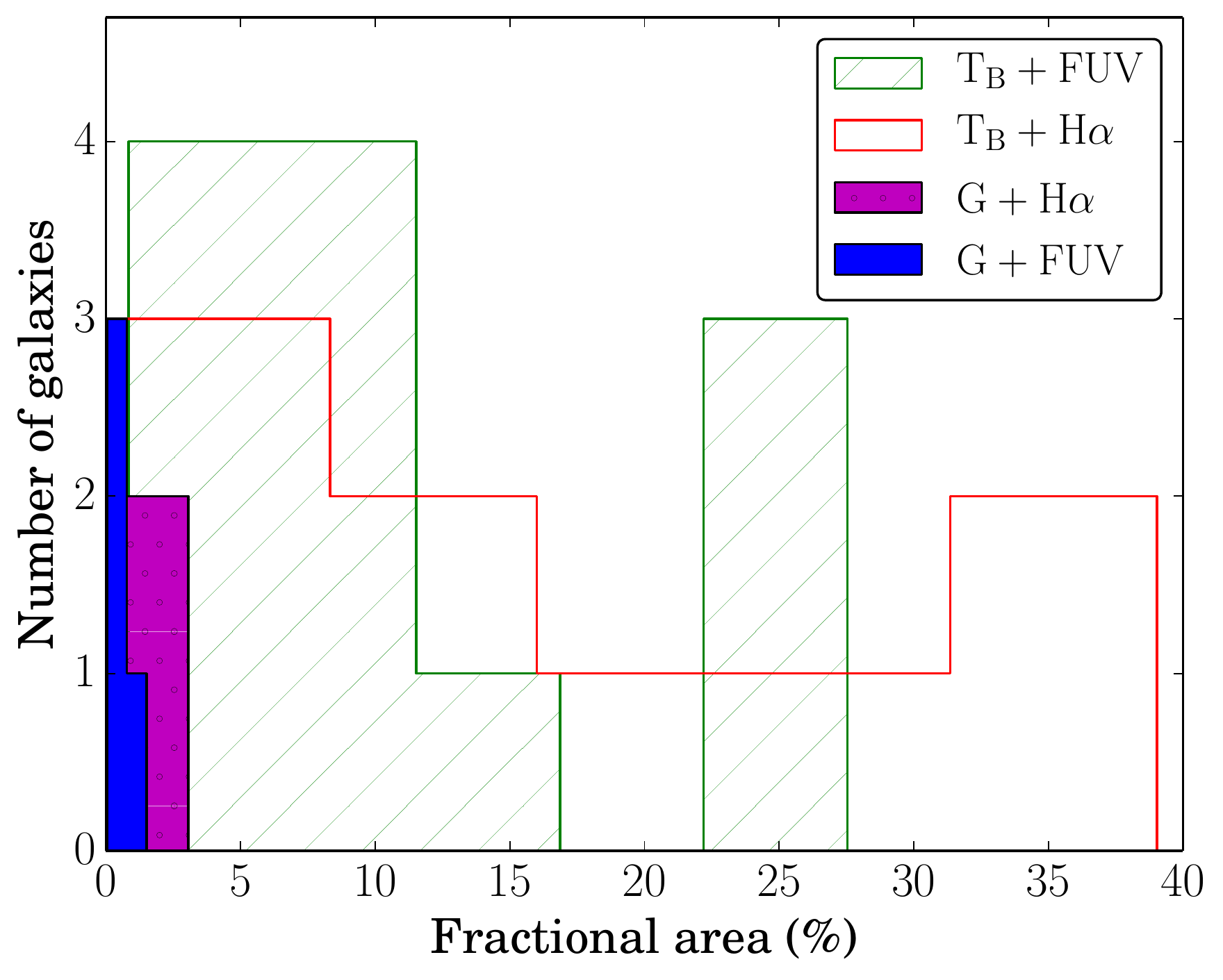}}
\end{center}
\caption{Histogram of areal fraction of recent star formation that is associated with cold \HI. The notations are similar to that of Fig.~\ref{gasfrac}. See text for more details.}
\label{sfrfrac}
\end{figure}

To examine the association of cold \HI with recent star formation, we compute the area covered by pixels with detected cold gas which also have a star formation rate density greater than the 3$\sigma$ level of the SFR map, and normalizing this area by the total area covered by cold gas alone. In Fig.~\ref{gasfrac} we plot the histogram of fractional cold gas with associated star formation for our sample galaxies. The filled blue histogram present the fractional cold \HI (detected in Gaussian decomposition method) associated with recent star formation as measured by the FUV emission. The filled magenta histogram represent the same but this time \ssfr~is measured using \ha map. The red empty step histogram represents fractional cold \HI (as detected by \tb method) associated with recent star formation as traced by \ha, whereas hatched green histogram presents the fractional cold \HI (detected in \tb method) associated with FUV star formation. From Fig.~\ref{gasfrac} it is implied that most of the cold \HI (detected by either method) has some associated star formation (as traced by either \ha or FUV). We note that the number of galaxies having detected cold \HI in Gaussian decomposition method and available \ha and FUV maps are 2 and 4 respectively. The median fractional area of cold \HI in \tb method with associated \ha SFR is $\sim$ 71\% and with FUV SFR is $\sim$ 90\%.

\begin{figure}
\begin{center}
\resizebox{90mm}{!}{\includegraphics{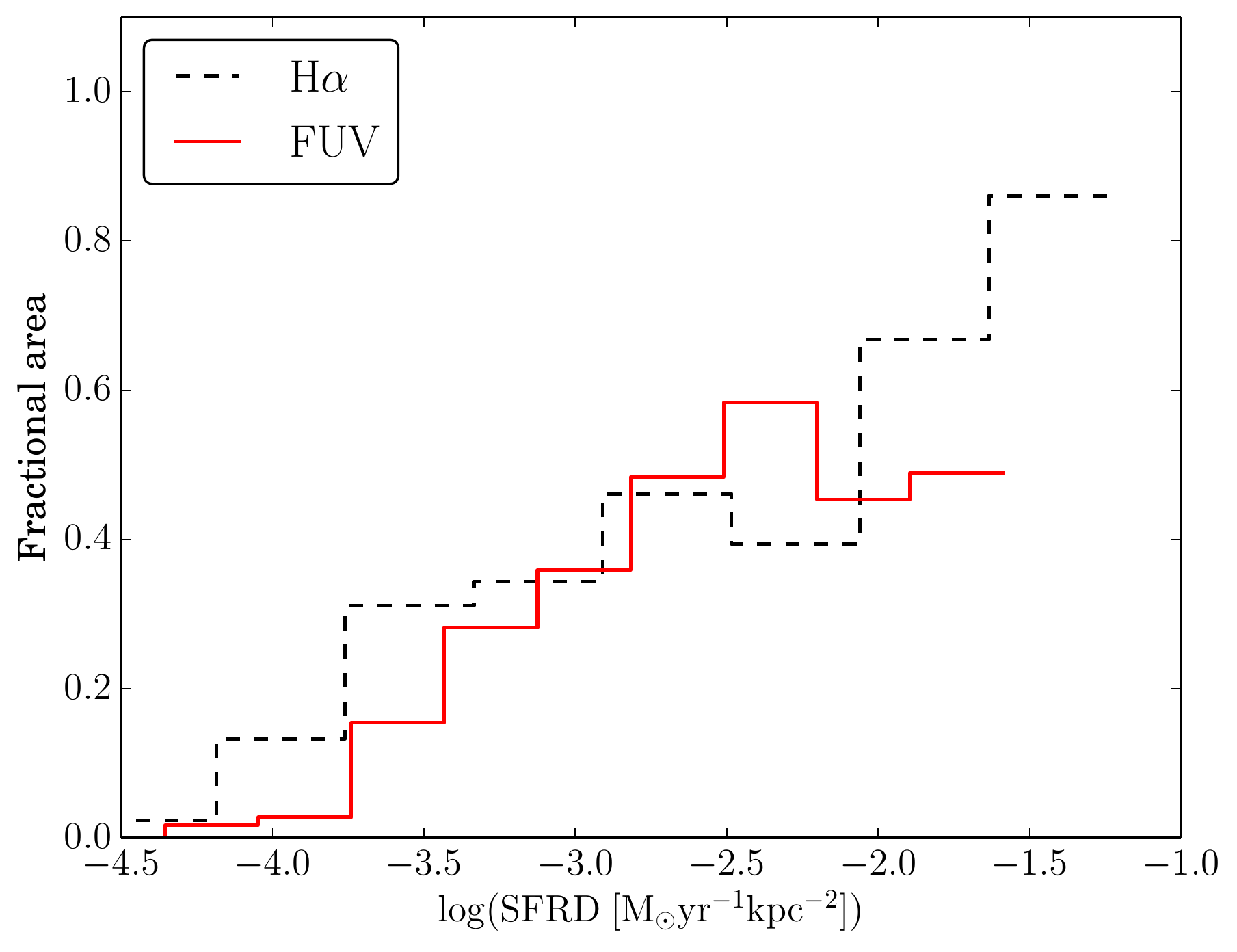}}
\end{center}
\caption{Histogram of areal fraction of recent star formation with associated cold \HI, as a function of the star formation rate. The red solid histogram is for FUV SFR, whereas the black dotted histogram is for \ha SFR. At the highest star formation rates as traced by \ha (black dotted histogram), almost all of the star forming regions ($\sim 80\%$) have associated cold gas whereas $\sim 40-50\%$ star forming regions as traced by FUV (red solid histogram) have associated cold \HI.}
\label{sfr_frac_hist}
\end{figure}

It is also interesting to ask the reverse question, i.e. if all star forming regions have associated cold \HI or not. In Fig.~\ref{sfrfrac}, we show the fractional area with both cold \HI and recent star formation normalised by total area covered by regions with recent star formation. The notations of the histograms are the same as in Fig.~\ref{gasfrac}. As the detection efficiency of Gaussian decomposition method is very poor, it is expected that most of the cold \HI remains undetected and hence the fractional area of cold \HI with associated recent star formation (both \ha and FUV) found to be very less ($\lesssim 5\%$). But in \tb method, the fraction varies between few percent to 40\% with a median value of $\sim$ 14\% for \ha SFR and $\sim$ 8\% for FUV star formation. So, although most of the regions with cold \HI have some associated recent star formation, the converse is not true. It is interesting to note that in  the case of the recent star formation as estimated from the \ha emission (which traces much more recent star formation than FUV) a larger fraction ($\sim 14\%$) of the star forming regions have associated cold \HI. We show in Fig.~\ref{sfr_frac_hist} how this fraction varies as a function of the star formation rate. The red solid line represents FUV star formation whereas the black dashed line represents \ha star formation. As can be seen, at the highest star formation rates, as traced by \ha emission, almost all of the star forming regions ($\sim 80\%$) have associated cold gas. But at highest FUV star formation rates, $\sim 40 - 50 \%$ star forming regions have associated star formation. As FUV traces an ongoing star formation for longer period of time, it is possible that star formation feedback destroys cold gas at highest FUV star forming cites. More sensitive data for larger number of galaxies is required for any firm conclusion.

\subsubsection{The Kennicutt-Schmidt law for cold \HI}

\begin{figure}
\begin{center}
\resizebox{90mm}{!}{\includegraphics{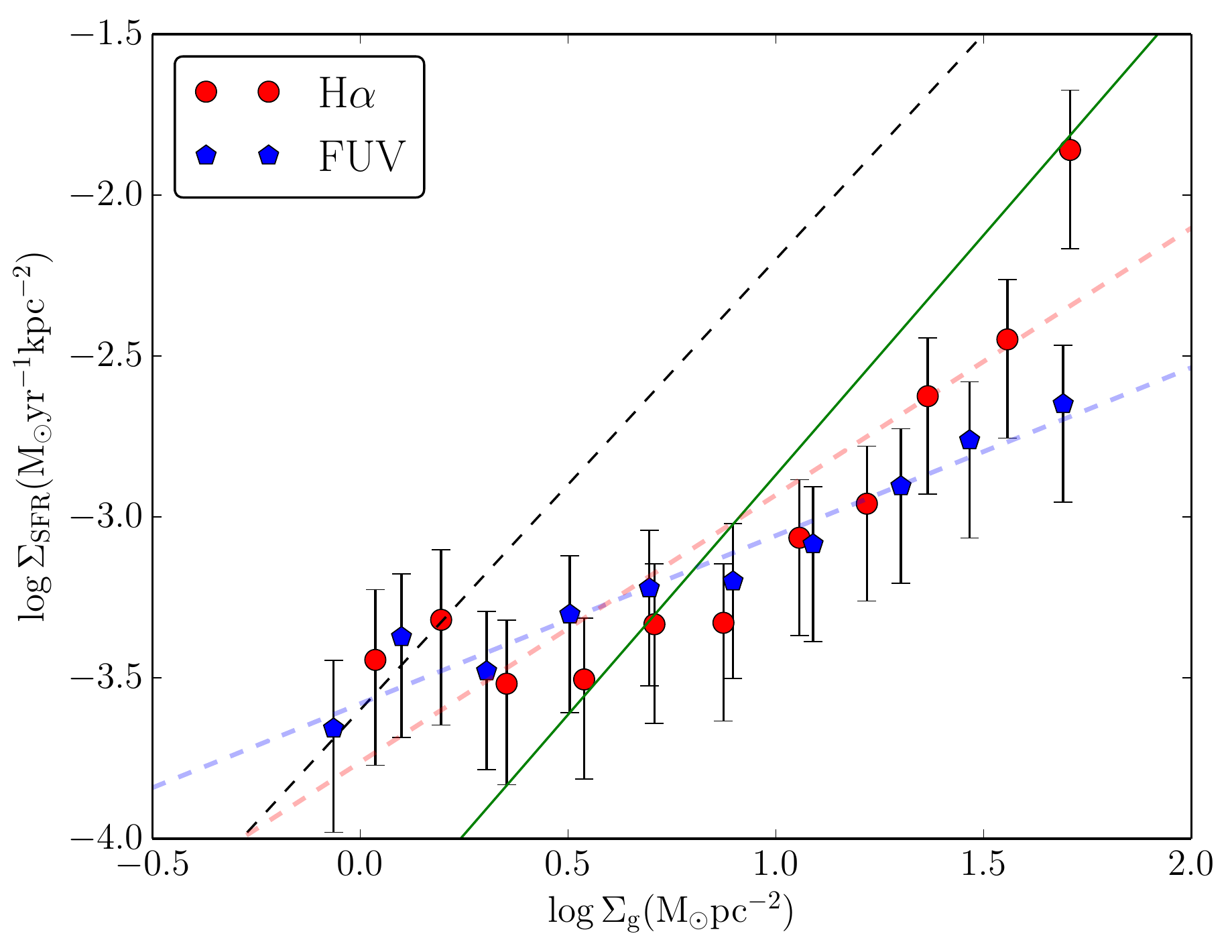}}
\end{center}
\caption {The Kennicutt-Schmidt law for cold \HI. The red filled circles with error bars represent data points for our sample galaxies using the star formation rate as measured from the \ha emission and the red thick dashed line is the straight line fit to these data points. The blue filled pentagons represent the data points for which star formation rate density is measured using FUV maps and the thick blue dashed line represents a straight line fit to these points. The green solid line represents K-S law for FIGGS galaxies \citep{roychowdhury09}, measured using the total gas content (i.e. not just the cold gas) at 400~pc resolution. The thin black dashed line represents the canonical K-S law \citep{kennicutt98b}.}
\label{kslaw}
\end{figure}

In Fig.~\ref{kslaw} we plot the star formation rate surface density as a function of cold gas surface density. In Gaussian decomposition method, cold \HI is detected only in 5 galaxies with the gas distributed across only  $\sim$ a few independent beams per galaxy. Further only two of these galaxies have available \ha maps and only four have available FUV maps. In total we have only 5-10 independent beams over which both the star formation rate and cold \HI surface density measurements are available. Hence we exclude cold \HI detected by Gaussian decomposition method for this study.
 
In Fig.~\ref{kslaw} we show the \ha as well as FUV star formation rate density of our sample galaxies as a function of cold gas surface density (red circles and blue pentagons respectively). The error bars
are the quadrature sum of the rms of the data in a given bin and the
errors discussed in Sec.~\ref{sec:ssfrerr}. The thick dashed red line represents a linear fit to the \ha data. It has  a slope of 0.83 $\pm$ 0.15 and an intercept of -3.76 $\pm$ 0.15.  The thick dashed blue line represents the fit to the FUV data. It has a slope of 0.52 $\pm$ 0.04 and an intercept of -3.58 $\pm$ 0.04. The black dashed line represents the canonical K-S law \citep{kennicutt98b} and the green solid line represents K-S law for FIGGS galaxies at 400 pc resolution (and using the total gas content, not just the cold gas) taken from \citep{roychowdhury09}. It can be seen that at the highest column densities the \ha data points approach to the KS law determined using the total gas (green solid line). Further, although the \ssfr~correlates with the cold gas column density, the slope is significantly flatter than that found using the total gas column density \citep[e.g.][]{kennicutt97, bigiel08, roychowdhury09, roychowdhury14}. This suggests that the total \HI may be a better tracer of the star formation rate, (and the total molecular gas column density) than the cold \HI alone. We note also that recent models of star formation in \HI dominated region also find a flatter relationship between the \ssfr~and the \HI column density \citep{hu15}.

\section{Summary}

We used two different methods to identify cold \HI in a sample of dwarf irregular galaxies selected from the FIGGS and FIGGS2 survey. 
In the first method, line-of-sight \HI spectra with SNR greater than 10 were decomposed into multiple Gaussian components. Following earlier studies, Gaussian components having velocity dispersion $\sigma \ \textless \ 6$ \kms were identified as cold \HI.  For a galaxy that is in common between the two samples, we compare narrow velocity dispersion \HI detected using this method with that detected using the same technique (but with a different data set, and a different software routine) by \citep{warren12}. In our sample we detect narrow velocity dispersion components in 5 galaxies out of the total sample of 12. In common with previous studies, we find that the narrow velocity dispersion components do not overlap with the locations of the highest column density \HI or the locations of the brightest $\rm H\alpha$ emission. This is somewhat surprising if the narrow velocity dispersions are accurately tracing the cold gas content of the galaxies. 

We hence used a different method to try and identify cold atomic gas. Since the WNM has low opacity, there is a limit to the brightness temperature expected from physically plausible path lengths through the WNM. Gas with higher \tb than this threshold must be associated with cold gas. Since the CNM is clumpy, high velocity and spatial resolution is required to identify the high \tb gas. We re-imaged our sample galaxies at 100~pc and identified all regions with \tb~$> 50$~K as being associated with the CNM. We find that cold gas identified in this way is associated with regions of high \HI column density as well as regions with on going star formation. The amount of gas identified as being cold by the \tb method is also significantly larger than the amount of gas identified as being cold using Gaussian Decomposition.  We compare the peak \HI column density that we observe (at 100~pc resolution) with the largest column density expected in low metallicity gas, and find good agreement with theoretical models.

We study the relationship between the cold \HI detected using the \tb method and recent star formation. For our sample of galaxies we find that regions with cold \HI are almost always have associated recent star formation. $\sim 70\%$ of the cold gas has some associated recent star formation as traced by \ha emission and $\sim 90\%$ of the cold gas are associated with star formation as measured by FUV emission. The converse however is not true, only $\sim 10-15\%$ of the area covered by regions with recent star formation is associated with cold \HI. In the case of \ha emission (which traces more recent star formation) the fraction of star forming regions that are associated with cold \HI is significantly higher. In fact, at the highest star formation rates, almost all of the star forming regions are associated with cold \HI.  If we focus on the regions where there is overlap  between cold gas and recent star formation, we find that $\rm \Sigma_{SFR}$ correlates to $\rm \Sigma_{HI}$, albeit with a slope ($\sim 0.83 \pm 0.15 $ for \ha and $ 0.52 \pm 0.04 $ for FUV star formation) that is significantly flatter than that seen in earlier studies which used the total (as opposed to the cold) gas content.


\section{Appendix}
\begin{figure*}
\begin{center}
\begin{tabular}{ccc}
\resizebox{55mm}{!}{\includegraphics{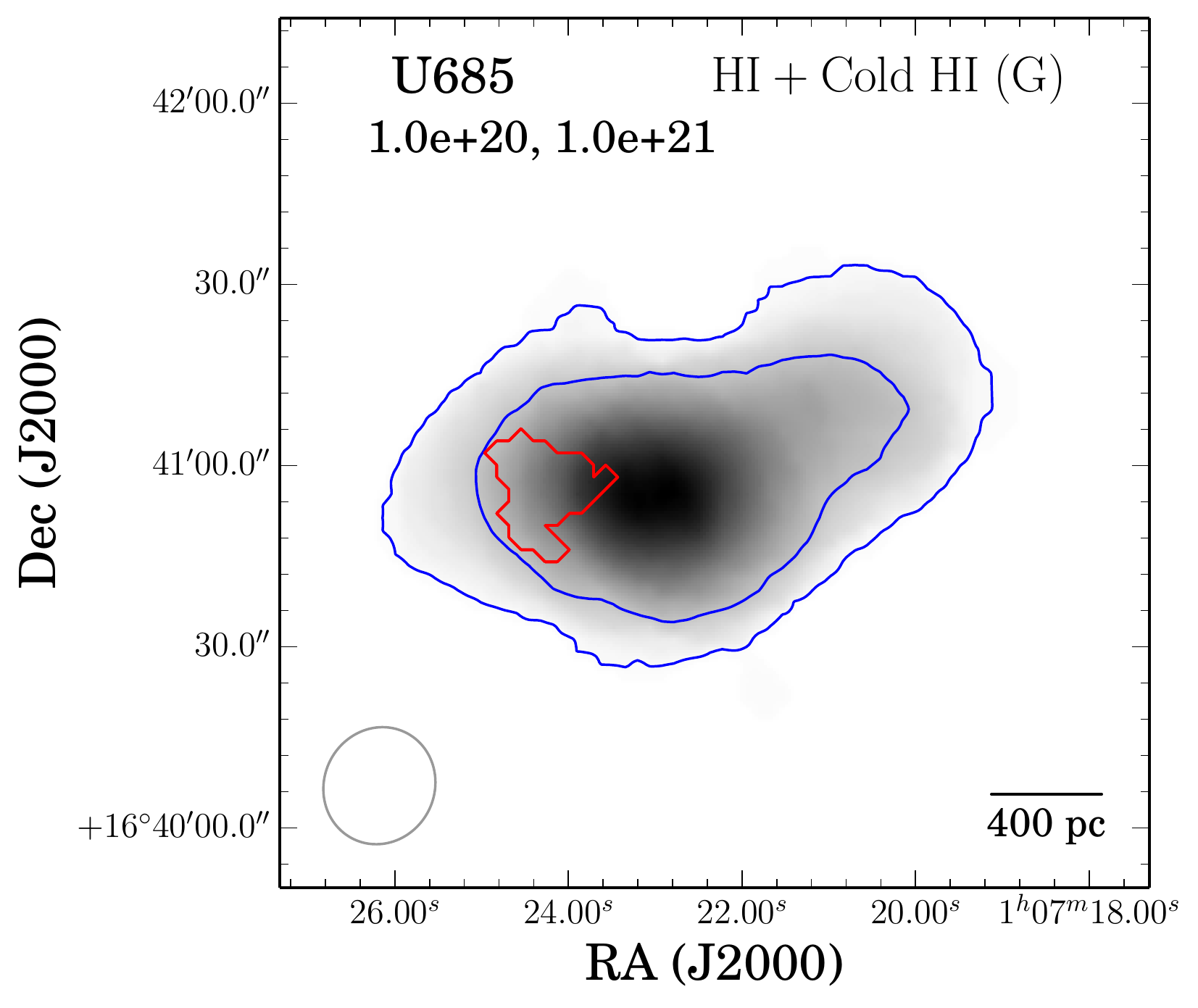}} &
\resizebox{55mm}{!}{\includegraphics{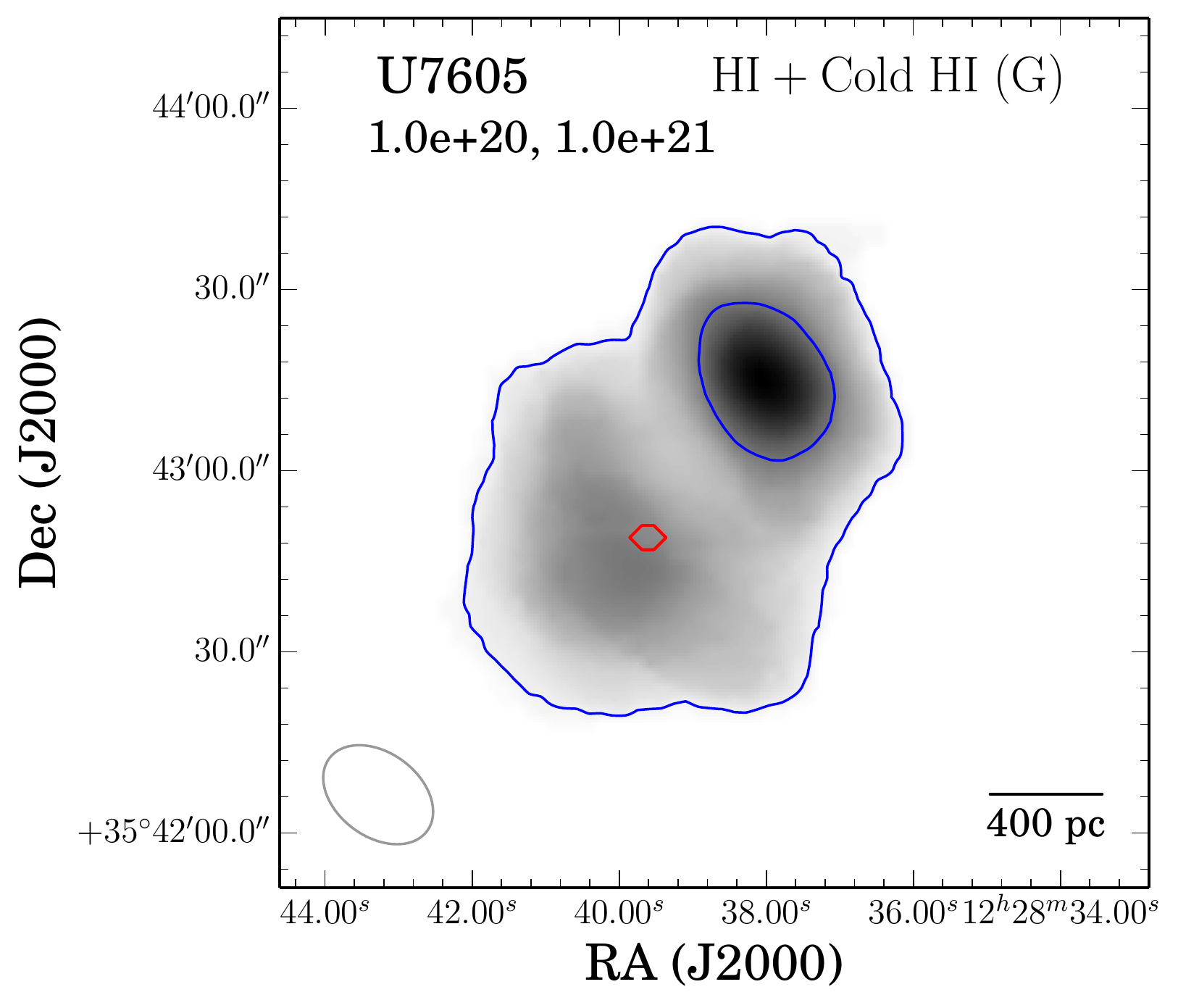}} &
\resizebox{55mm}{!}{\includegraphics{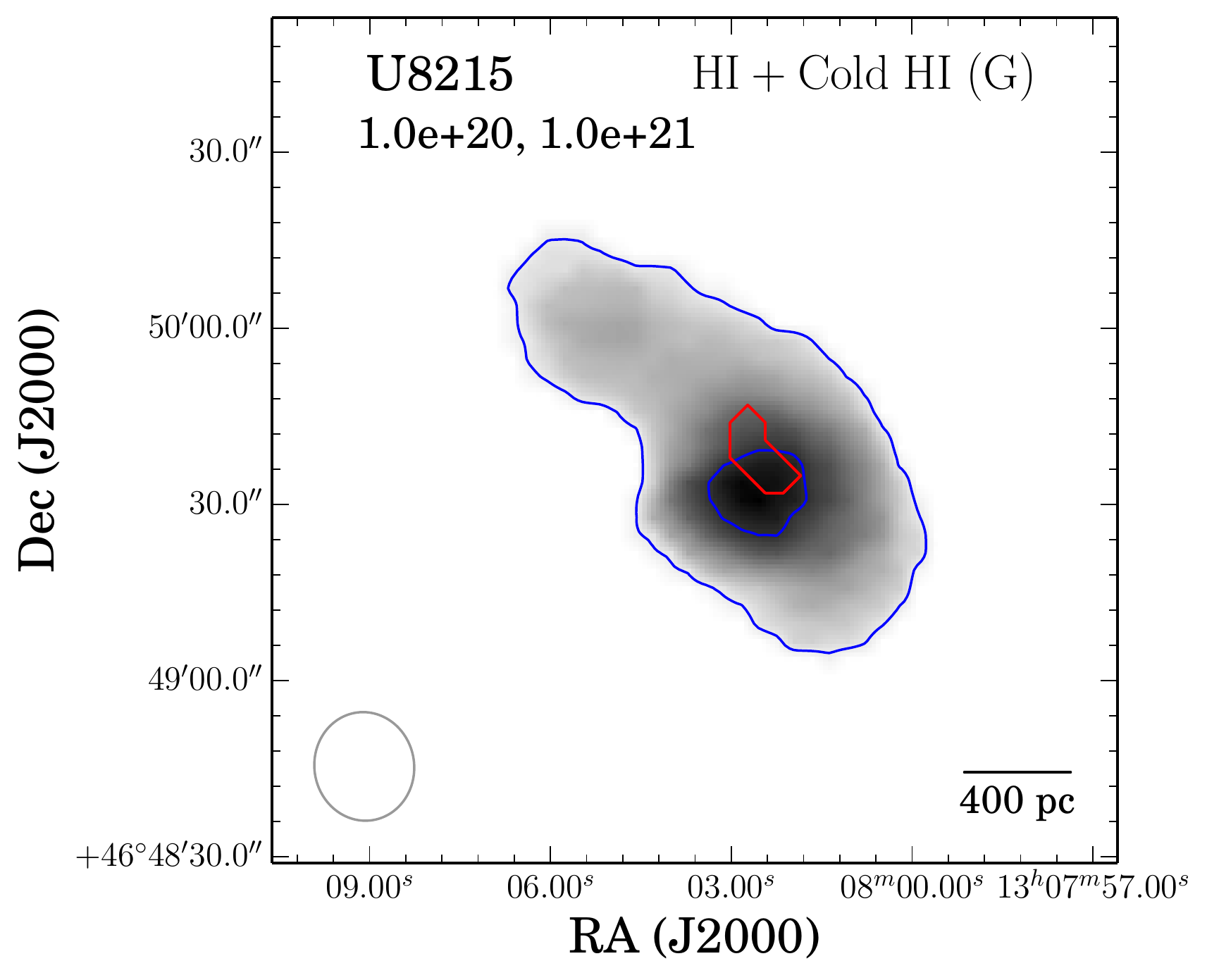}} \\

\resizebox{55mm}{!}{\includegraphics{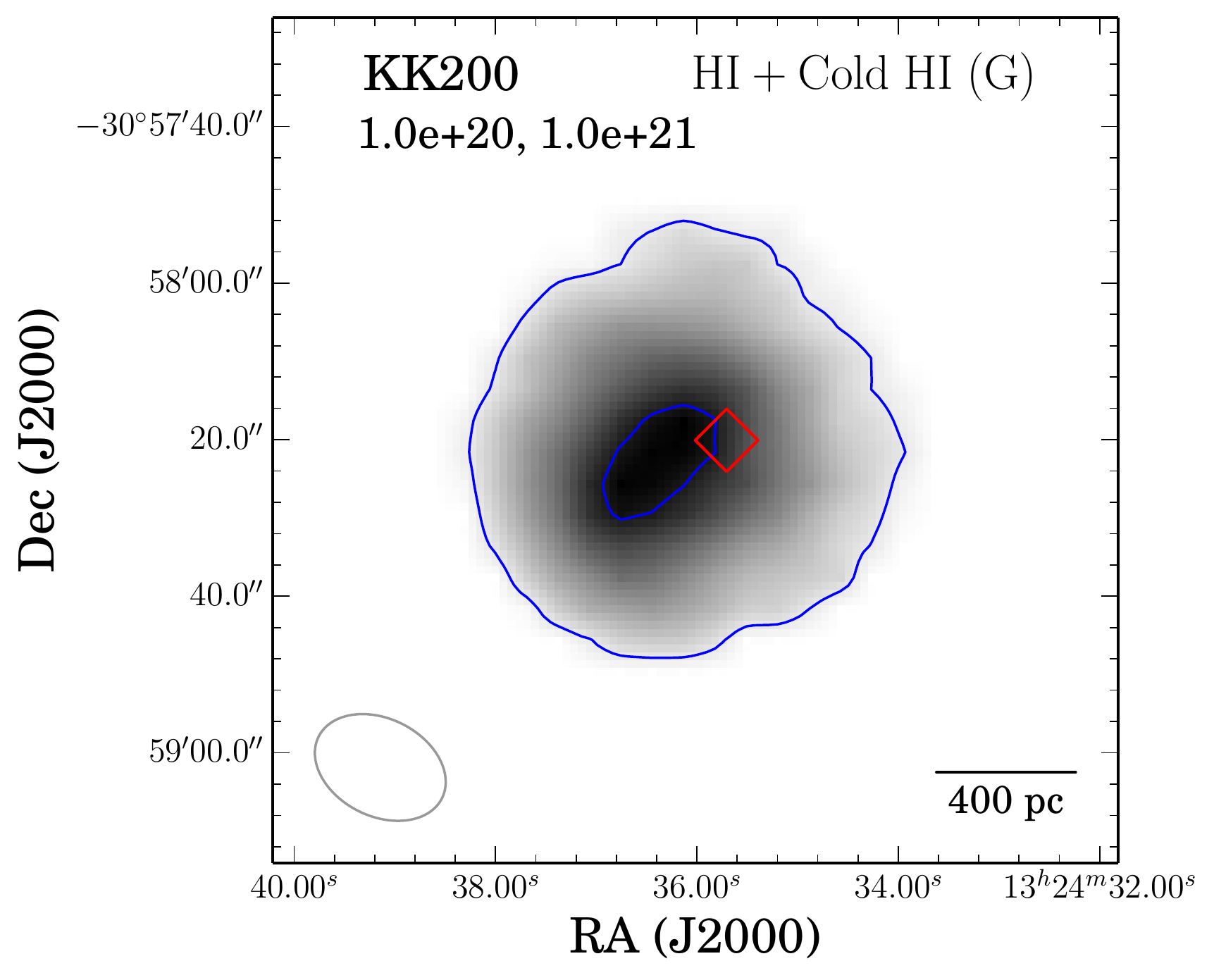}} &
\resizebox{55mm}{!}{\includegraphics{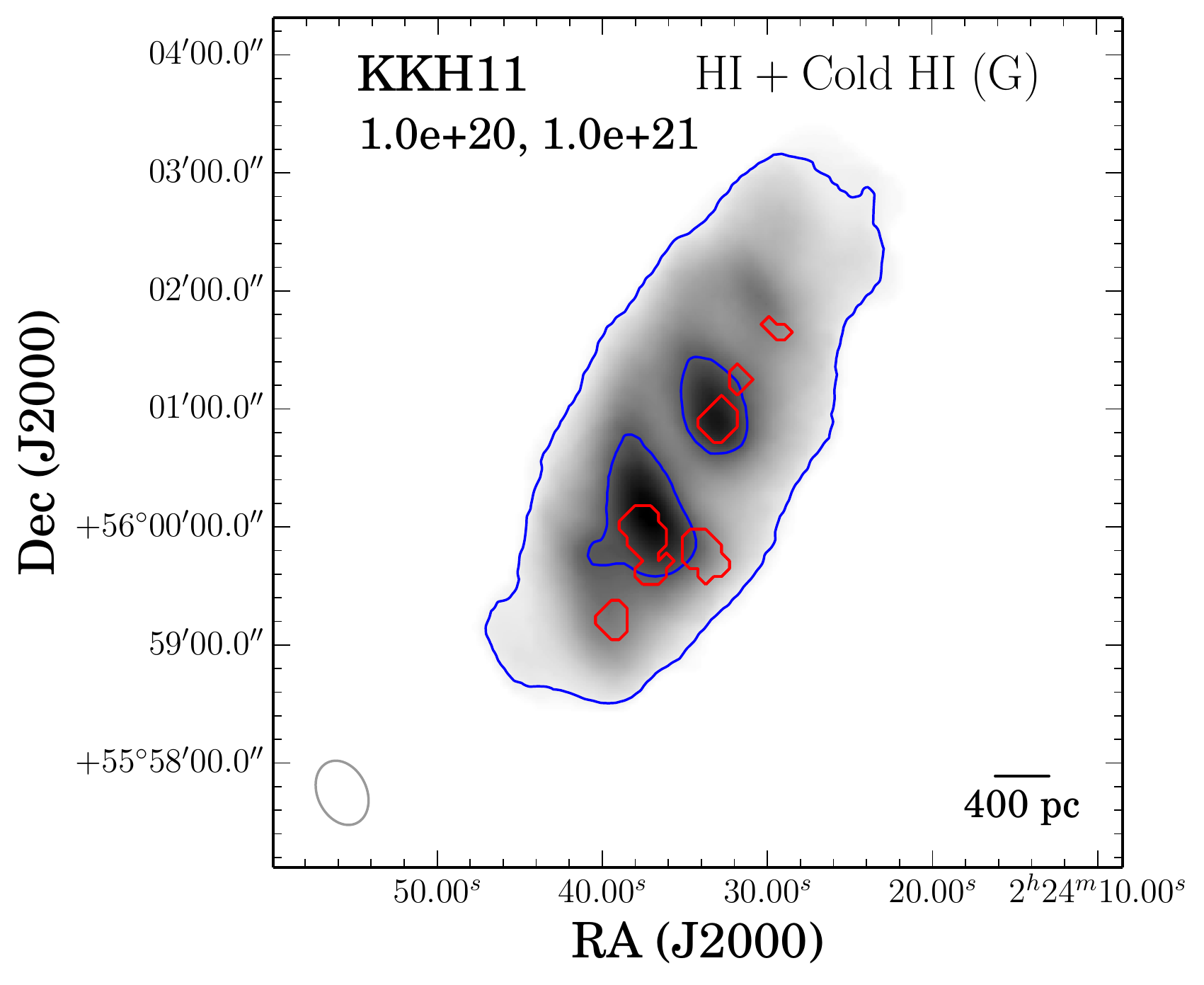}} \\
\end{tabular}
\end{center}
\caption{ Overlays of \HI (at 400 pc resolution, contours) with cold \HI as detected by Gaussian decomposition method. The total \HI is in gray scale with two blue contours representing two characteristic column densities as quoted on the top left corner. The red contour represents location of detected cold \HI in Gaussian decomposition method. As the cold \HI was not detected over many beams, the red contours are mostly representing same column density region.}
\label{ovr_h1_g}
\end{figure*}


\begin{figure*}
\begin{center}
\begin{tabular}{ccc}
\resizebox{50mm}{!}{\includegraphics{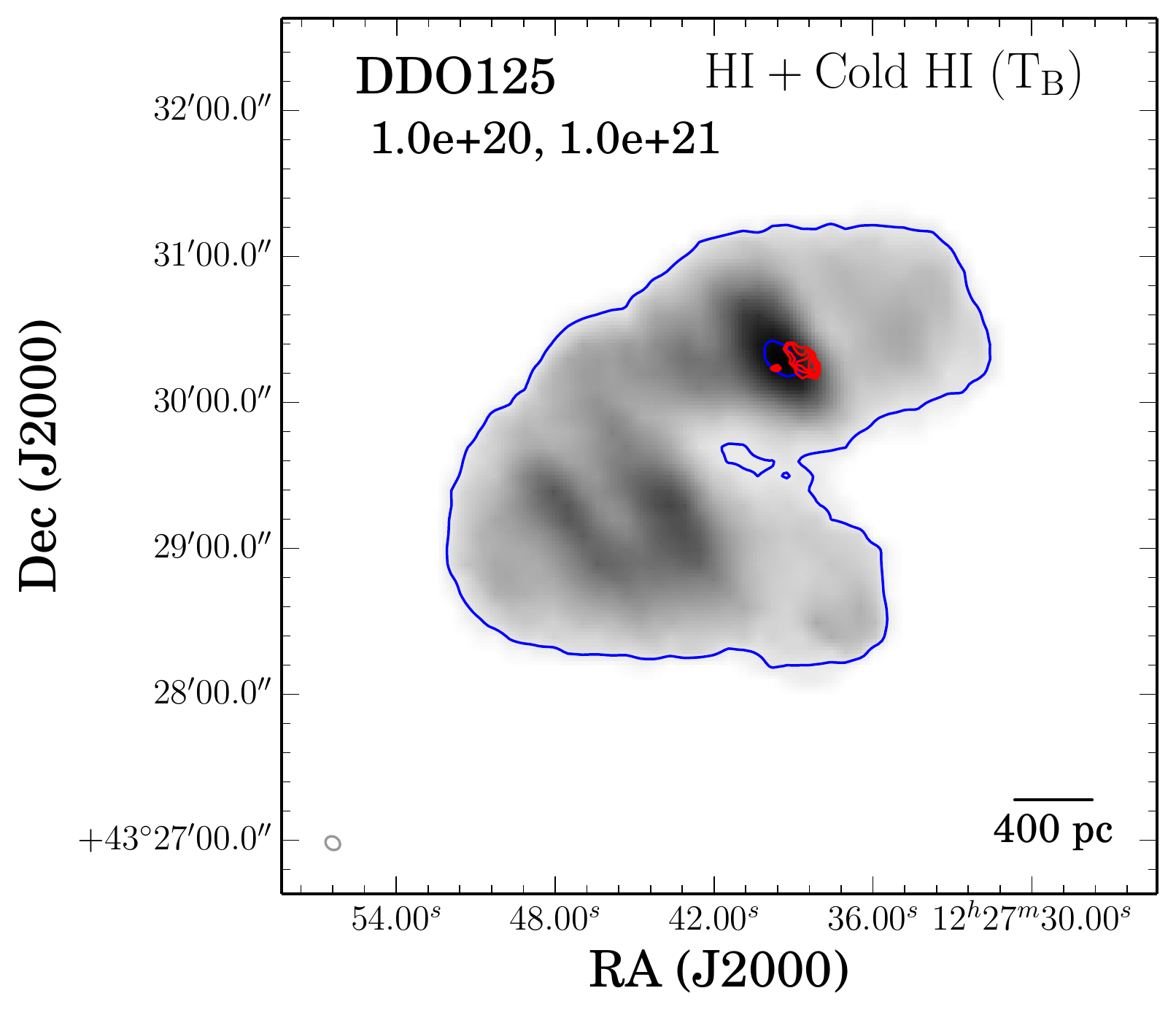}} &
\resizebox{50mm}{!}{\includegraphics{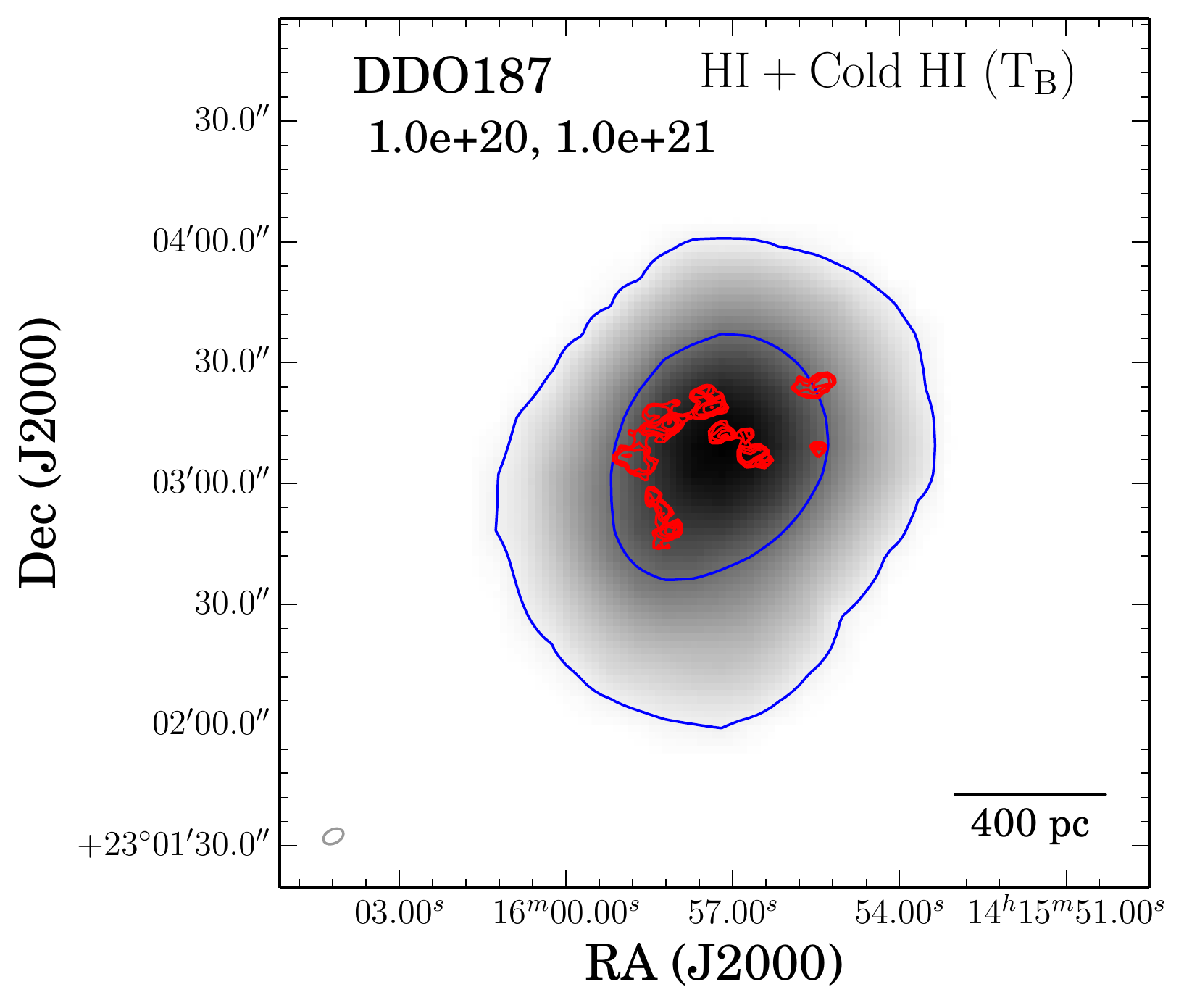}} &
\resizebox{50mm}{!}{\includegraphics{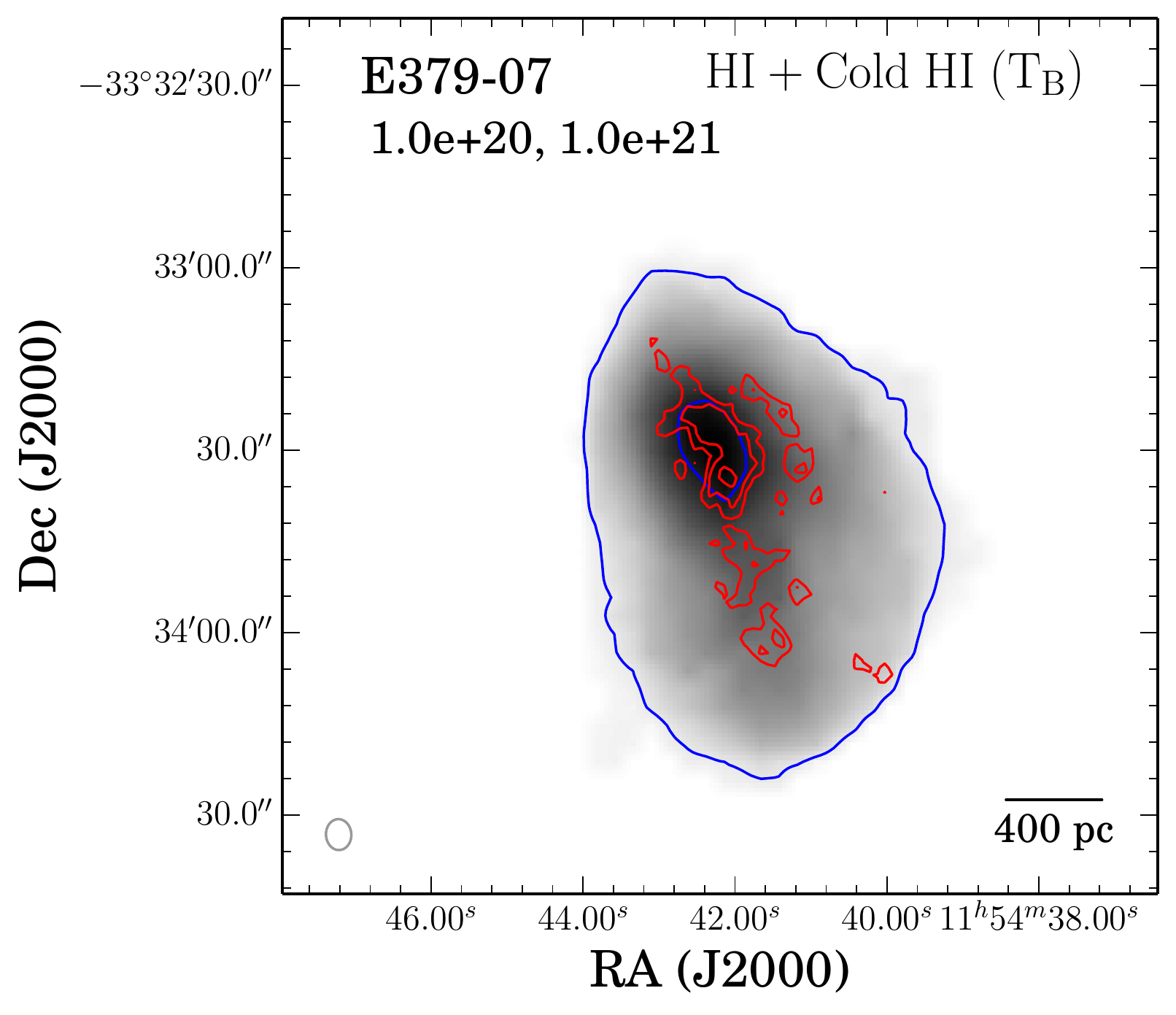}} \\

\resizebox{50mm}{!}{\includegraphics{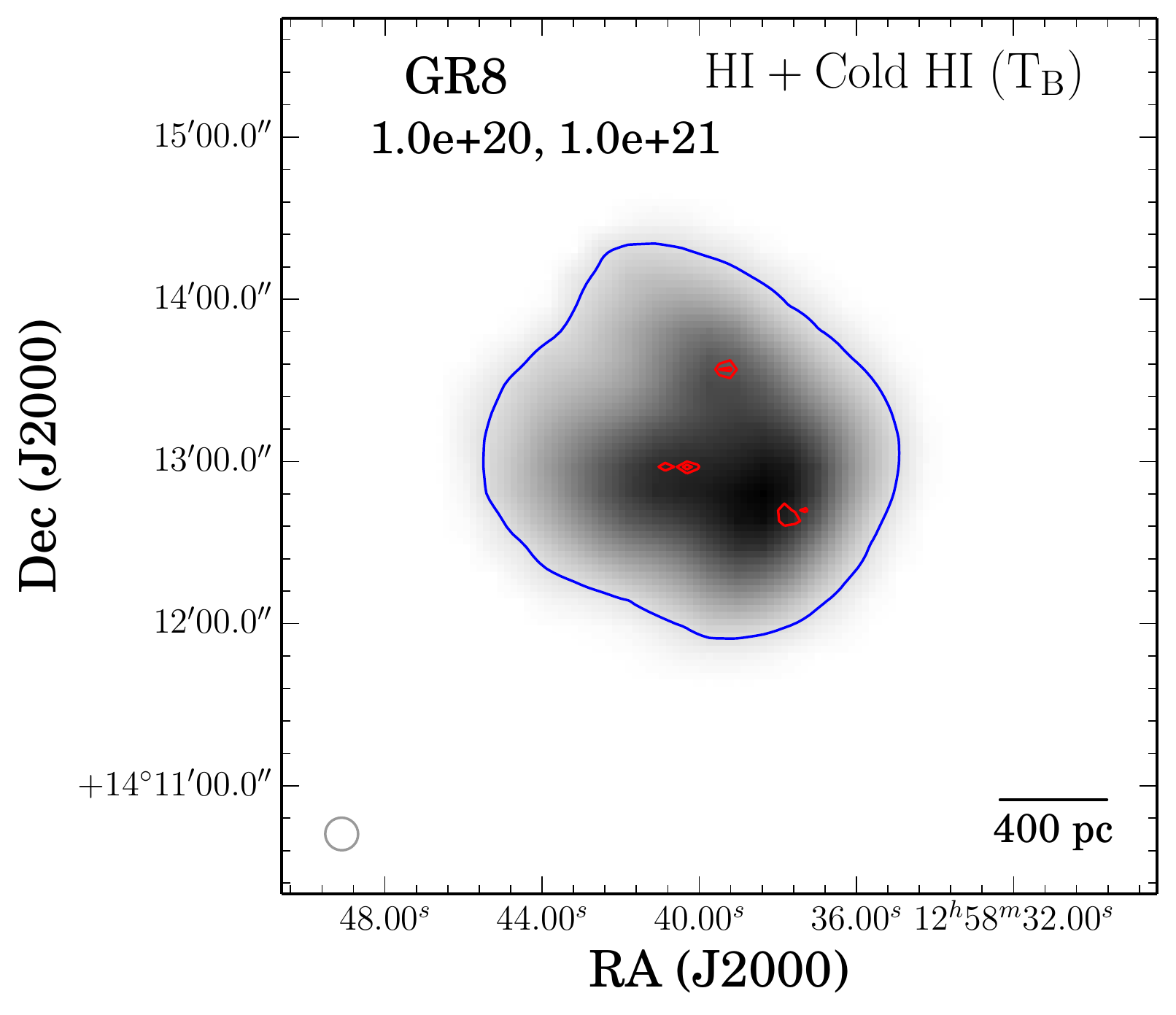}} &
\resizebox{50mm}{!}{\includegraphics{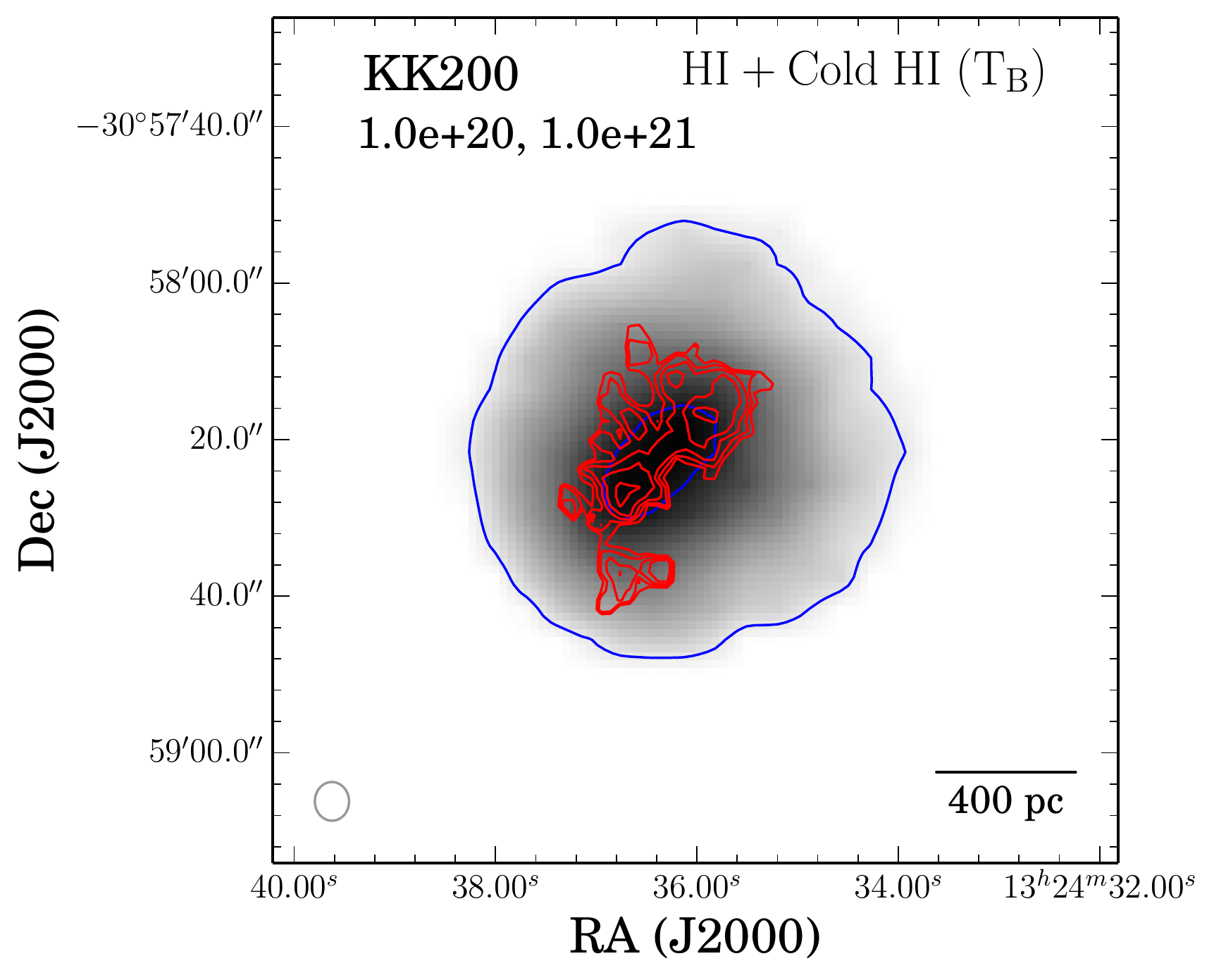}} &
\resizebox{50mm}{!}{\includegraphics{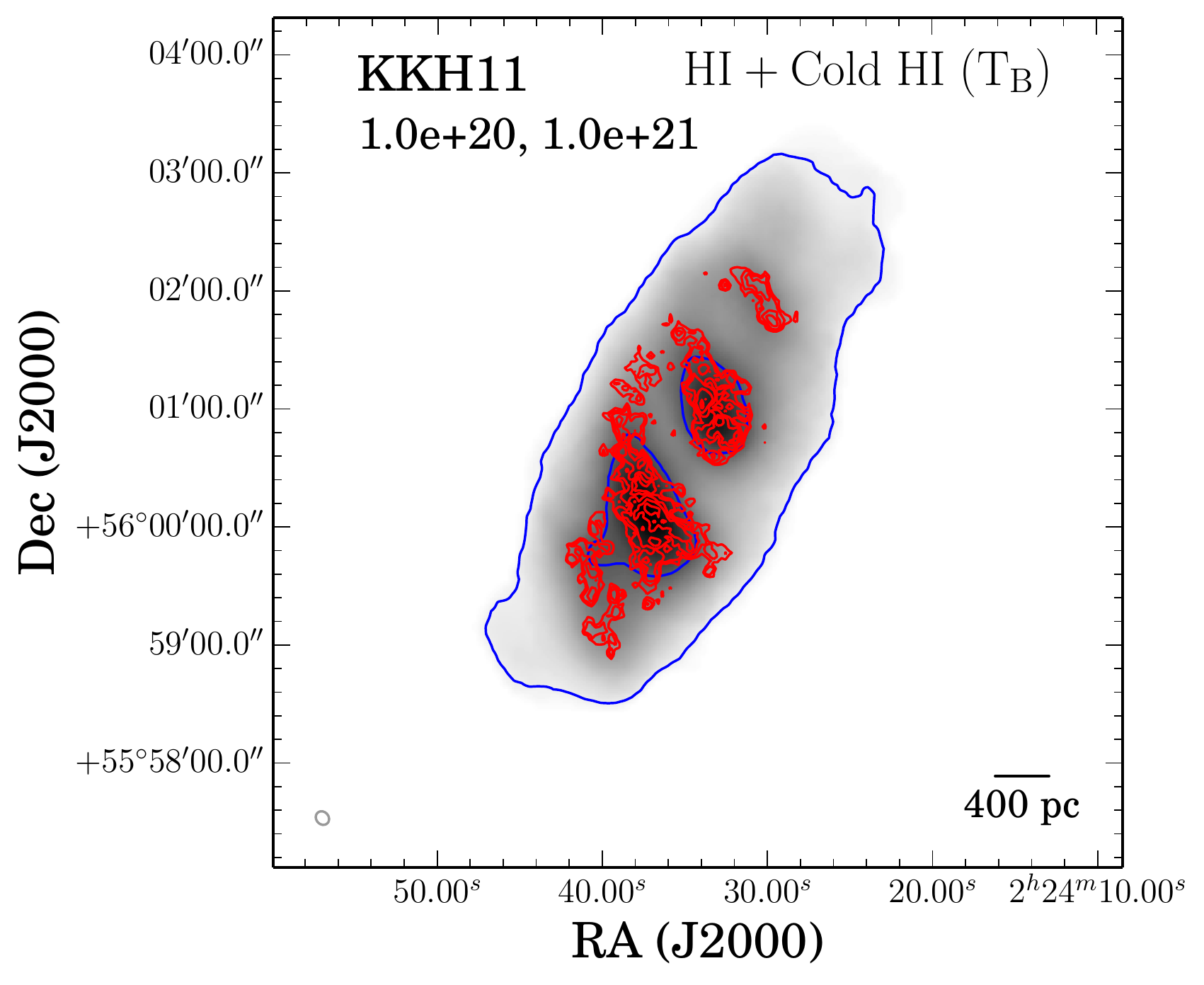}} \\

\resizebox{50mm}{!}{\includegraphics{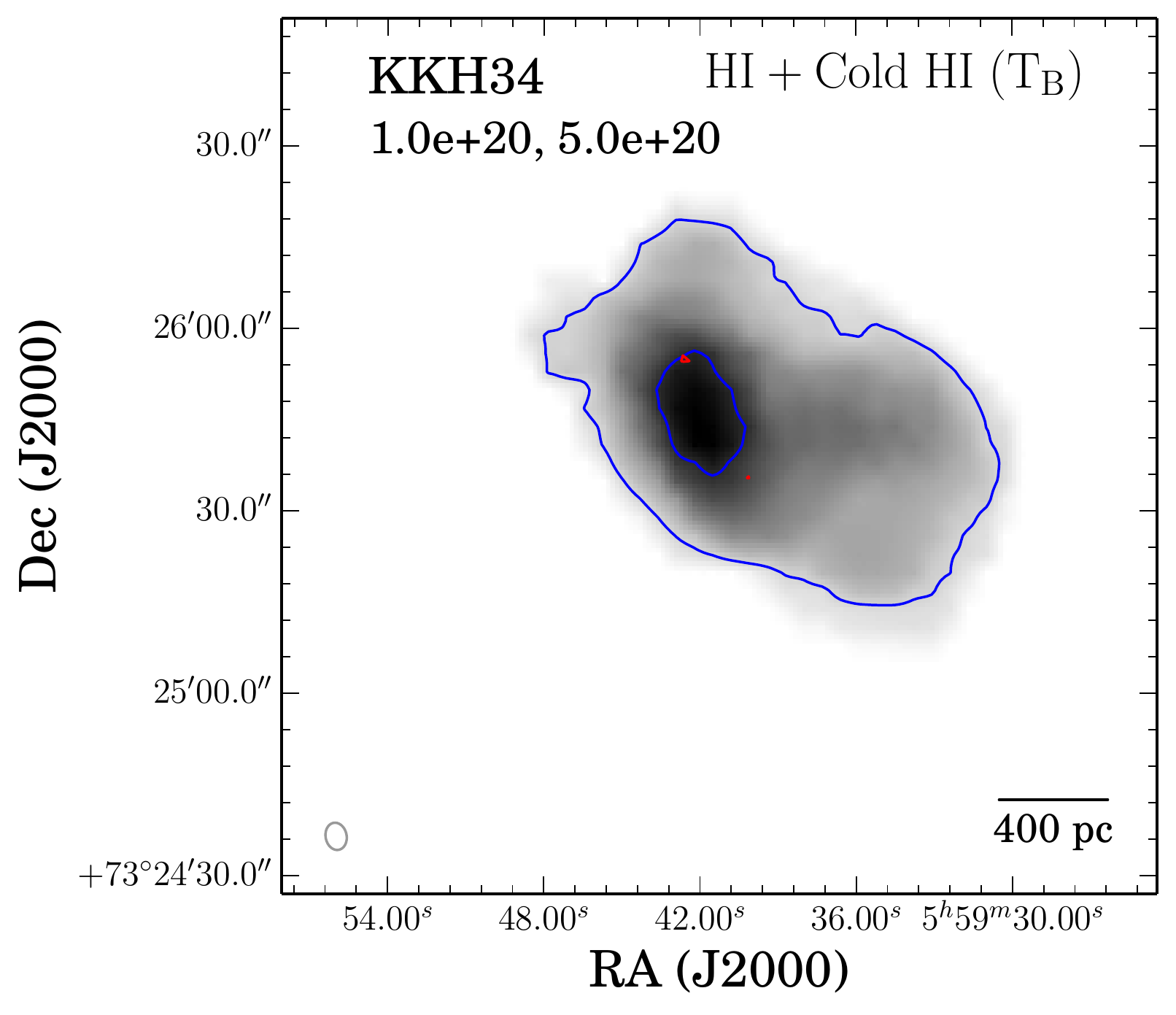}} &
\resizebox{50mm}{!}{\includegraphics{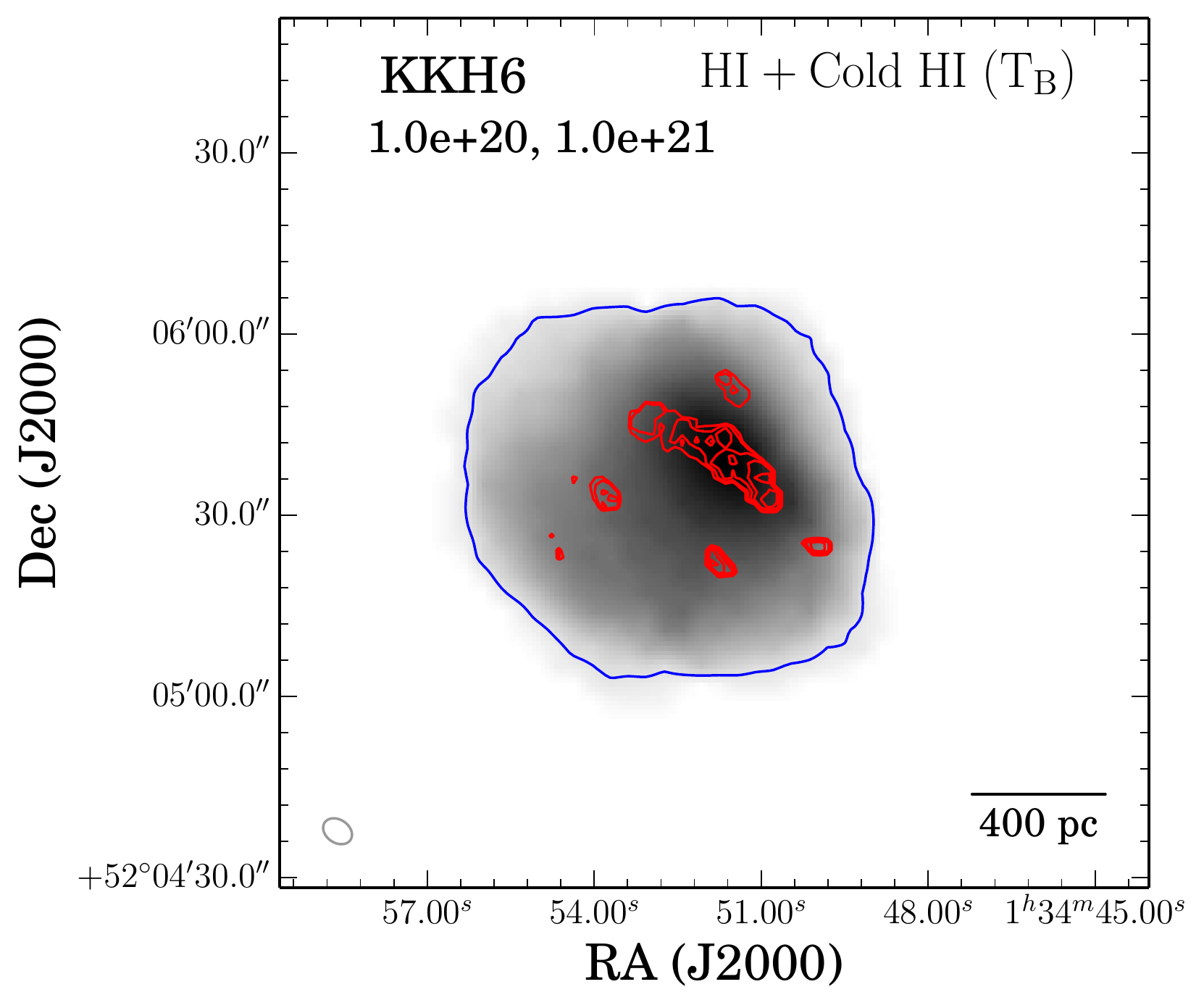}} &
\resizebox{50mm}{!}{\includegraphics{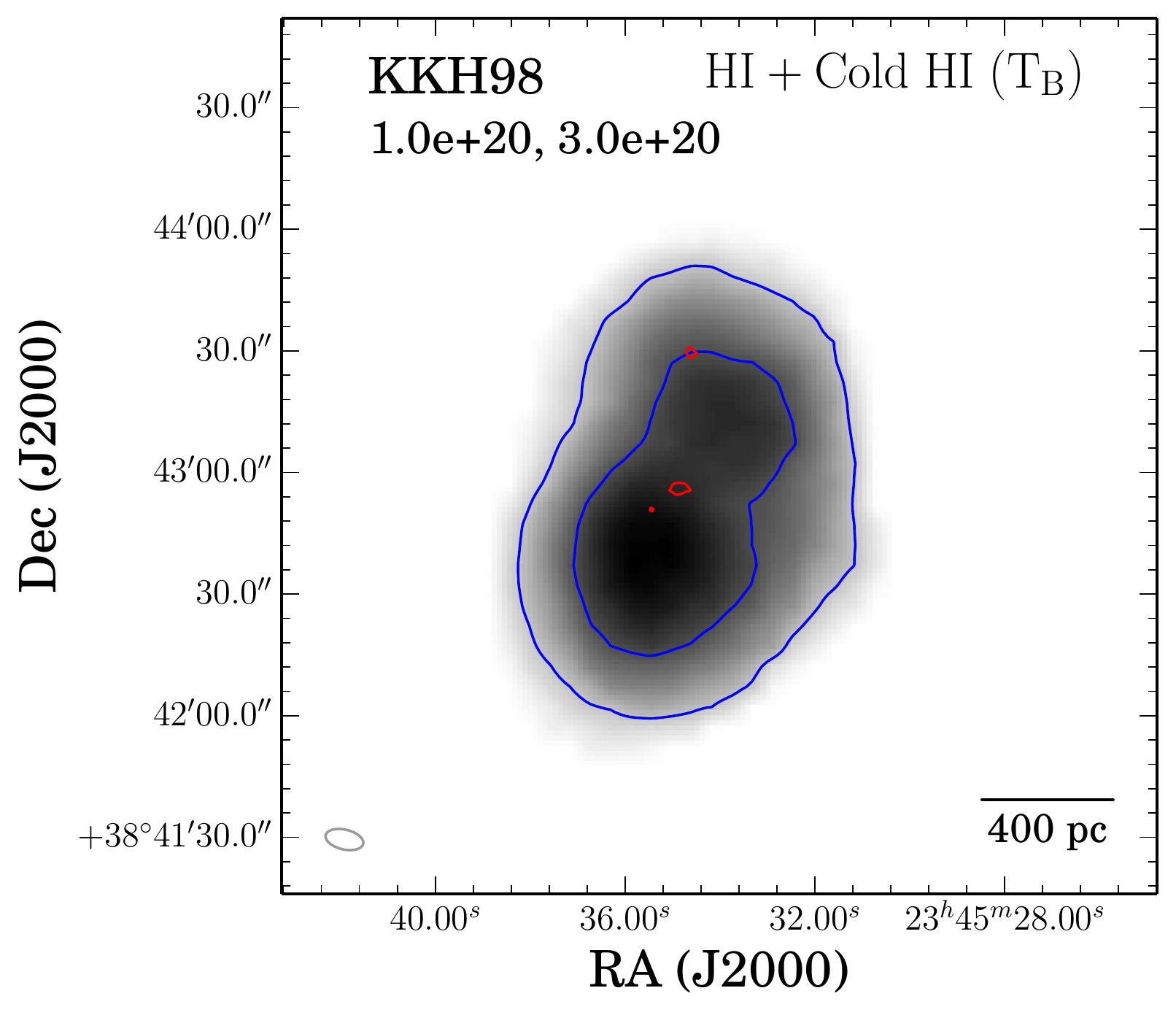}} \\

\resizebox{50mm}{!}{\includegraphics{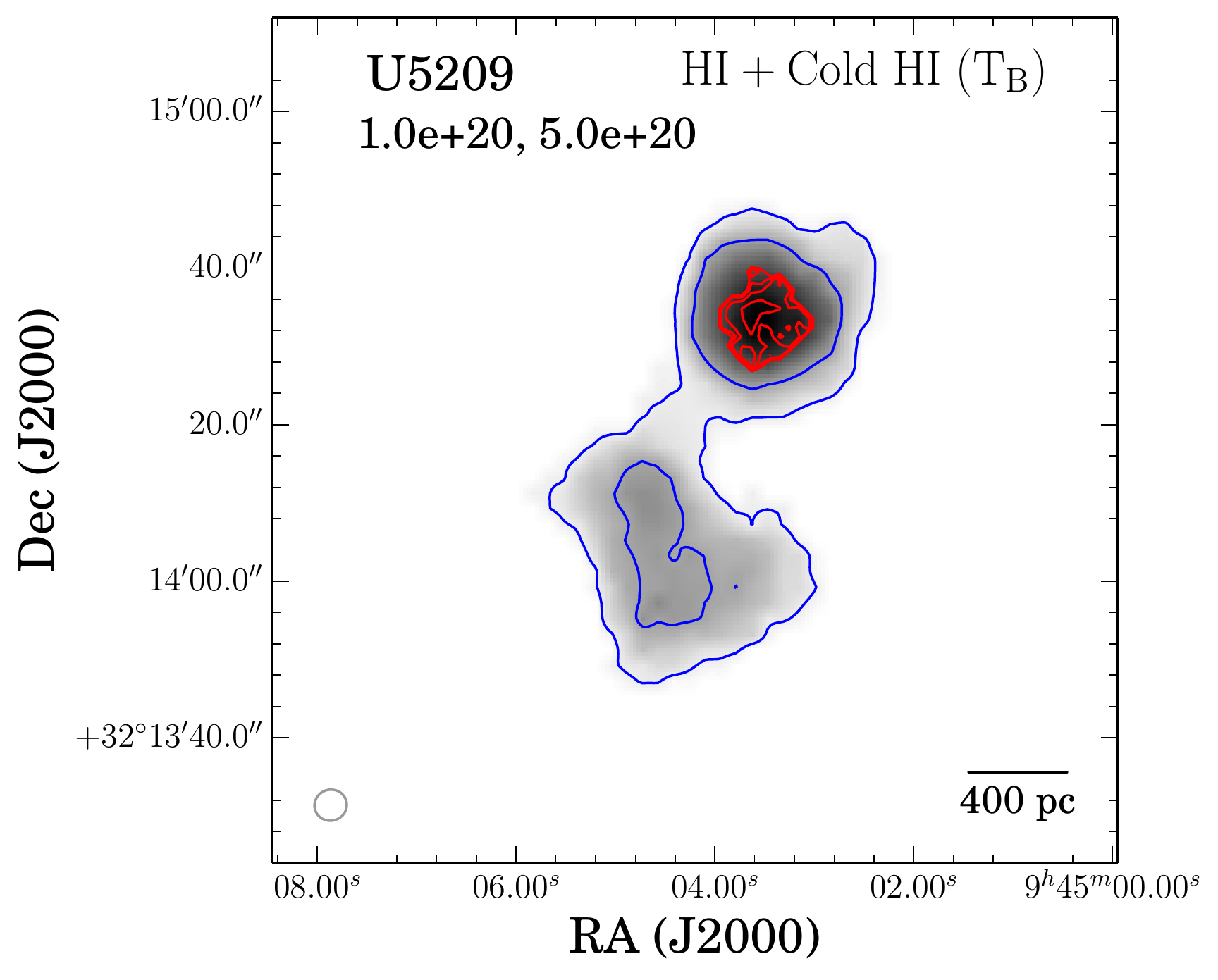}} &
\resizebox{50mm}{!}{\includegraphics{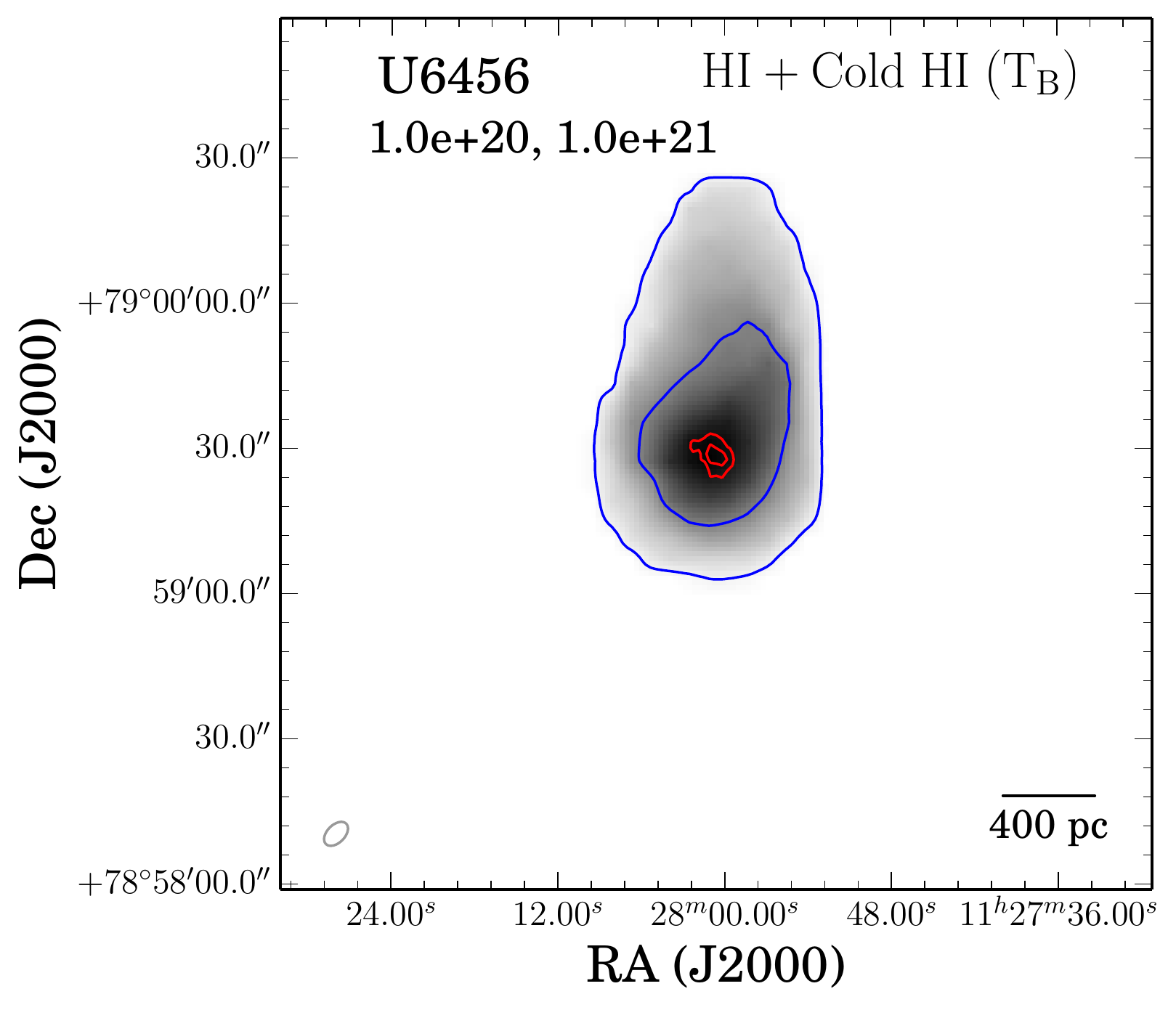}} &
\resizebox{50mm}{!}{\includegraphics{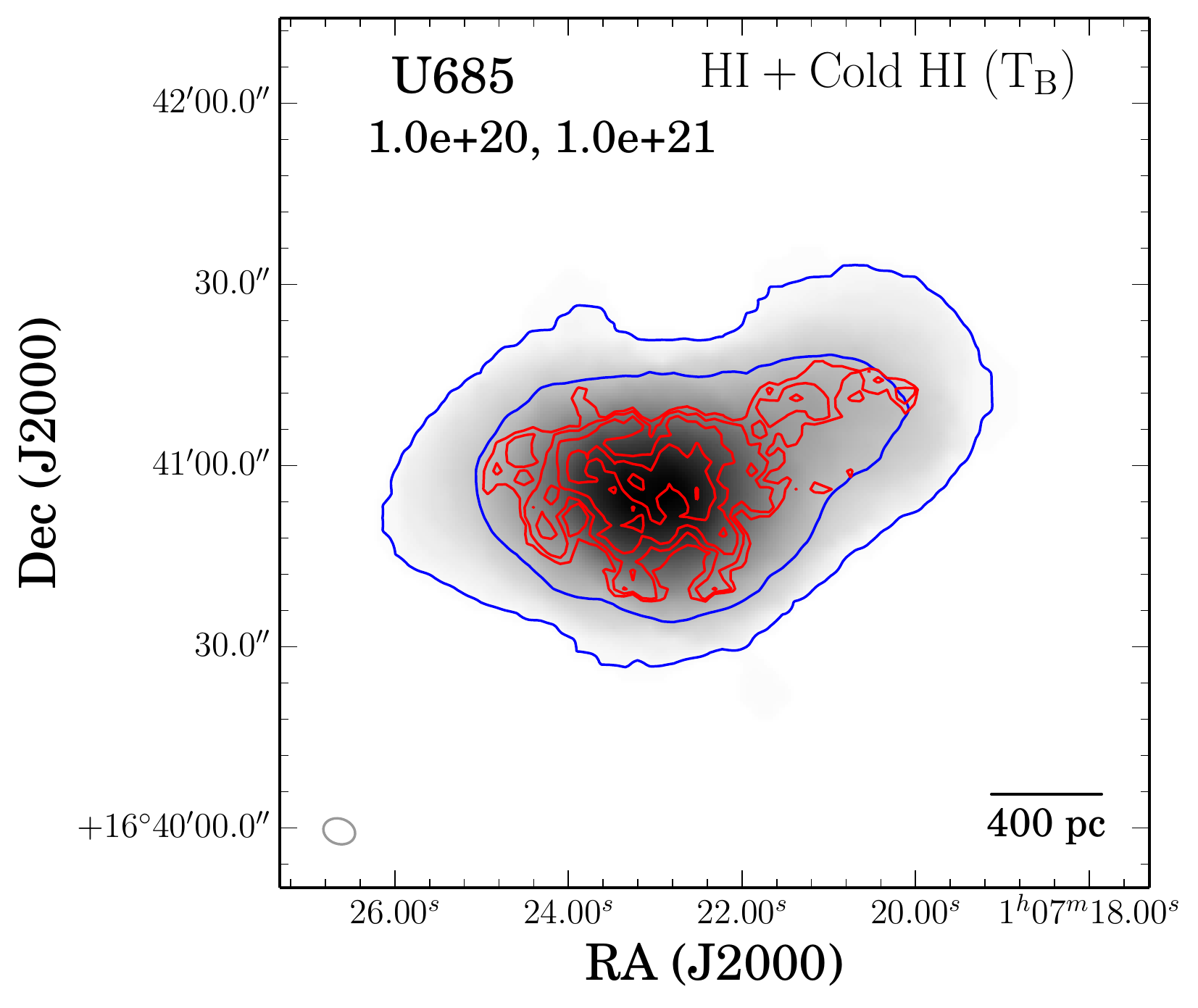}} \\

\resizebox{50mm}{!}{\includegraphics{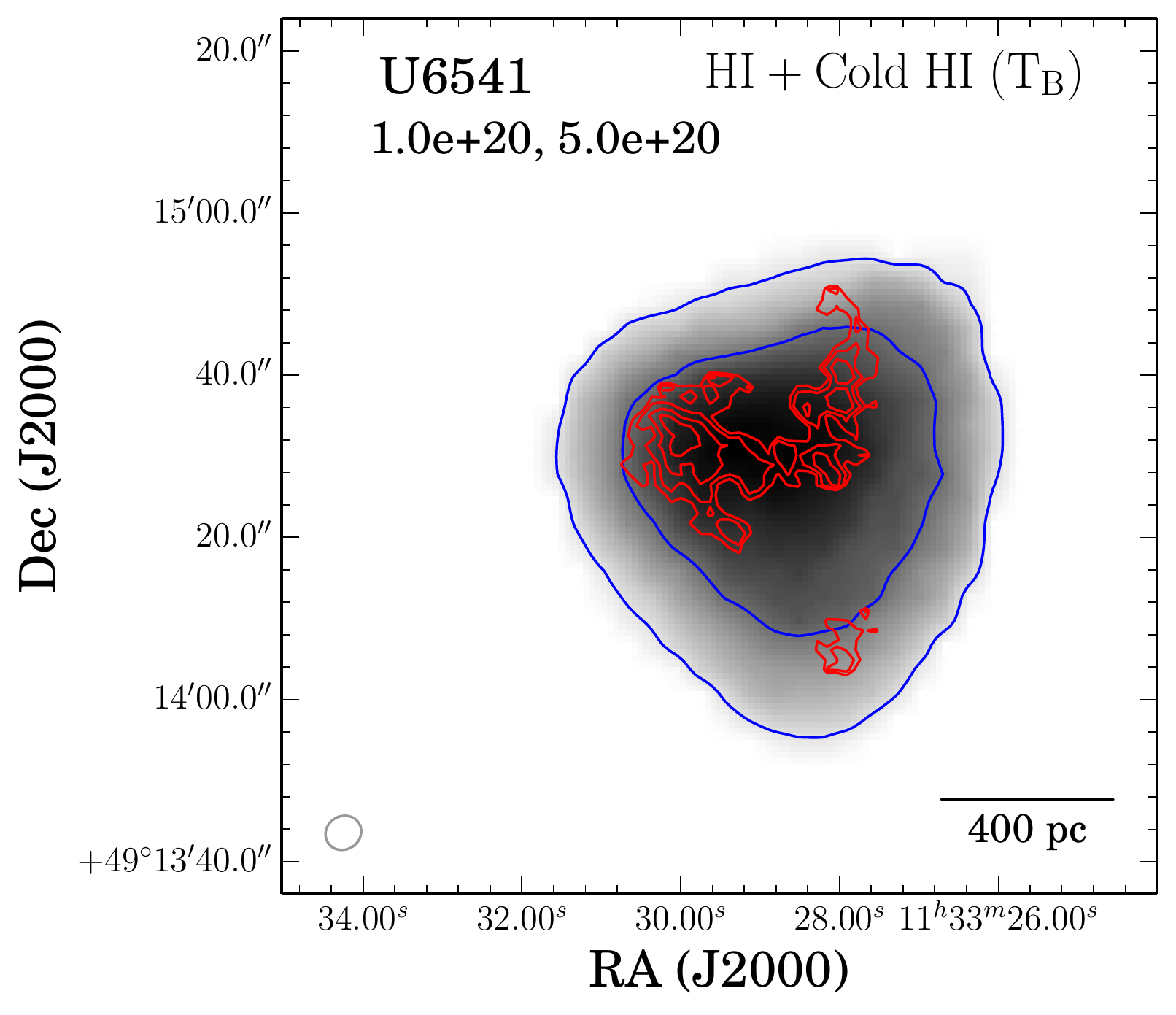}} &
\resizebox{50mm}{!}{\includegraphics{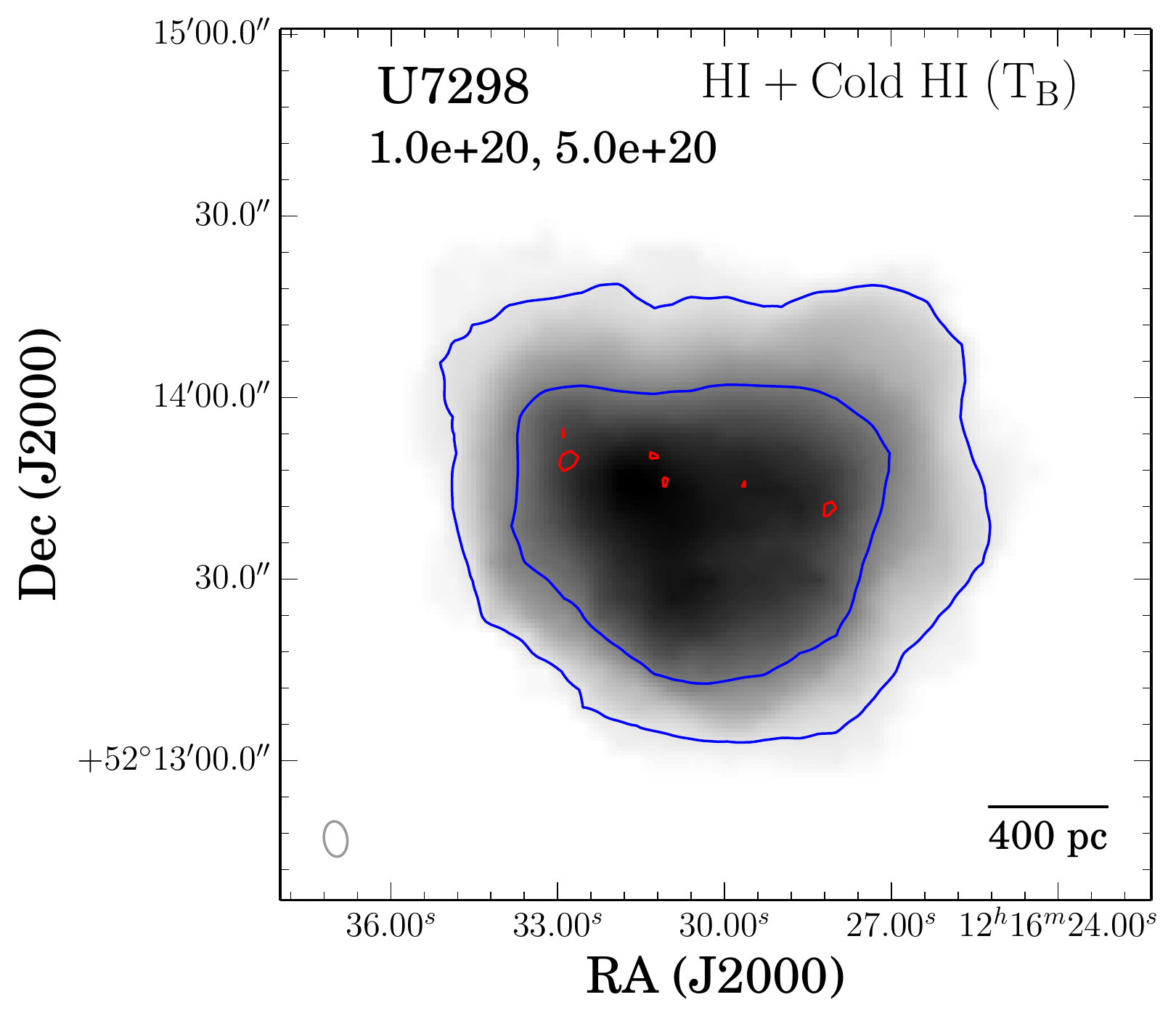}} &
\resizebox{50mm}{!}{\includegraphics{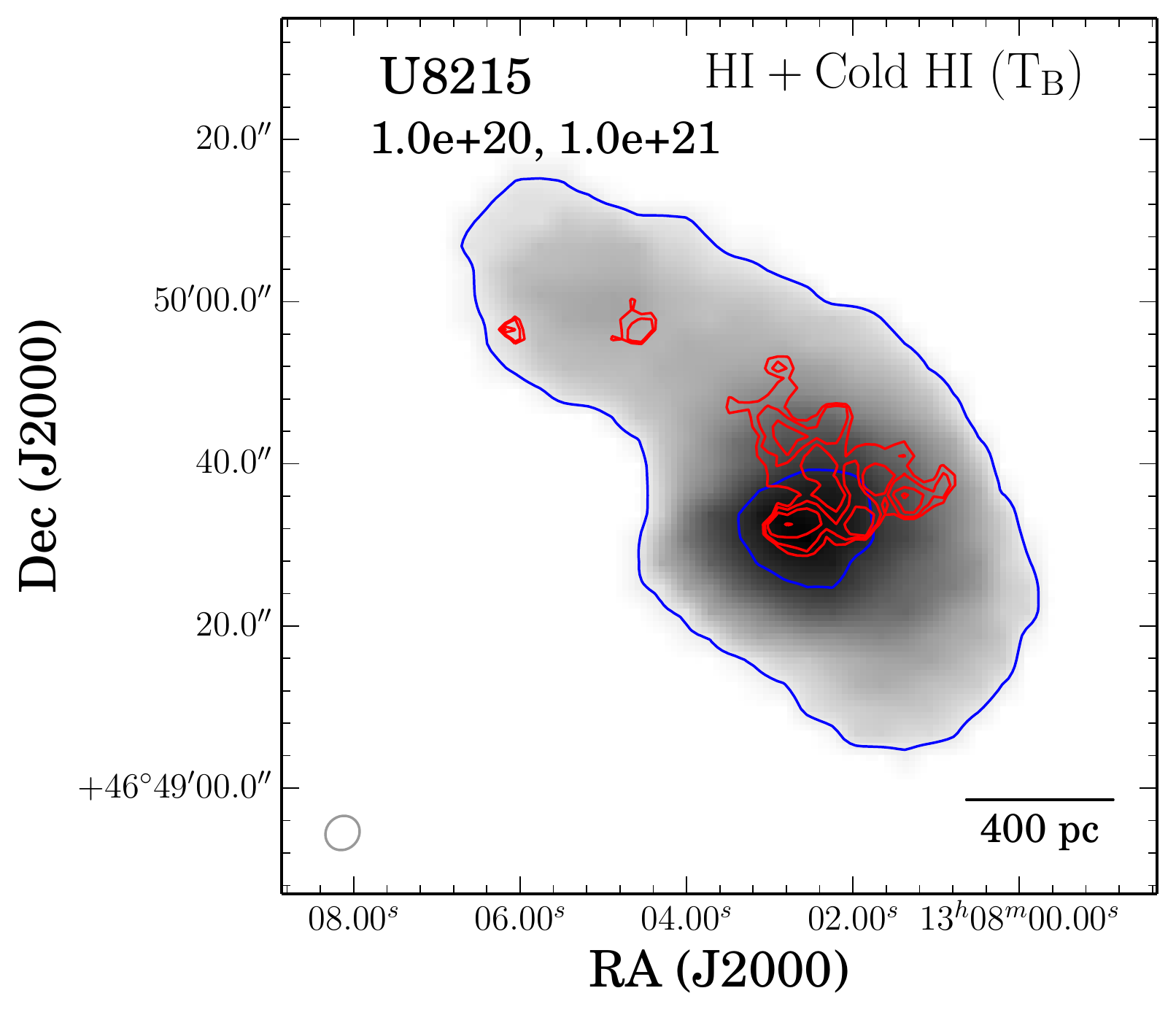}} \\

\end{tabular}
\end{center}
\caption{ Overlays of cold \HI detected using the \tb method (red contours) on the total \HI as traced by 400pc resolution \HI MOMNT maps (gray scale). The beam at the bottom left corner represents the resolution of the cold \HI recovered in \tb method. The conventions for the blue contours are the same as in Fig.~\ref{ovr_h1_g}.}
\label{ovr_h1_tb}
\end{figure*}

\begin{figure*}
\begin{center}
\begin{tabular}{ccc}
\resizebox{55mm}{!}{\includegraphics{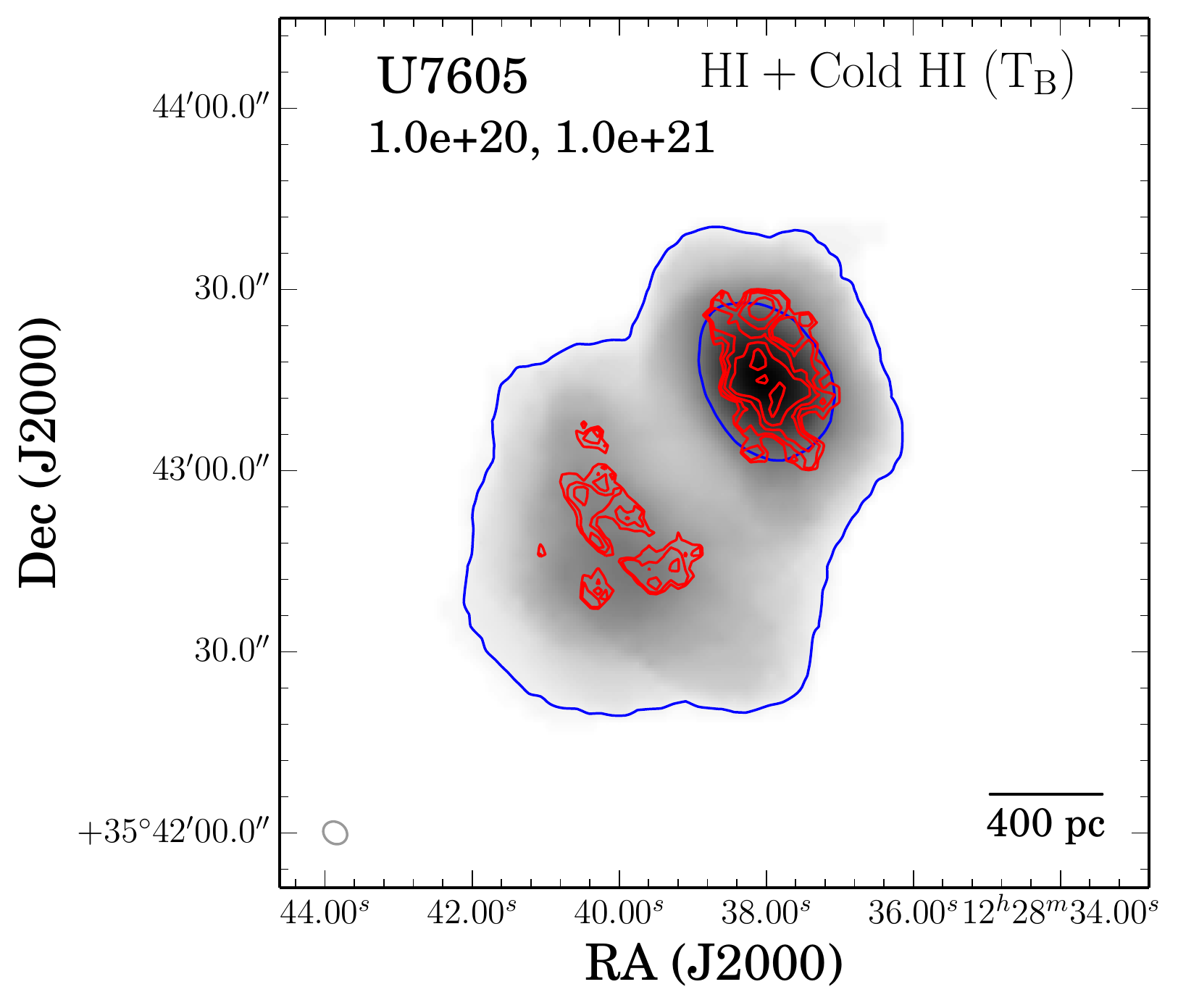}} &
\resizebox{55mm}{!}{\includegraphics{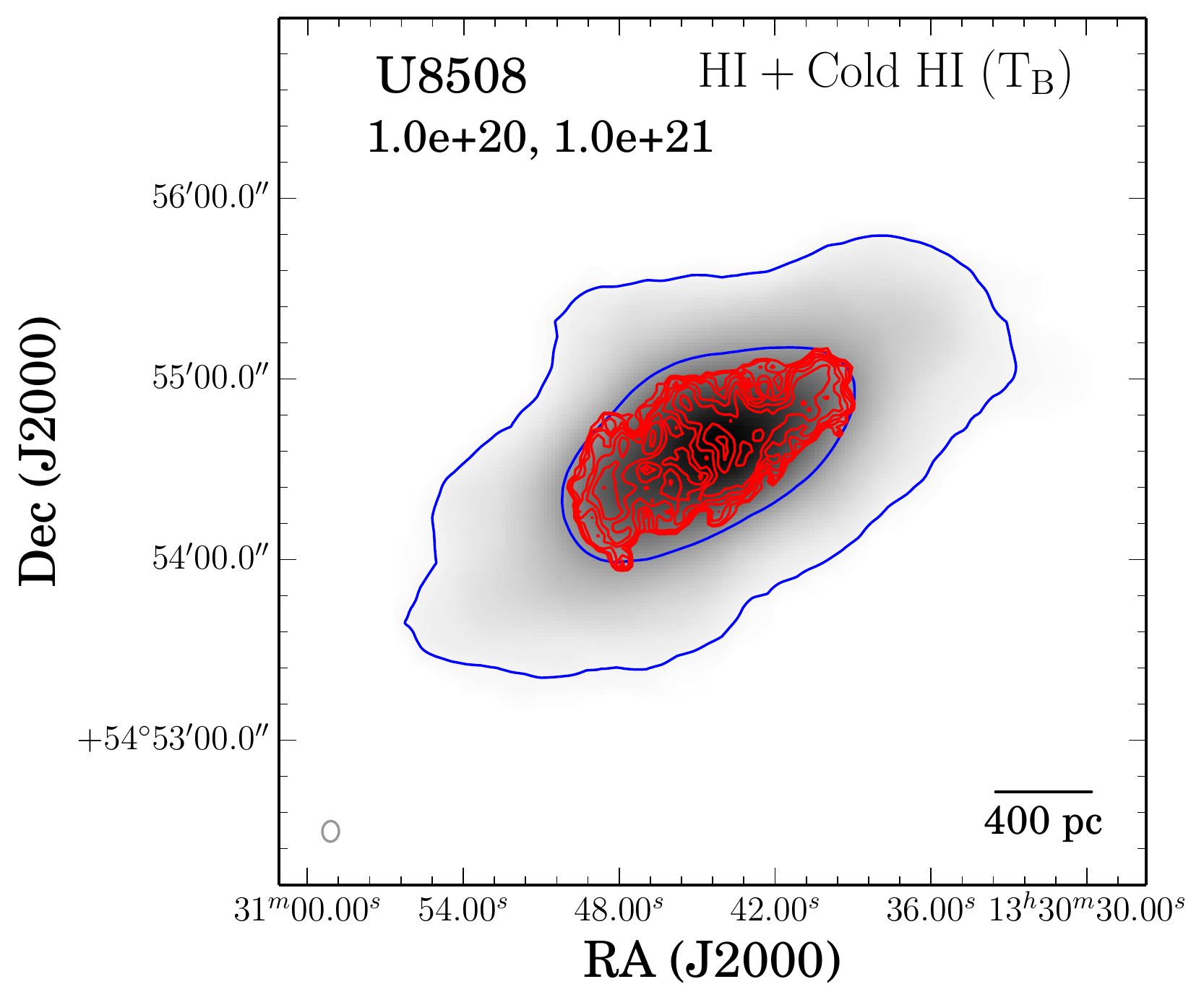}} &
\resizebox{55mm}{!}{\includegraphics{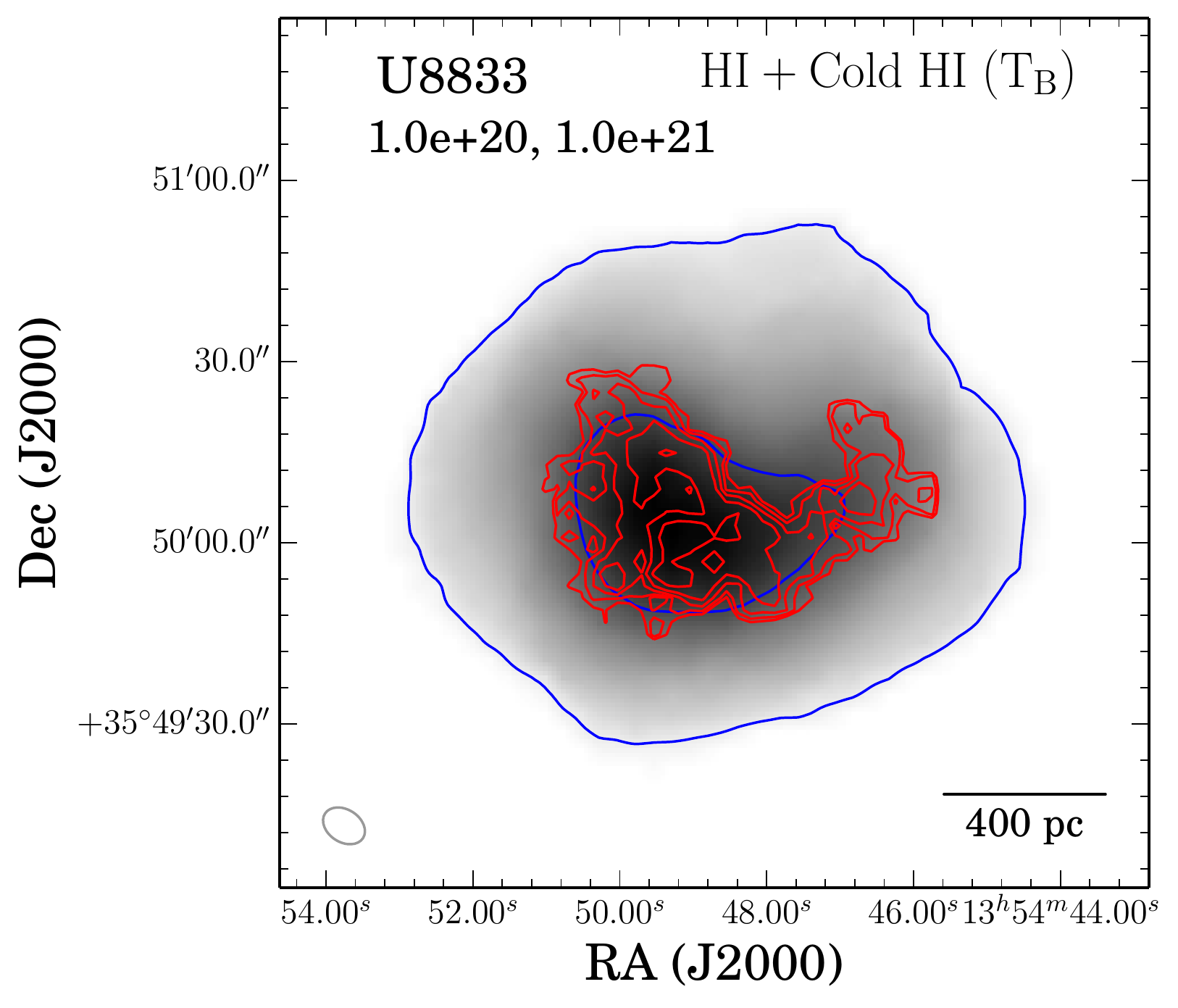}} \\
\resizebox{55mm}{!}{\includegraphics{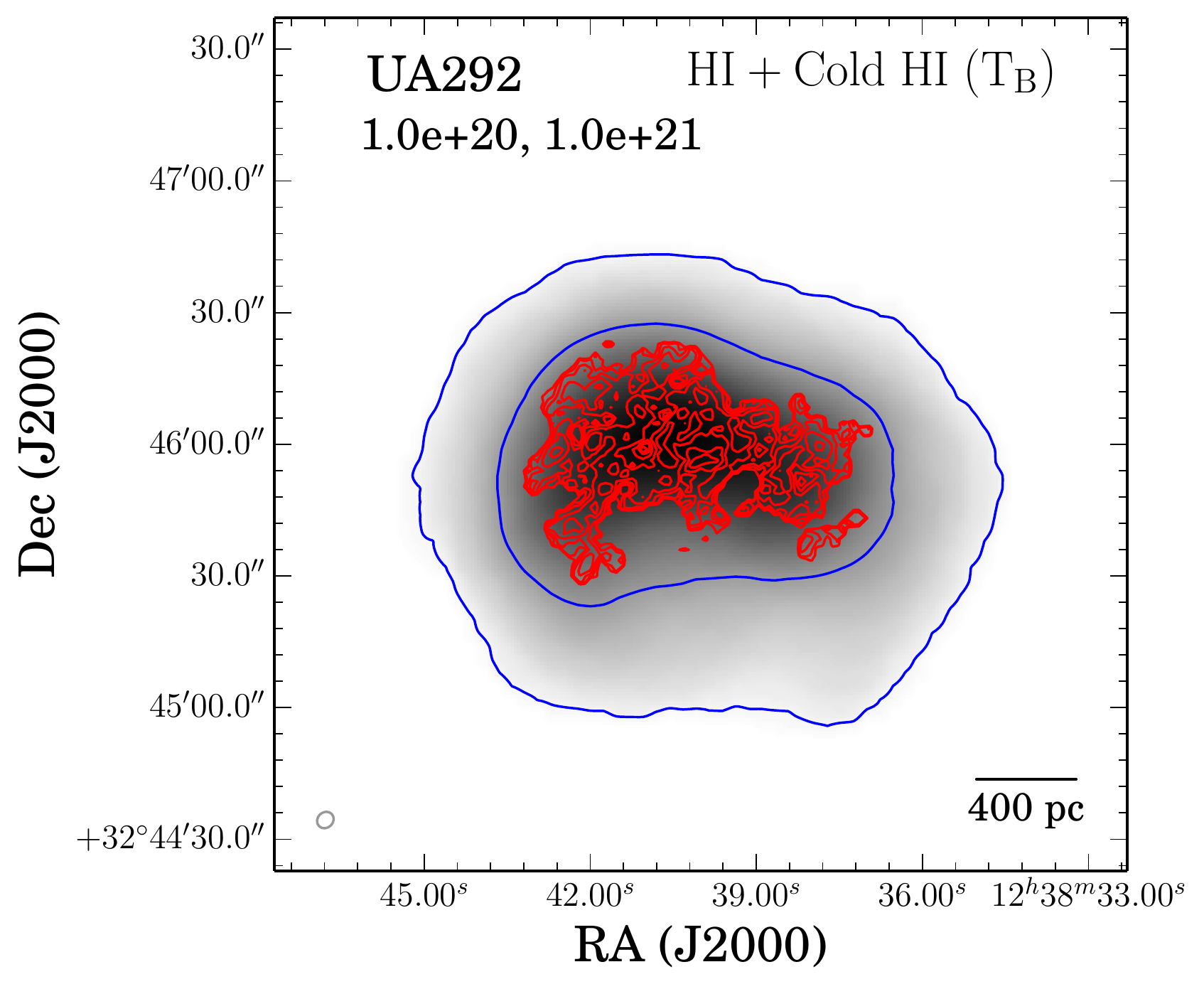}} \\
\end{tabular}
\end{center}
\caption{ Continuation of Fig.~\ref{ovr_h1_tb} }
\label{ovr_h1_tb2}
\end{figure*}


\begin{figure*}
\begin{center}
\begin{tabular}{ccc}
\resizebox{65mm}{!}{\includegraphics{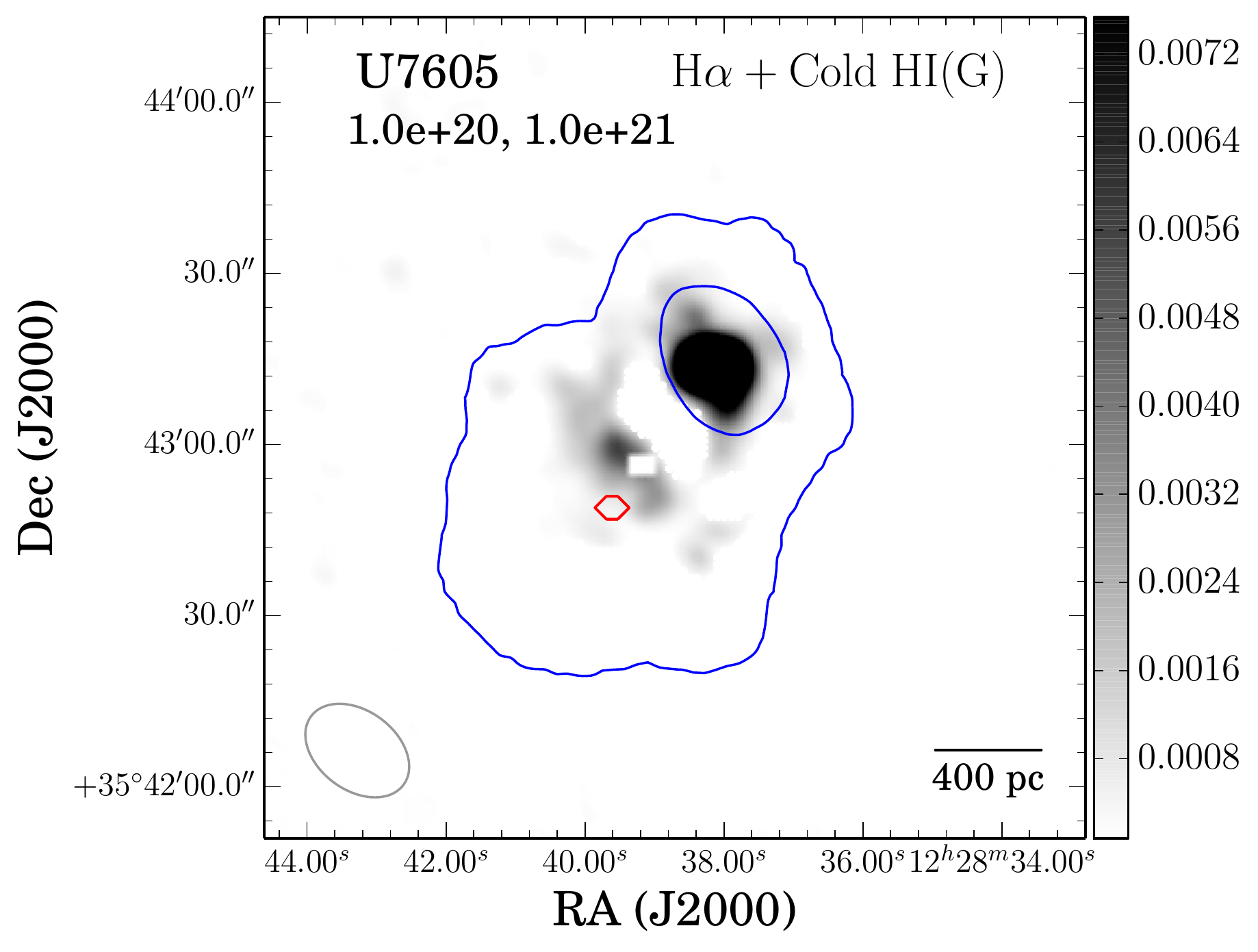}} &
\resizebox{65mm}{!}{\includegraphics{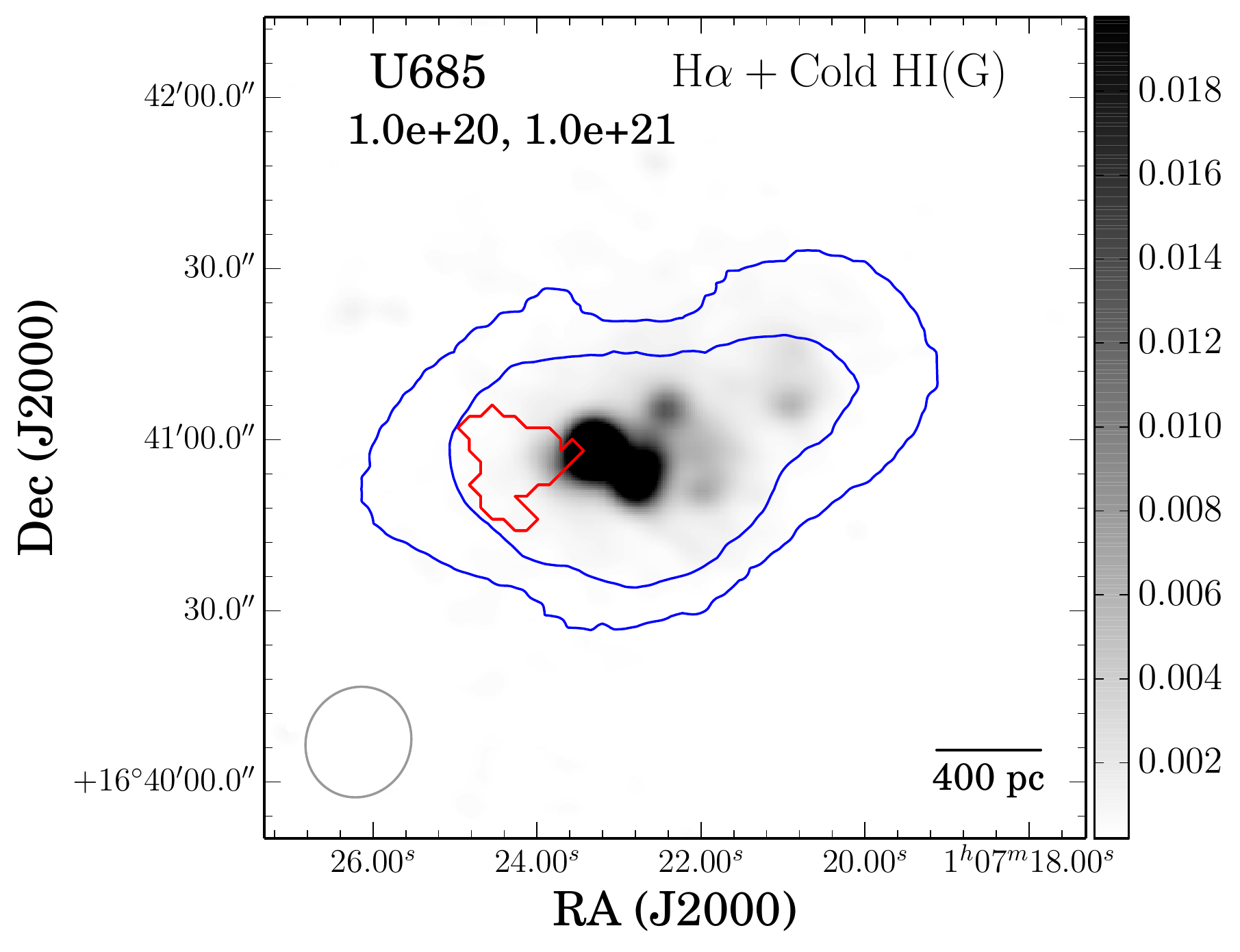}} \\
\end{tabular}
\end{center}
\caption{ Overlays of cold \HI detected using the Gaussian decomposition method (red contours) on \ha star formation in gray scale. The blue contours are the same as Fig.~\ref{ovr_h1_g}. The colorbar represents the \ha star formation rate density in the unit of \msyrkpc.}
\label{ovr_ha_g}
\end{figure*}

\begin{figure*}
\begin{center}
\begin{tabular}{ccc}
\resizebox{55mm}{!}{\includegraphics{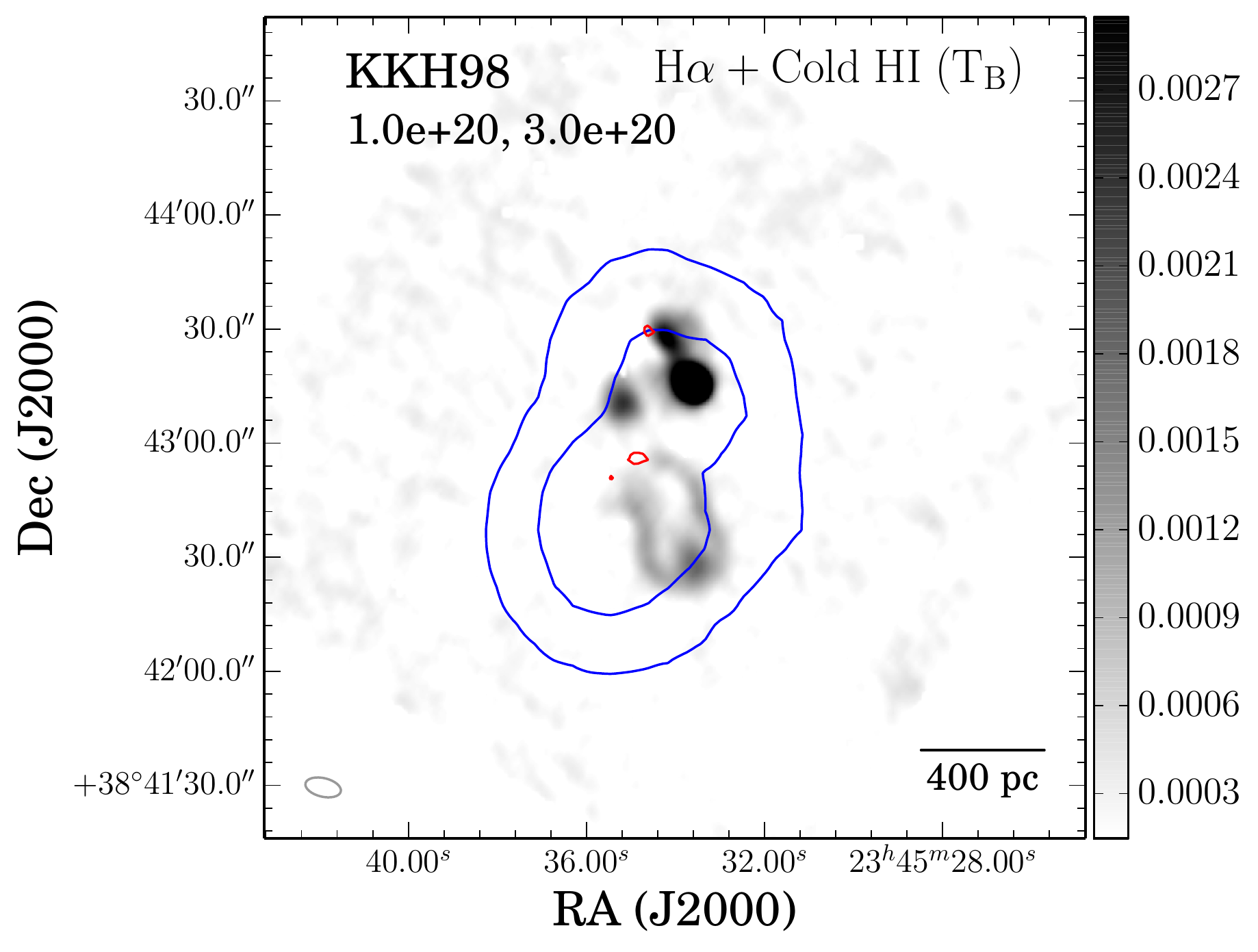}} &
\resizebox{55mm}{!}{\includegraphics{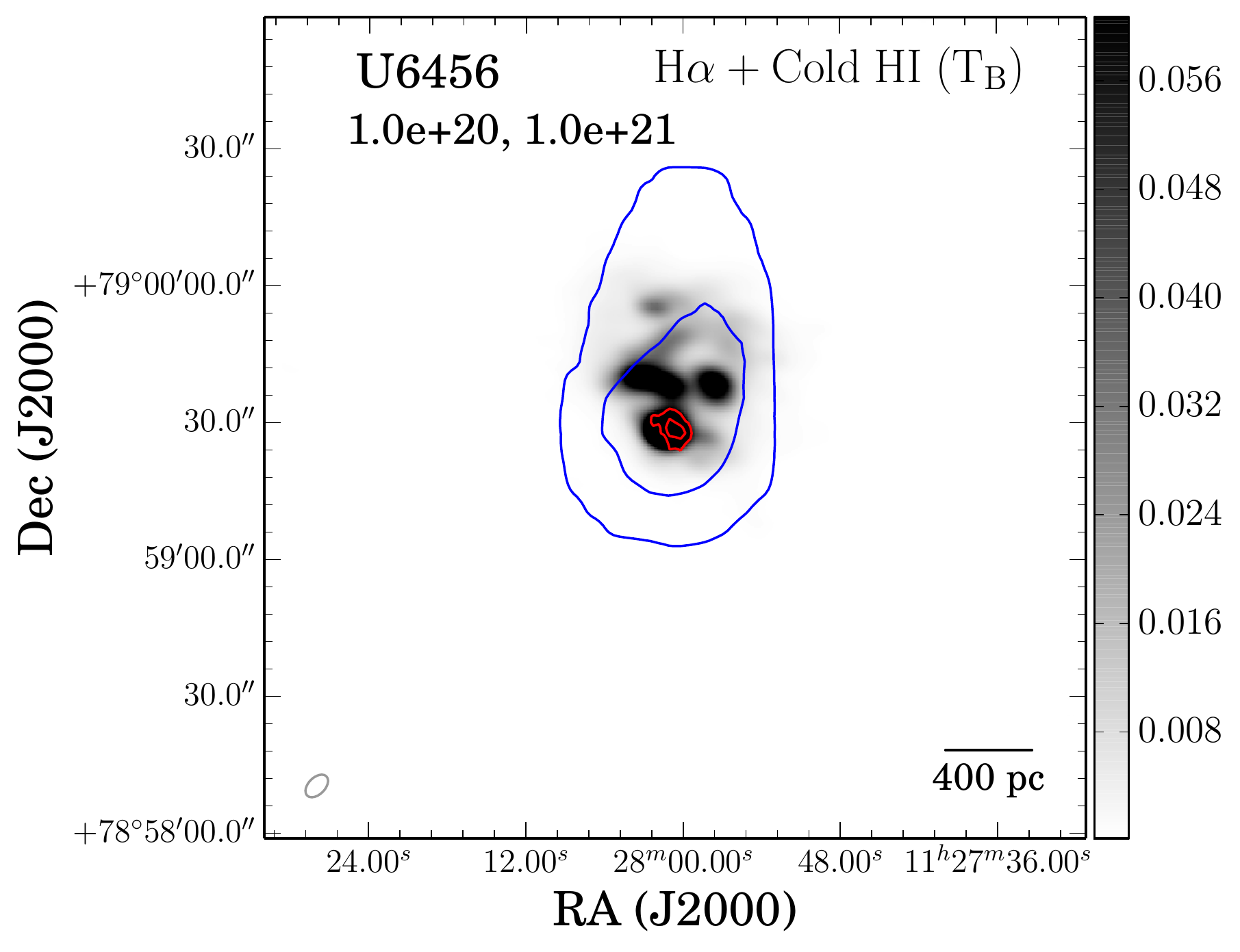}} &
\resizebox{55mm}{!}{\includegraphics{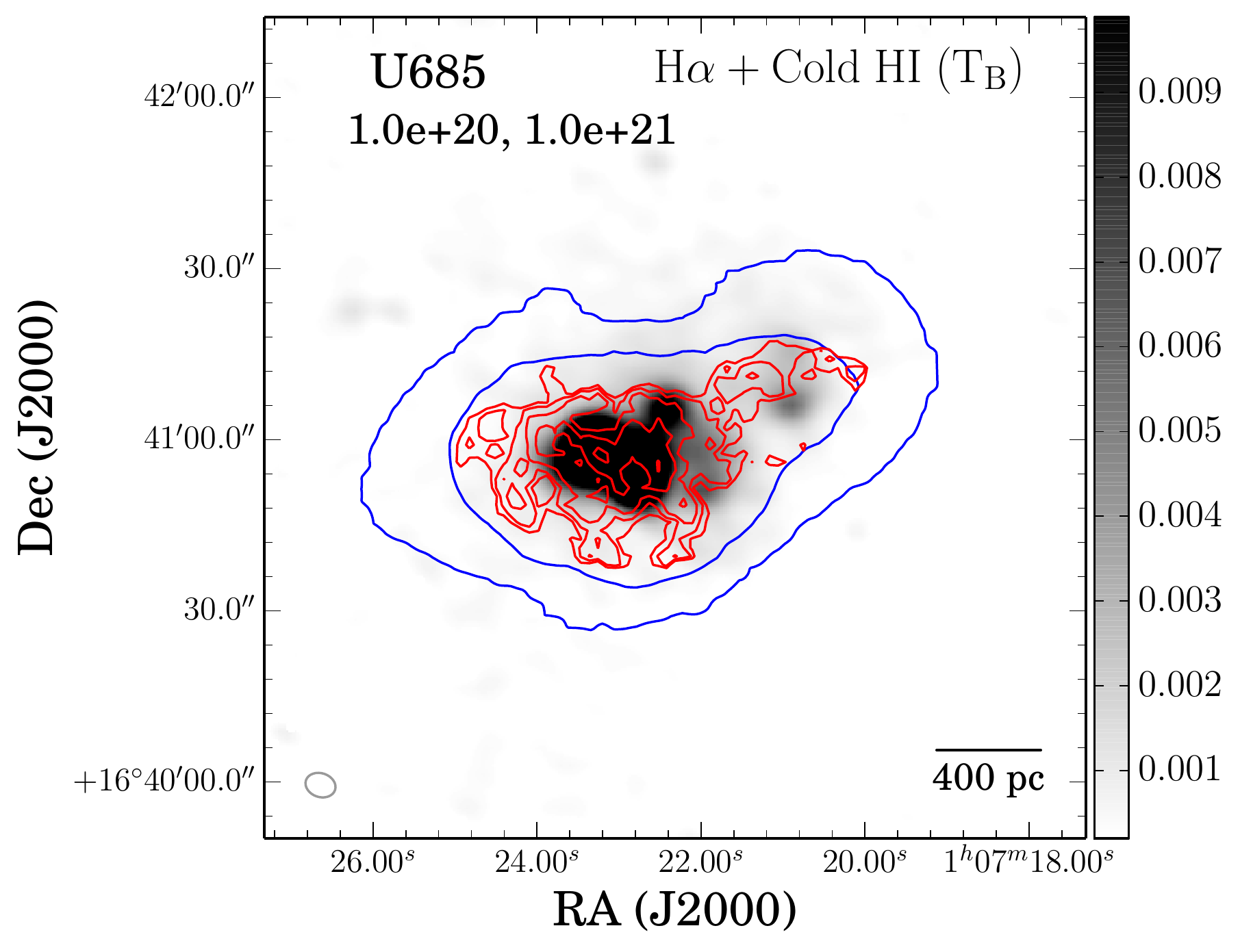}} \\

\resizebox{55mm}{!}{\includegraphics{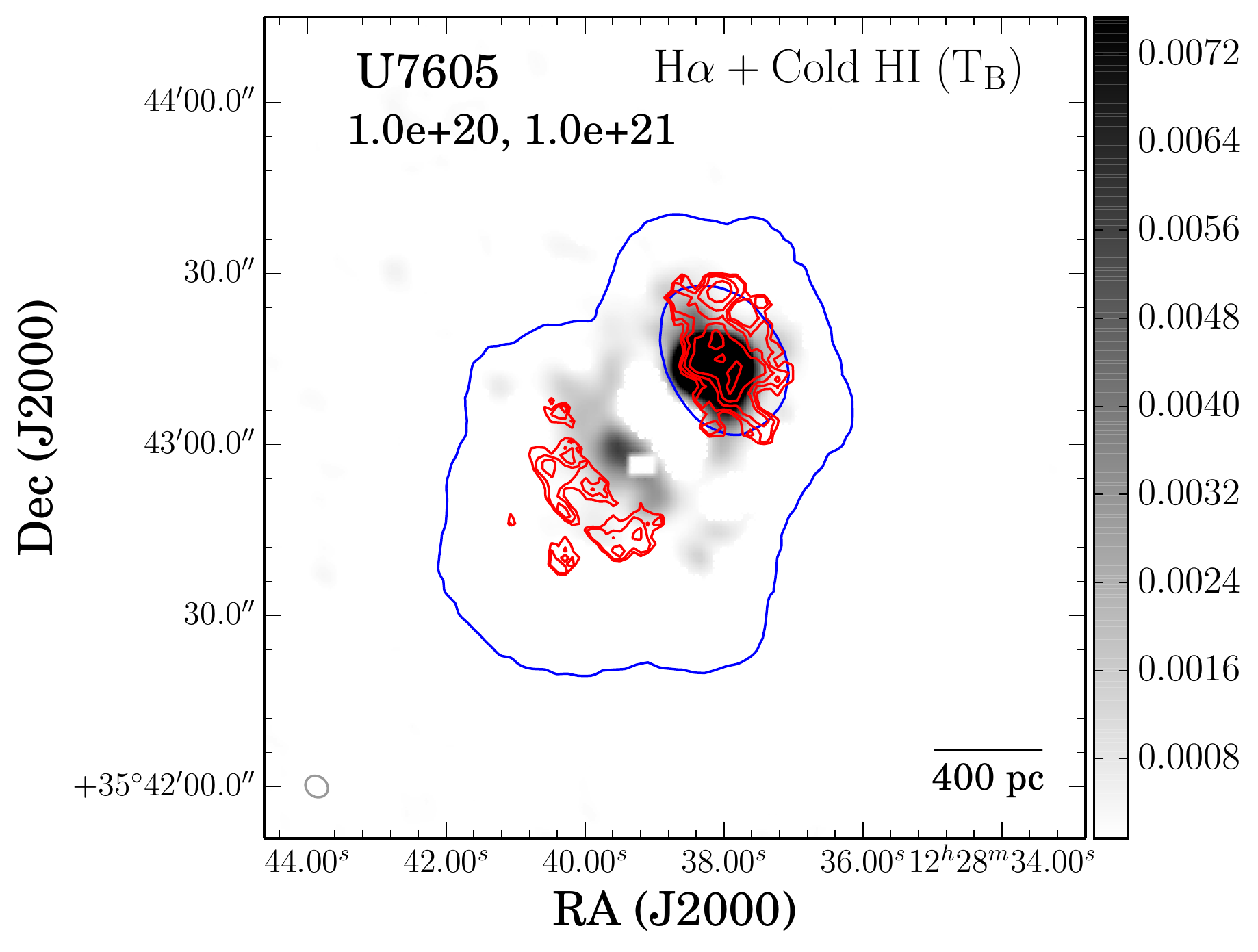}} &
\resizebox{55mm}{!}{\includegraphics{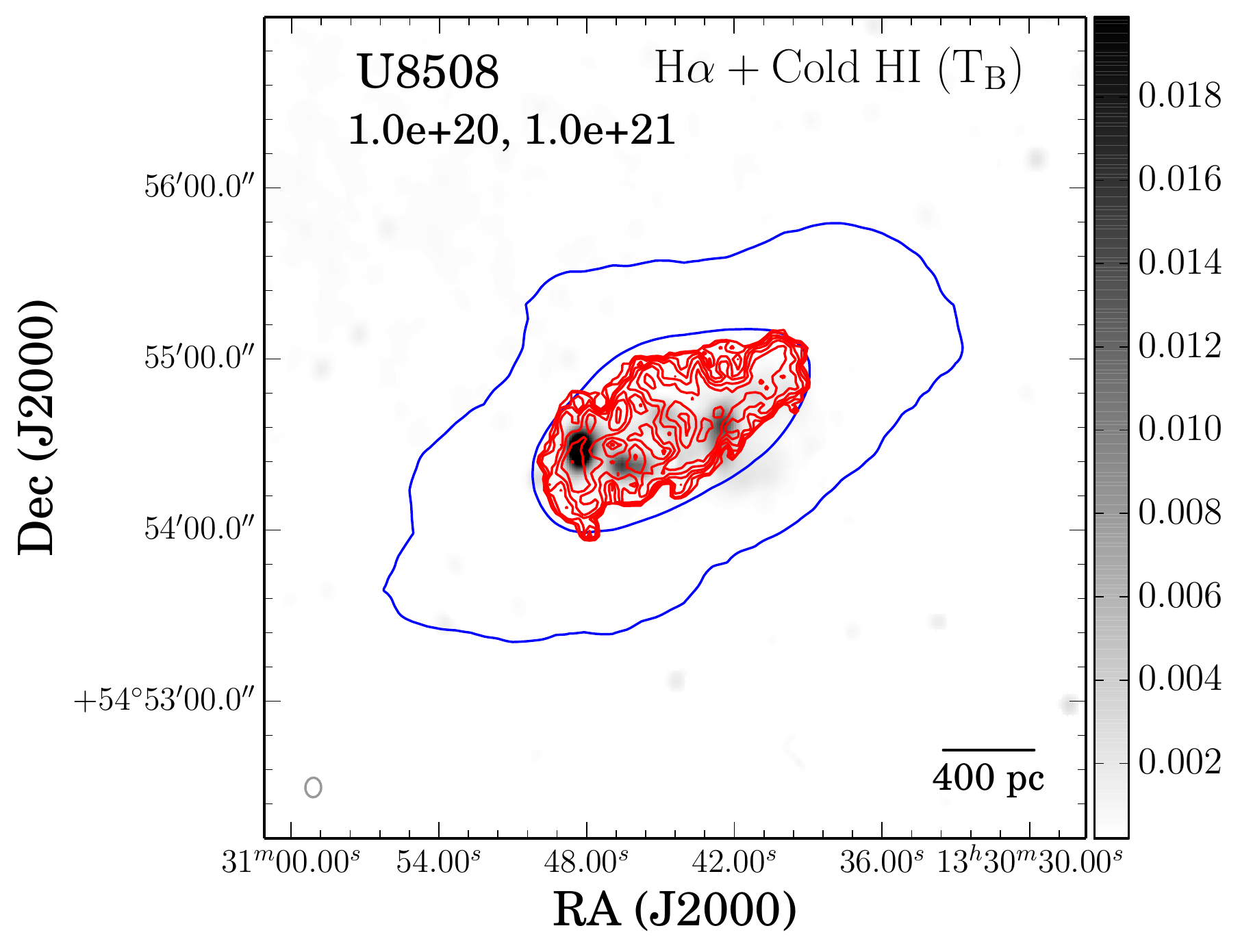}} &
\resizebox{55mm}{!}{\includegraphics{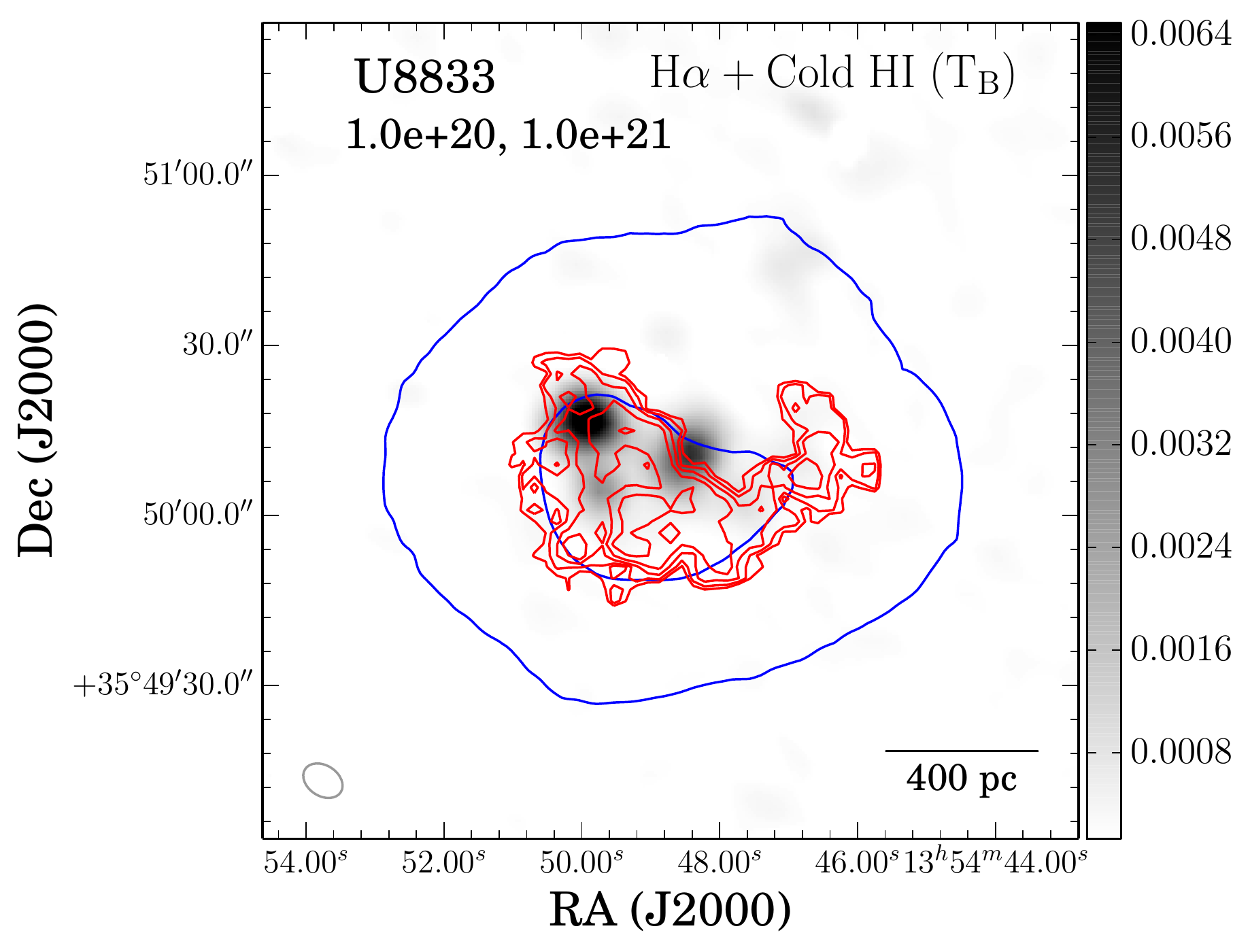}} \\

\resizebox{55mm}{!}{\includegraphics{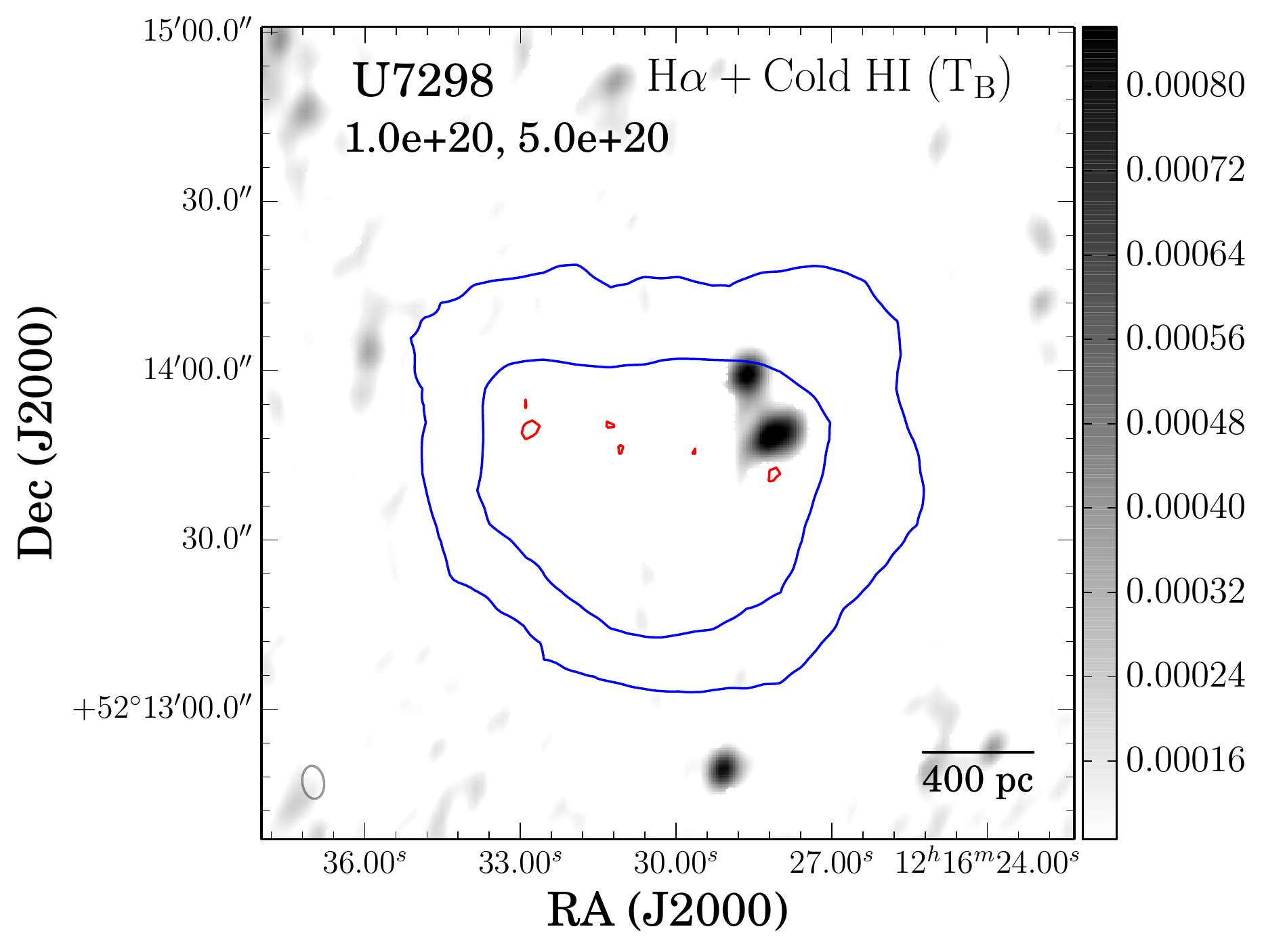}} &
\resizebox{55mm}{!}{\includegraphics{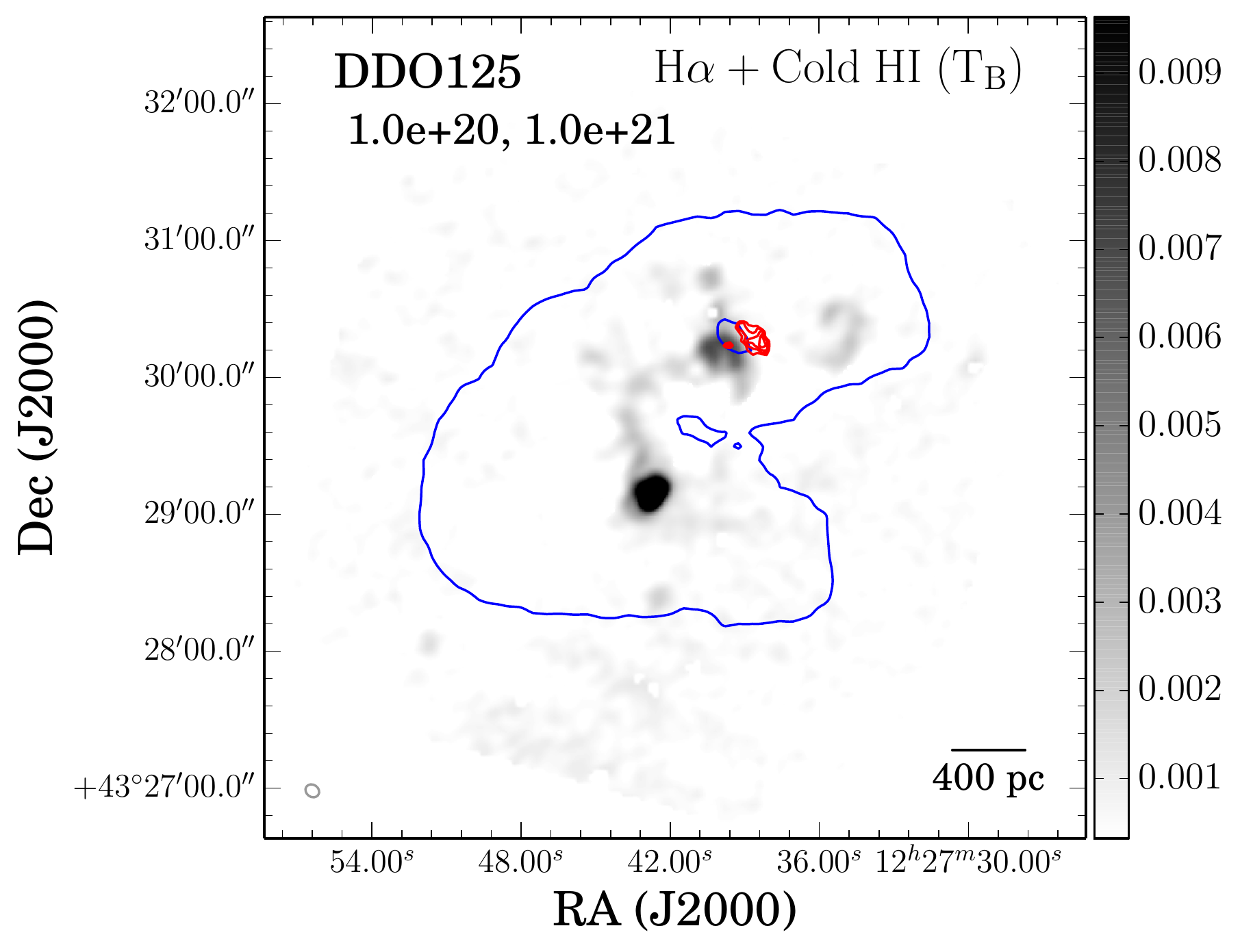}} &
\resizebox{55mm}{!}{\includegraphics{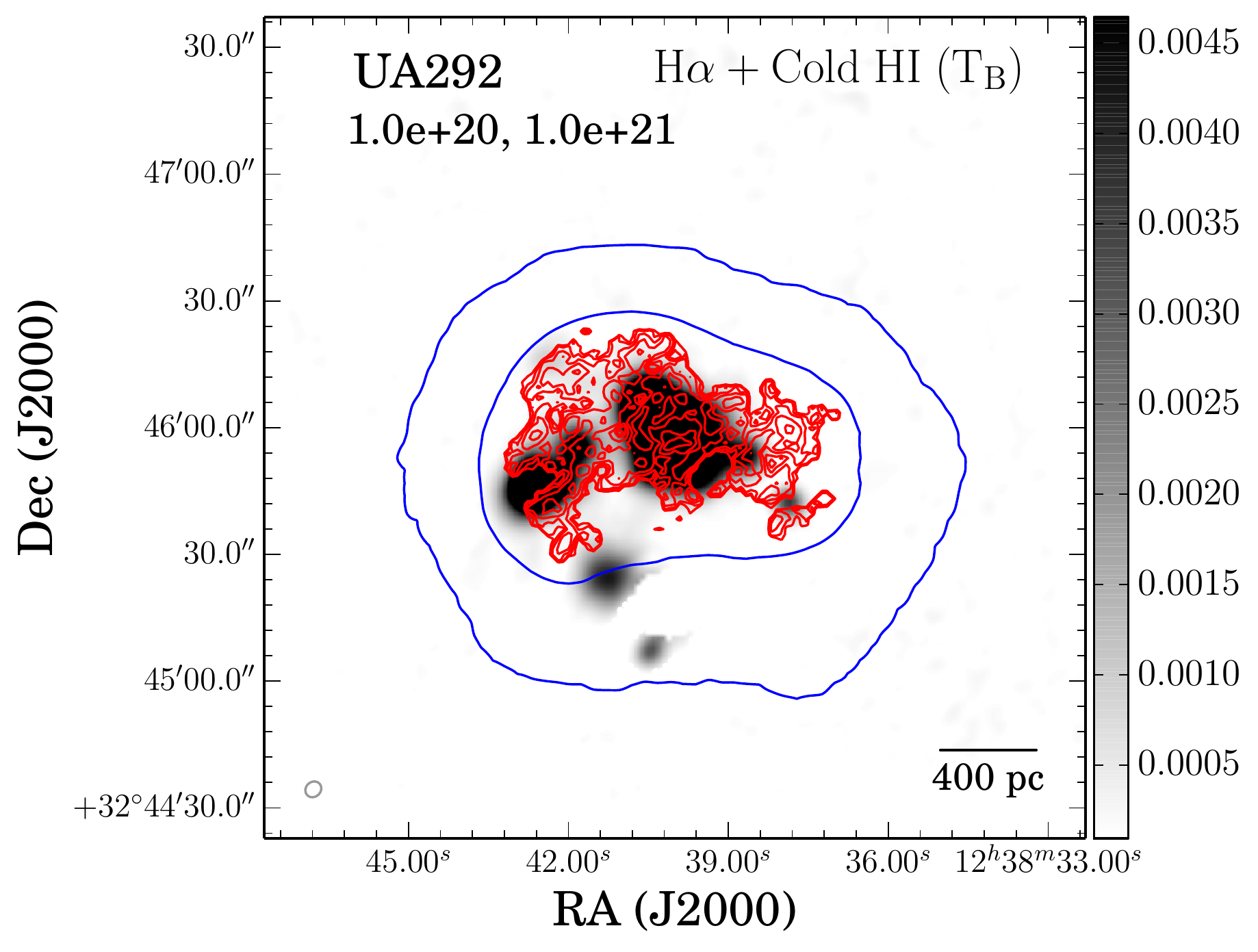}} \\

\end{tabular}
\end{center}
\caption{ Overlays of cold \HI detected using the \tb method (red contours) on \ha star formation rate density (gray scale). Color bars are in the units of \msyrkpc. The blue contours are the same as in Fig.~\ref{ovr_h1_g}. The beam at the bottom left corner of every panel represents the resolution of the cold \HI map.}
\label{ovr_ha_tb}
\end{figure*}

\begin{figure*}
\begin{center}
\begin{tabular}{ccc}
\resizebox{55mm}{!}{\includegraphics{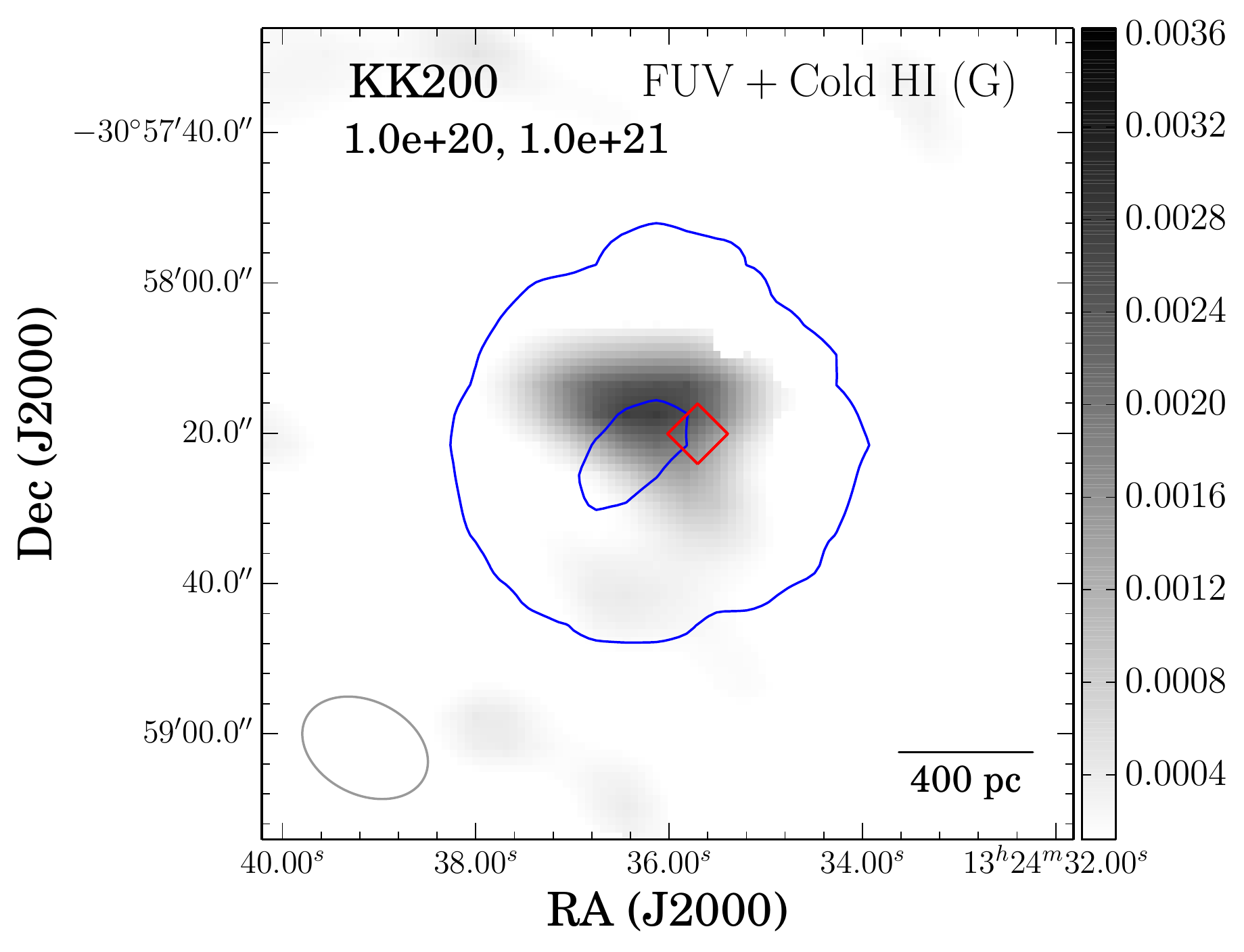}} &
\resizebox{55mm}{!}{\includegraphics{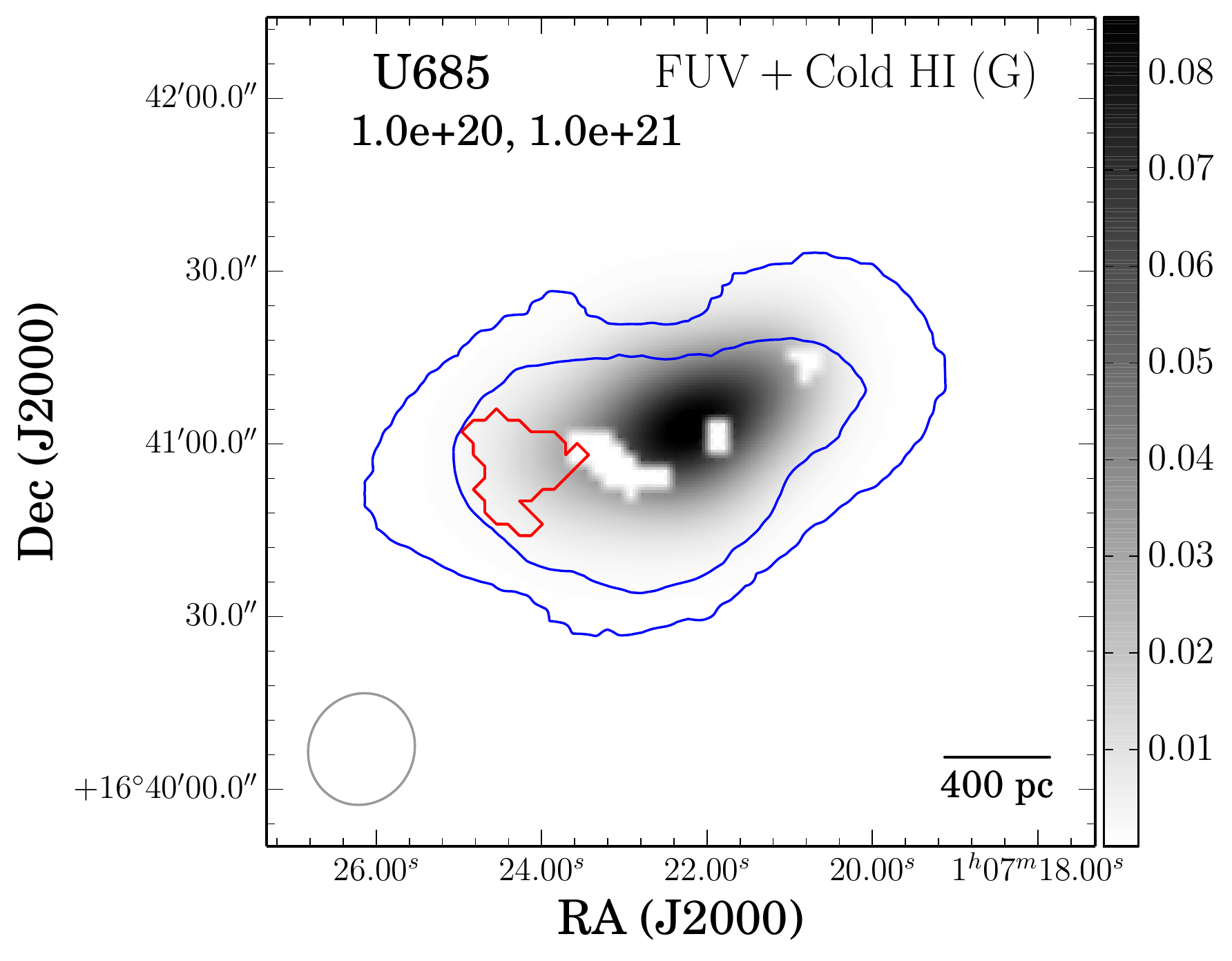}} &
\resizebox{55mm}{!}{\includegraphics{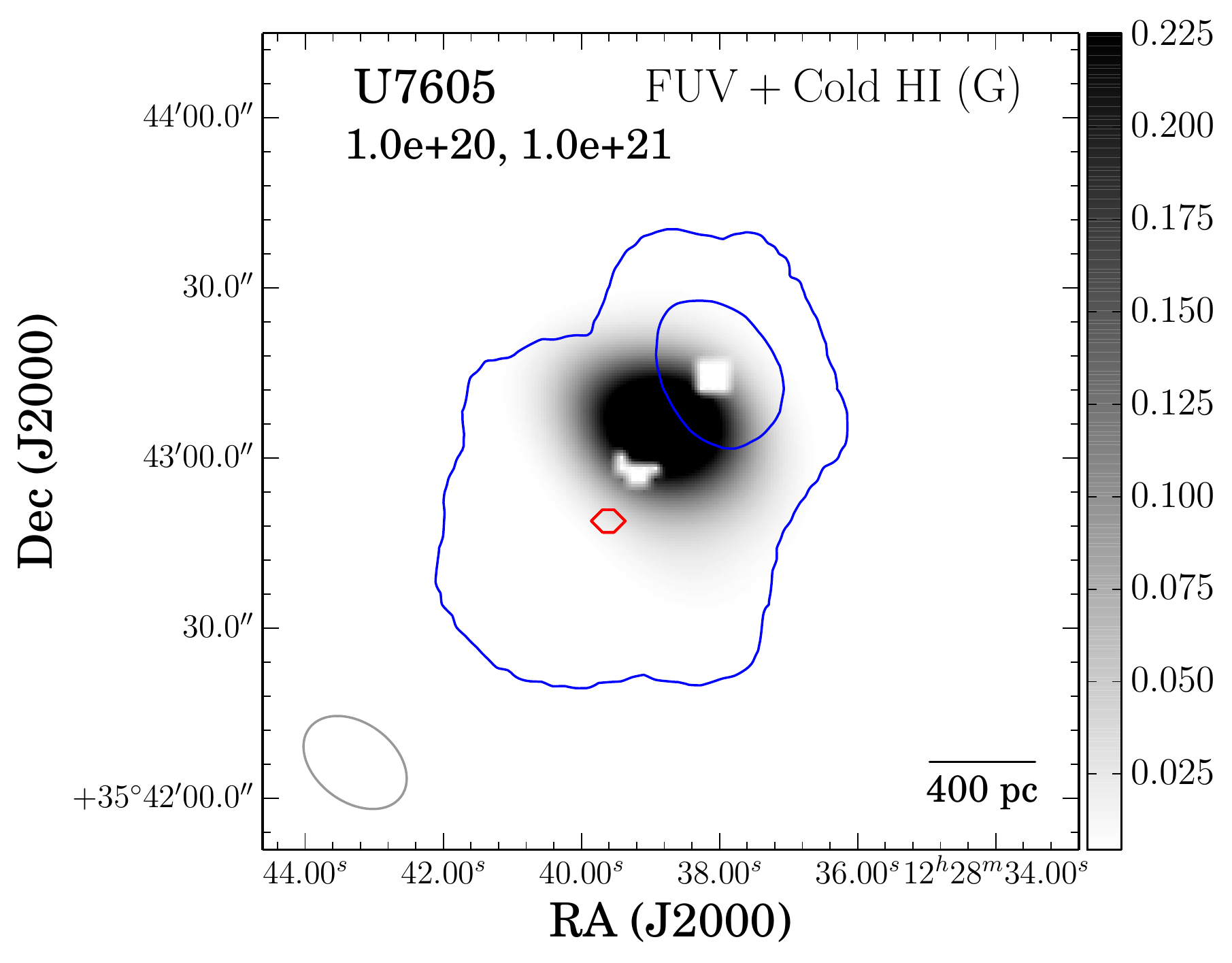}} \\

\resizebox{55mm}{!}{\includegraphics{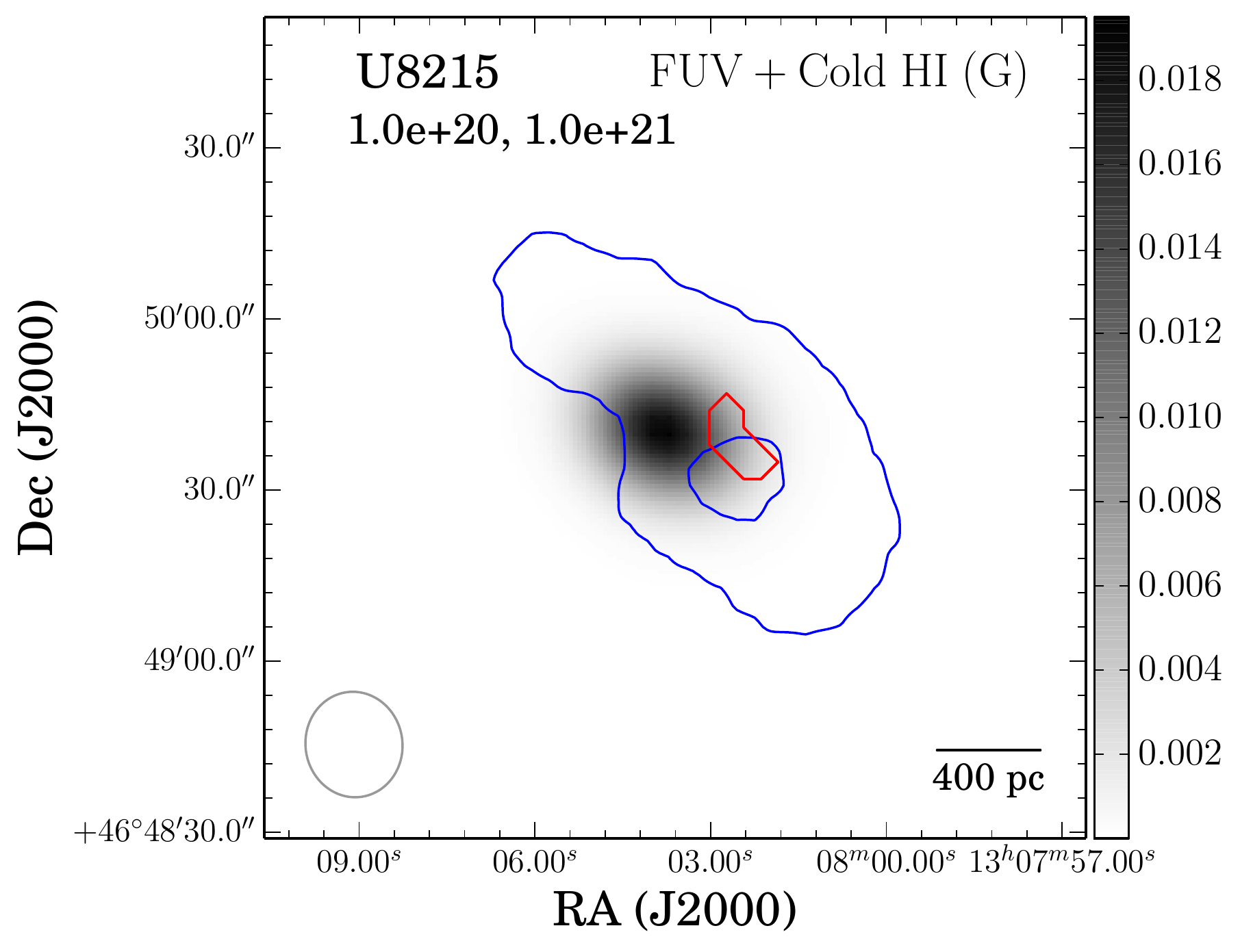}} \\

\end{tabular}
\end{center}
\caption{ Overlays of cold \HI detected using the Gaussian decomposition method (red contours) on FUV star formation rate density (gray scale). Color bars are in the units of \msyrkpc. The blue contours are the same as in Fig.~\ref{ovr_h1_g}.}
\label{ovr_fuv_gauss}
\end{figure*}

\begin{figure*}
\begin{center}
\begin{tabular}{ccc}
\resizebox{55mm}{!}{\includegraphics{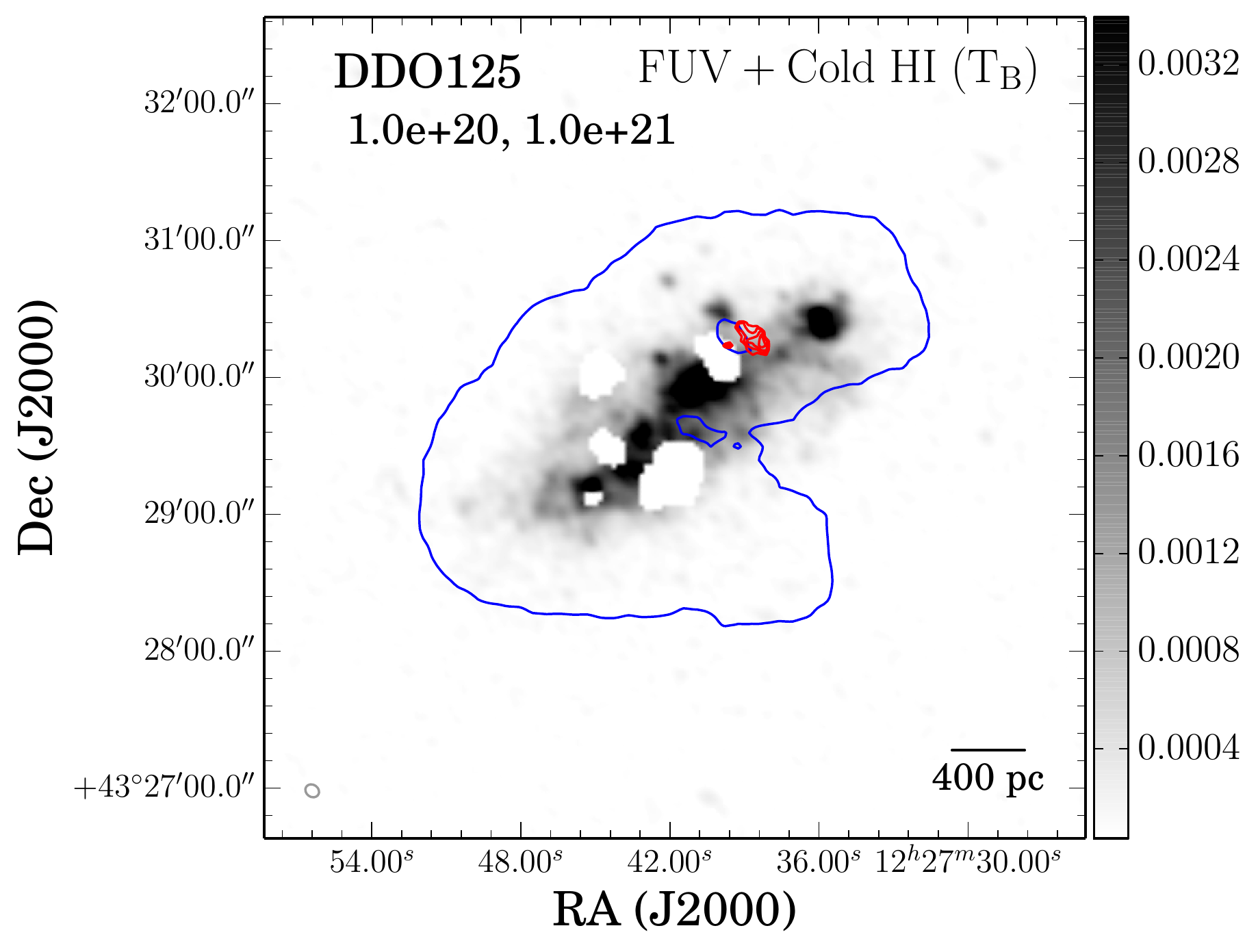}} &
\resizebox{55mm}{!}{\includegraphics{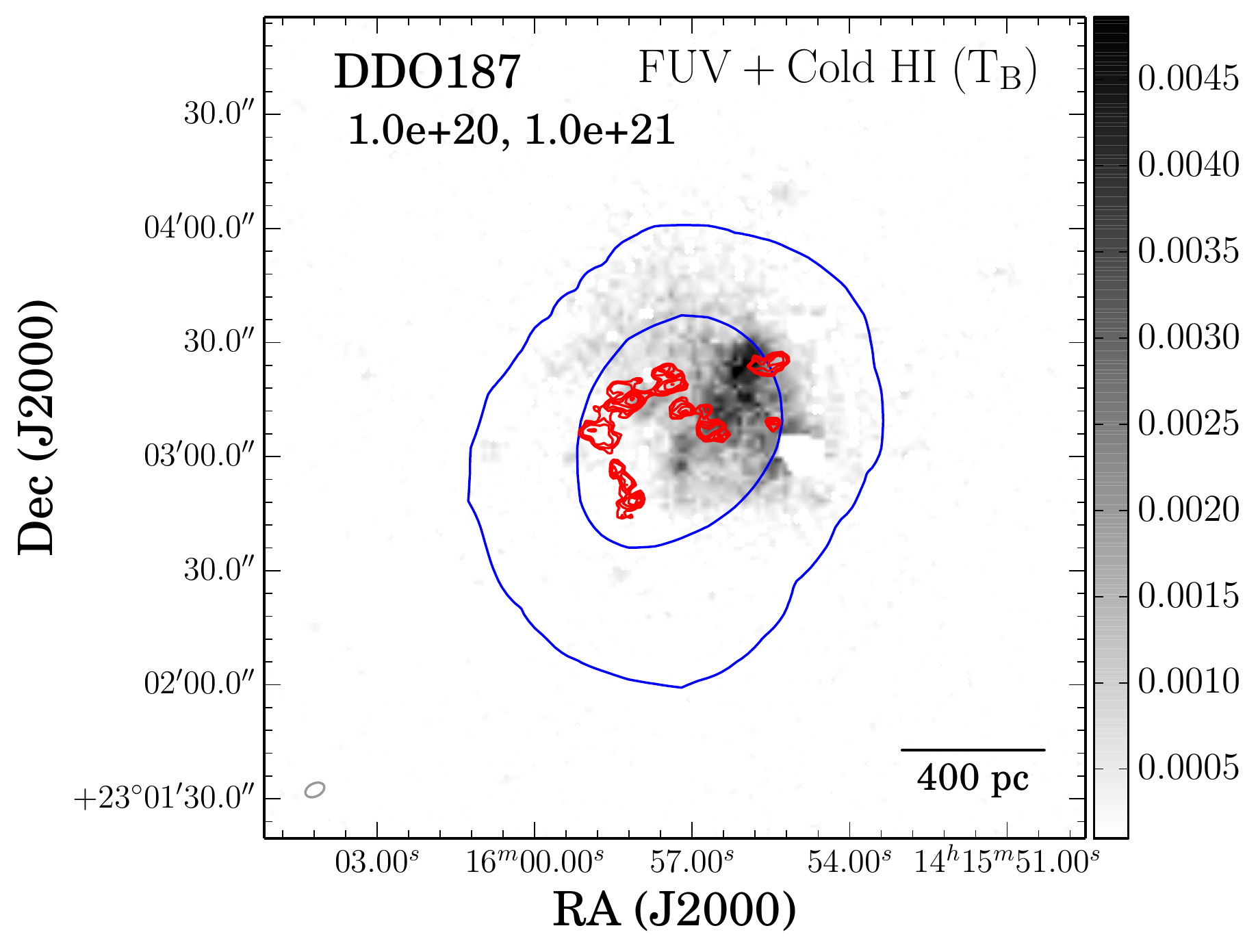}} &
\resizebox{55mm}{!}{\includegraphics{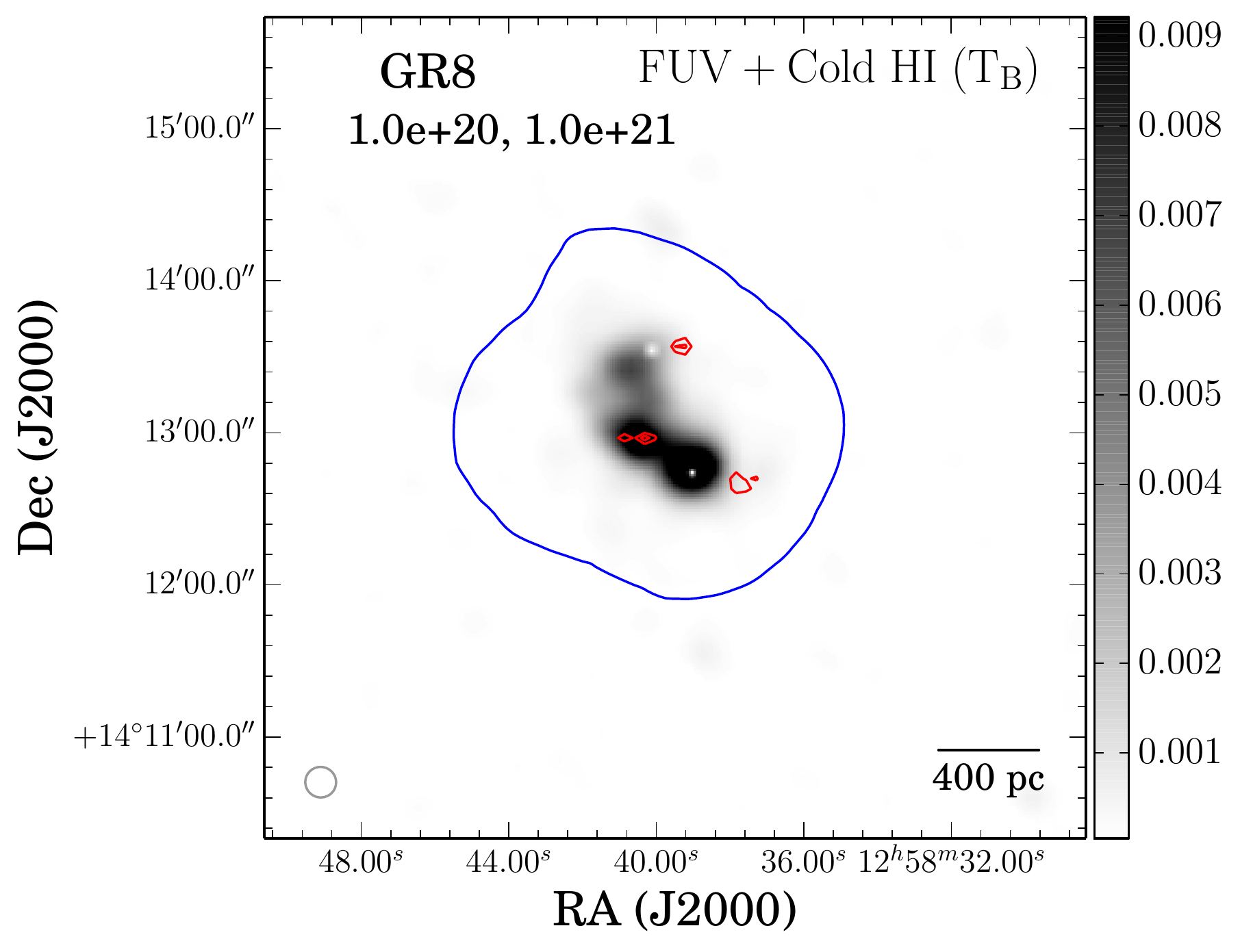}} \\

\resizebox{55mm}{!}{\includegraphics{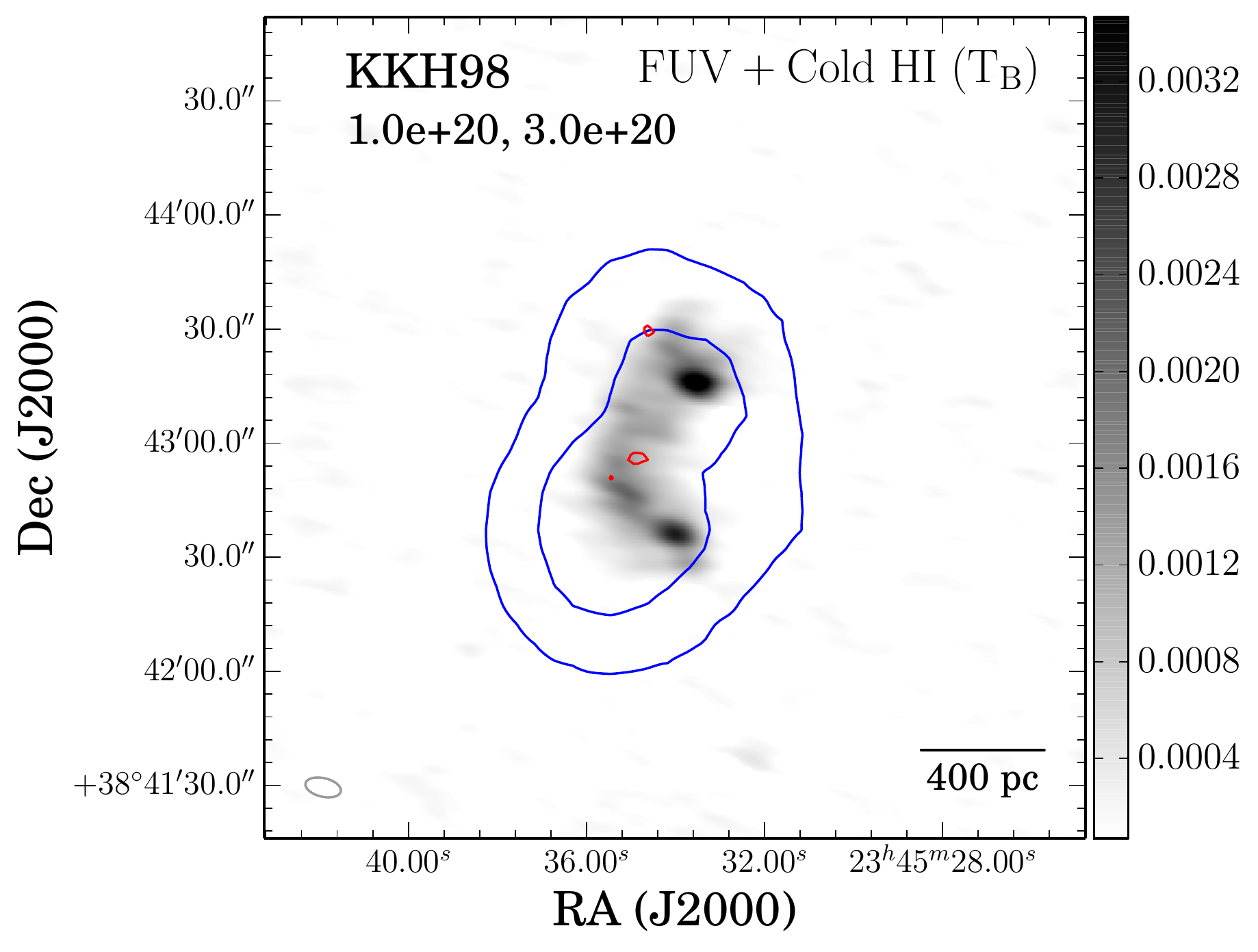}} &
\resizebox{55mm}{!}{\includegraphics{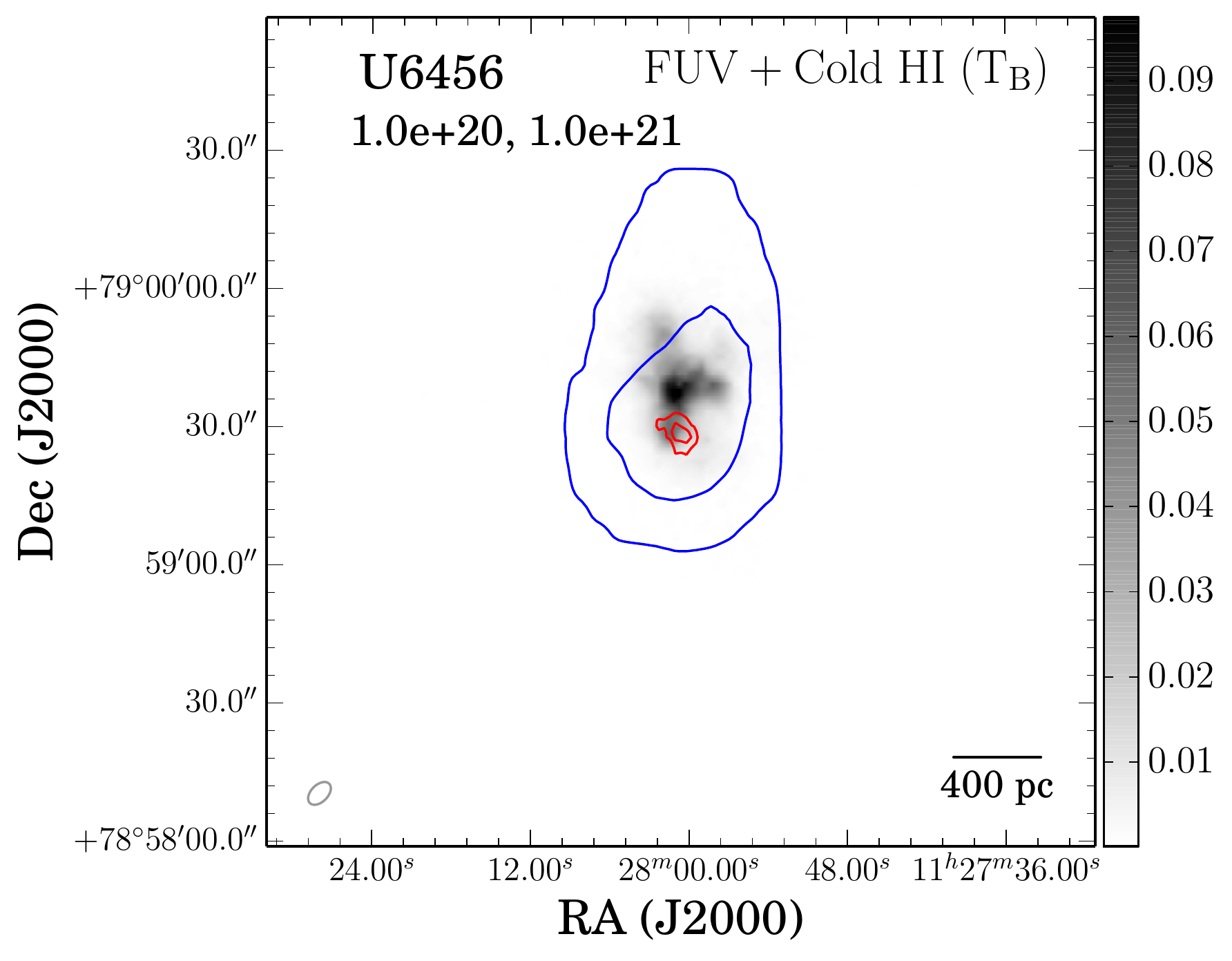}} &
\resizebox{55mm}{!}{\includegraphics{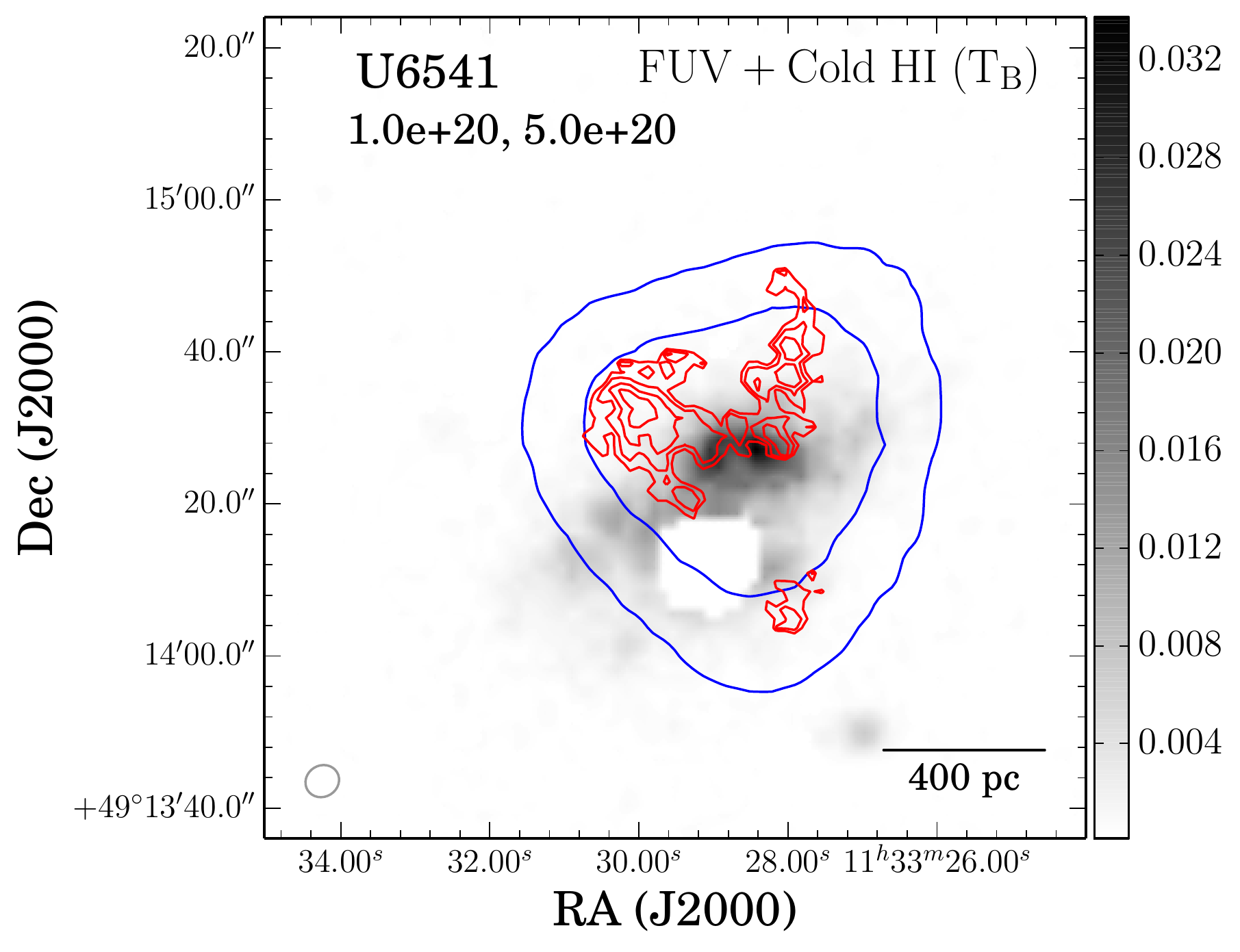}} \\

\resizebox{55mm}{!}{\includegraphics{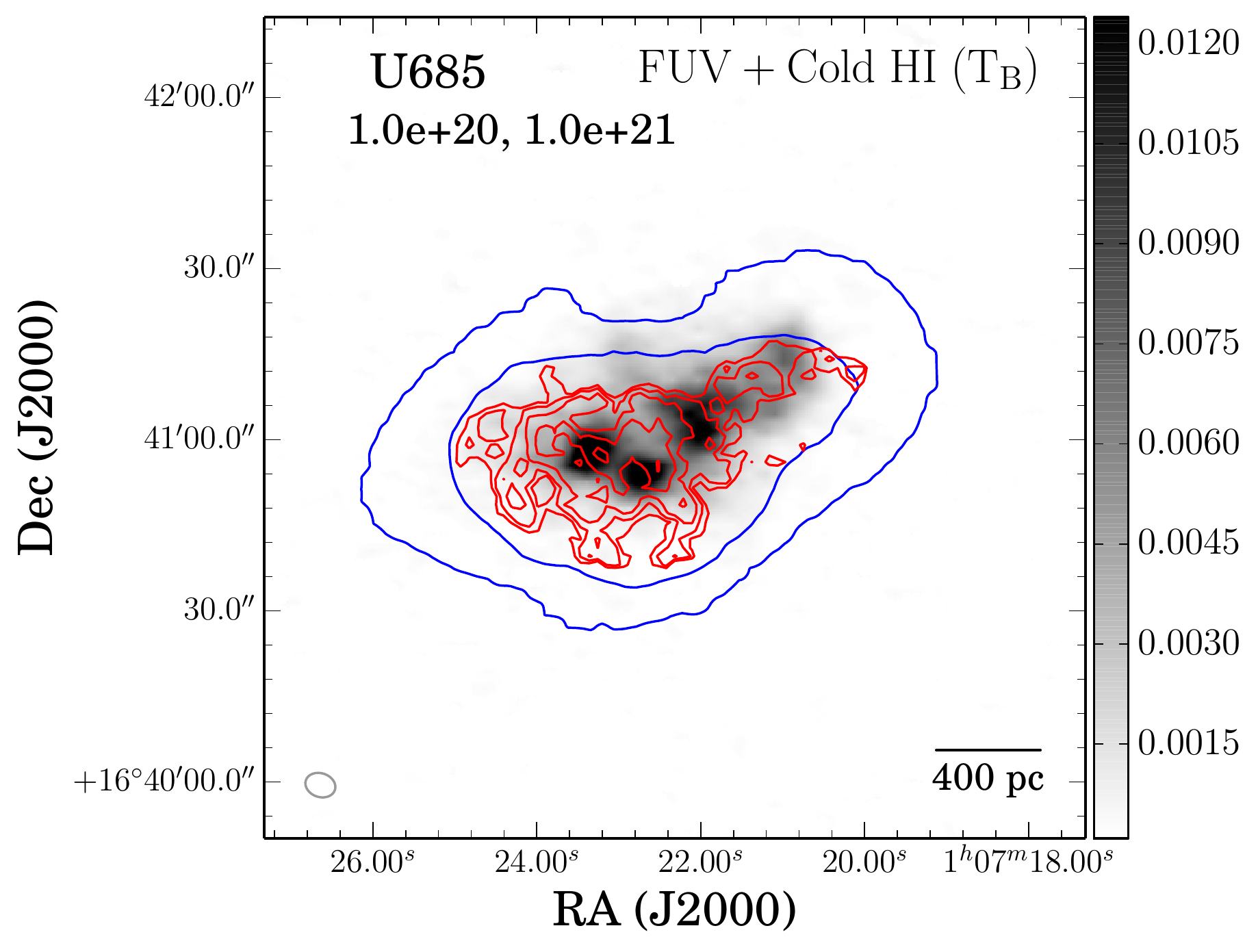}} &
\resizebox{55mm}{!}{\includegraphics{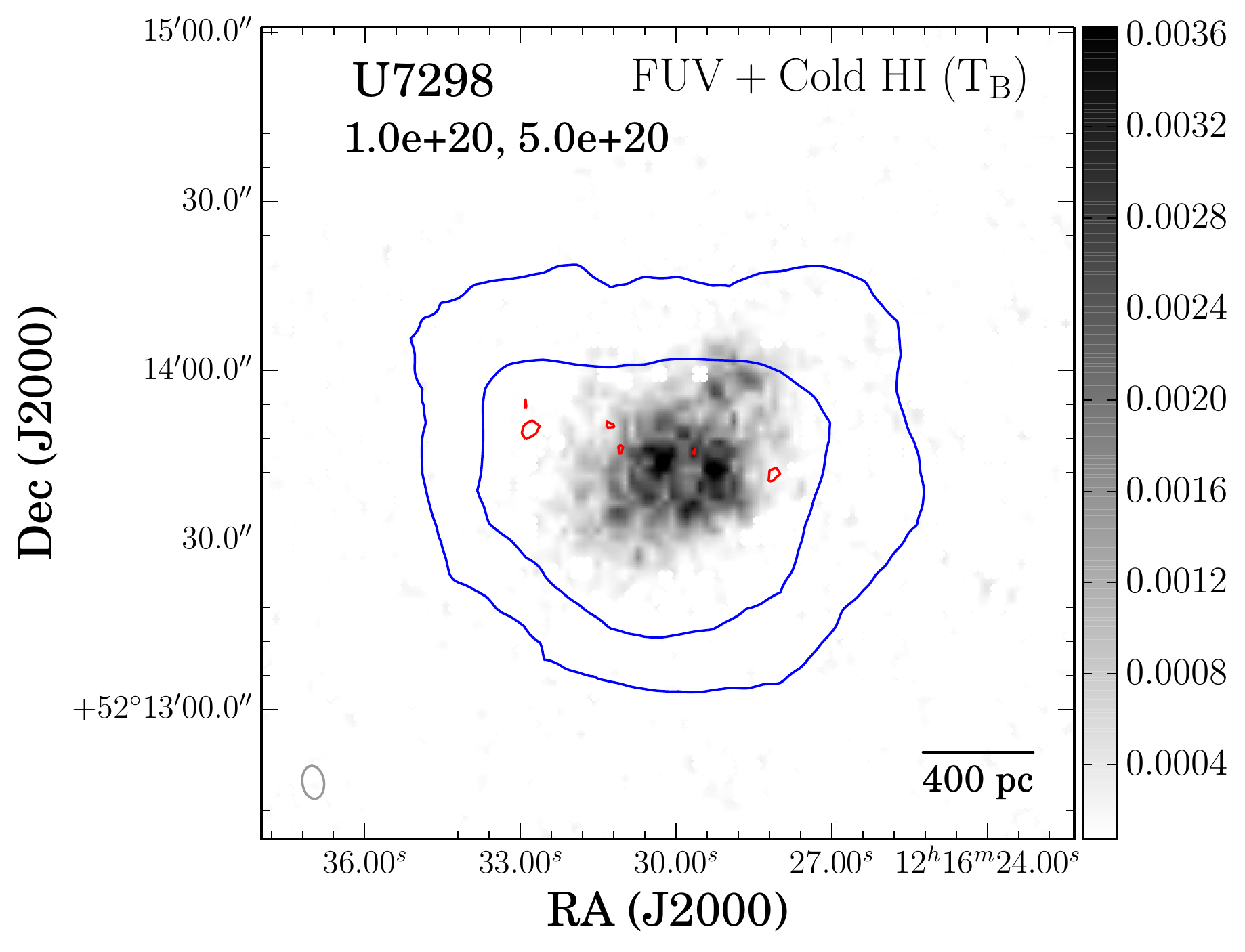}} &
\resizebox{55mm}{!}{\includegraphics{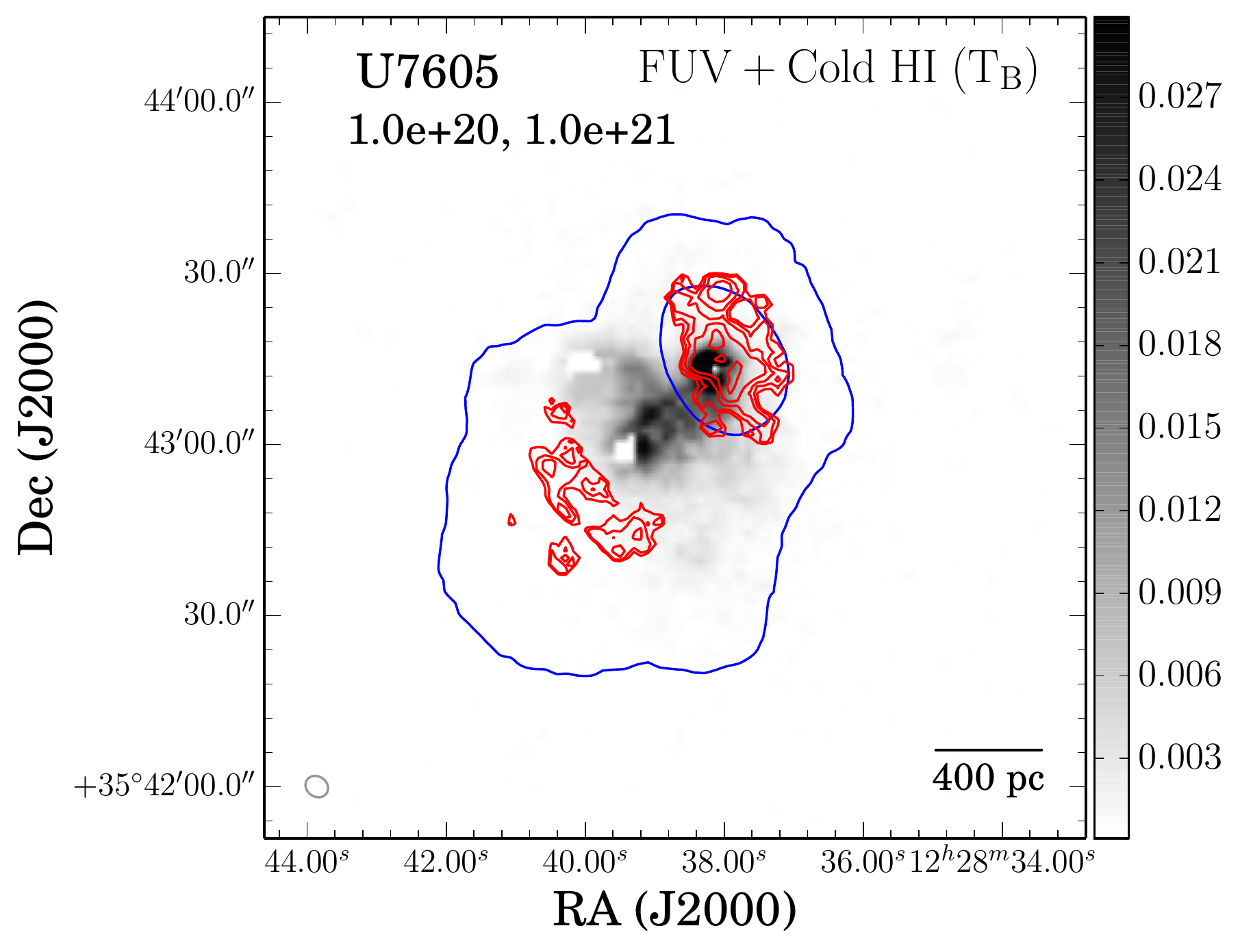}} \\

\resizebox{55mm}{!}{\includegraphics{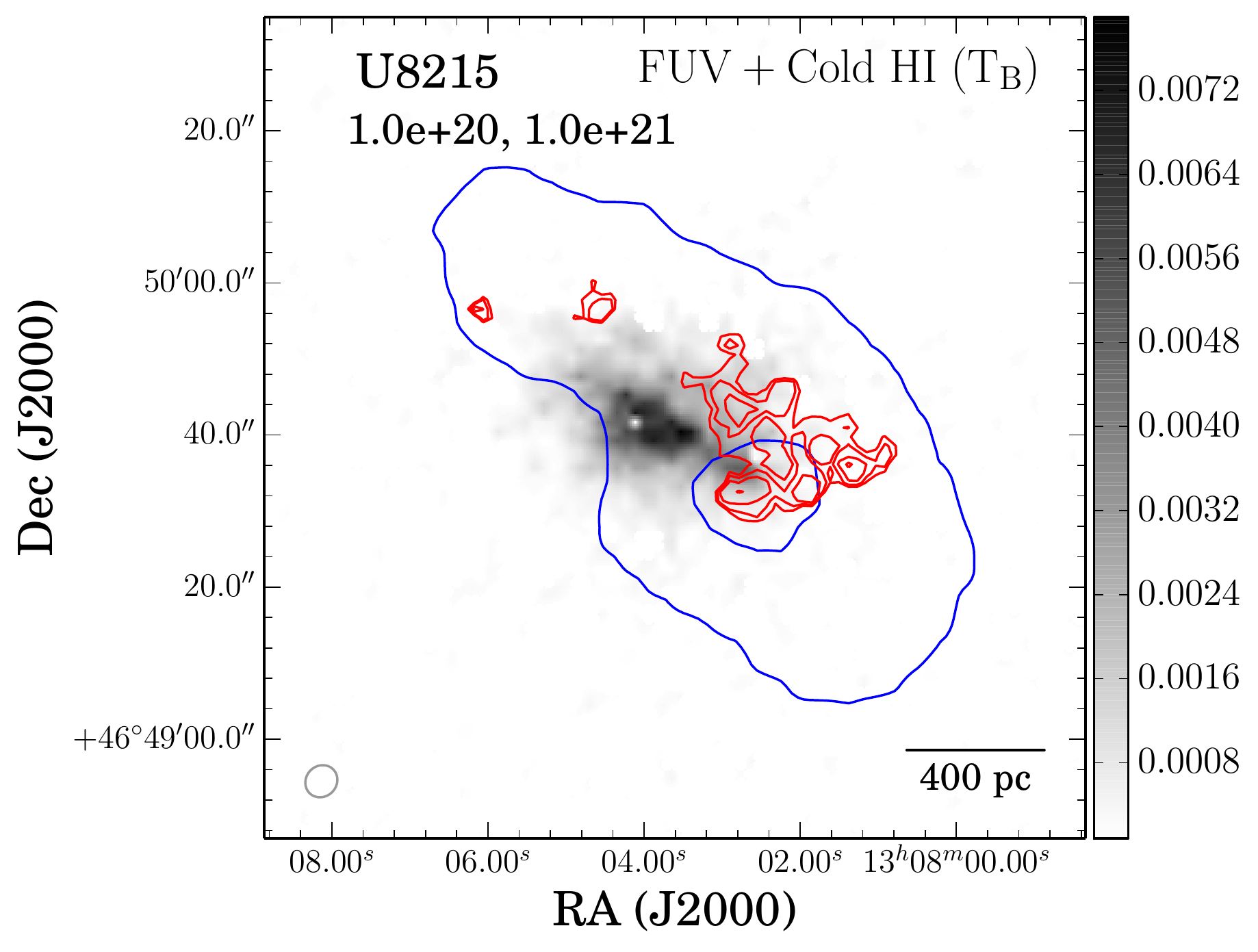}} &
\resizebox{55mm}{!}{\includegraphics{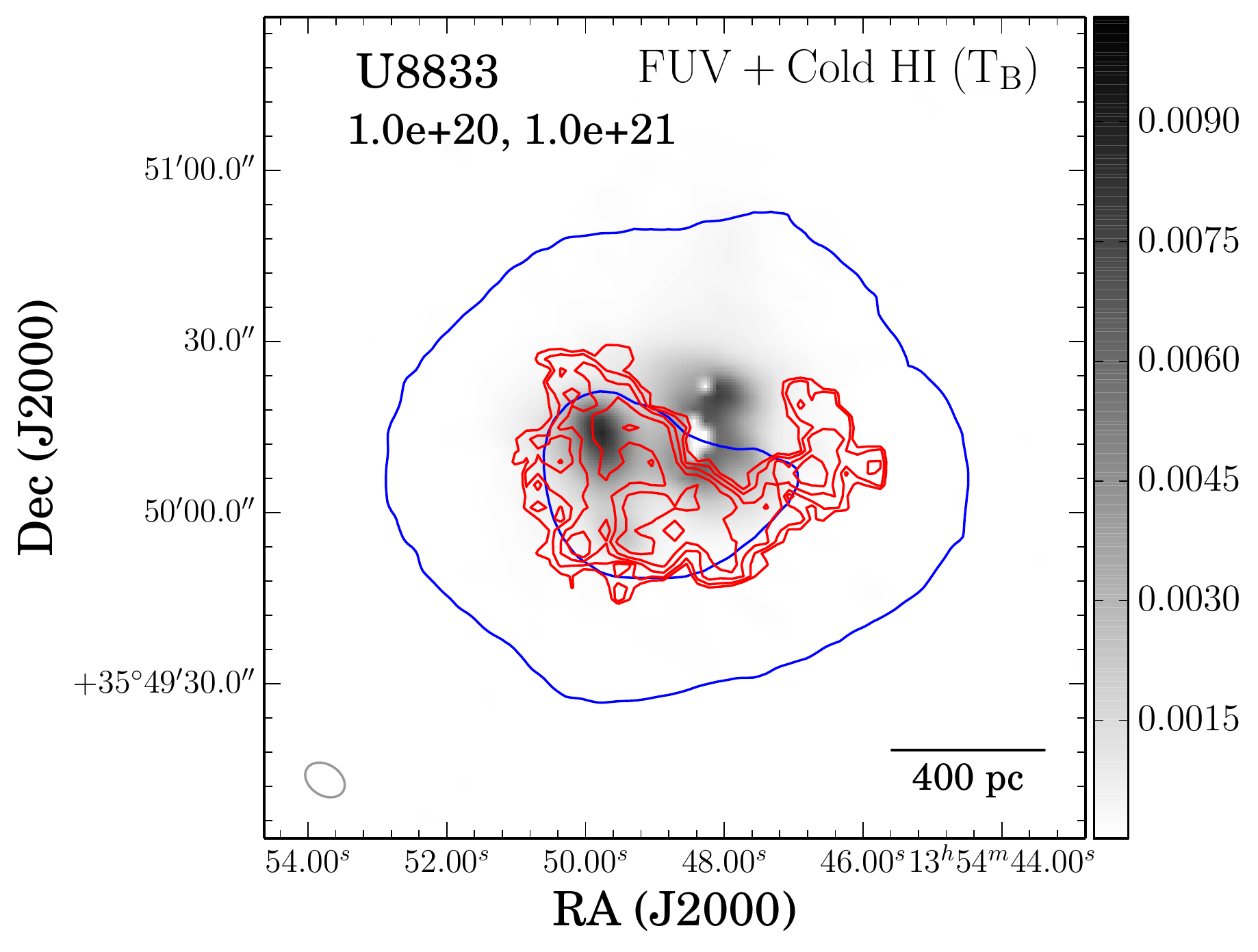}} &
\resizebox{55mm}{!}{\includegraphics{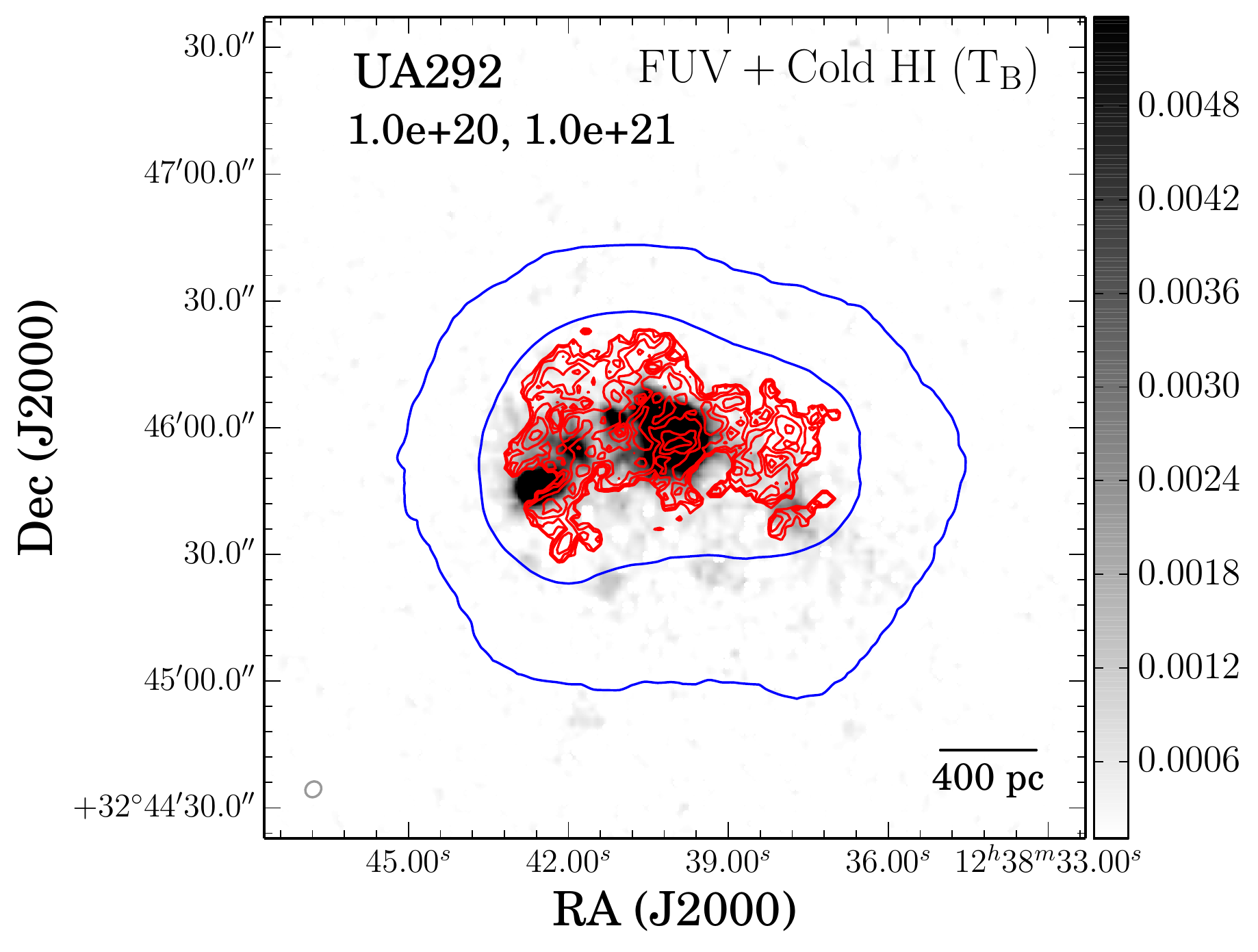}} \\

\end{tabular}
\end{center}
\caption{ Overlays of cold \HI detected using the \tb method (red contours) on FUV star formation rate density (gray scale). Color bars are in the units of \msyrkpc. The blue and red contours are the same as in Fig.~\ref{ovr_ha_tb}. The beam at the bottom left corner of every panel represents the resolution of the cold \HI map.}
\label{ovr_fuv_tb}
\end{figure*}

\end{document}